\documentclass[aps,prd,reprint,11pt,onecolumn,amsmath,amssymb,nofootinbib,groupedaddress,superscriptaddress,floatfix,tightenlines,eqsecnum]{revtex4-2}

\usepackage{macros}
\graphicspath{{./figs/}}

\begin{document}

\title{Dark forces suppress structure growth}

\author{Marco Costa}\email[]{mcosta1@perimeterinstitute.ca}
\affiliation{Perimeter Institute for Theoretical Physics, Waterloo, Ontario N2L 2Y5, Canada}

\author{Cyril Creque-Sarbinowski}\email[]{ccsb2025@pomona.edu}
\affiliation{Center for Computational Astrophysics, Flatiron Institute, 162 5th Avenue, New York, New York 10010, USA}
\affiliation{Department of Physics, University of Washington, Seattle, Washington, 98195, USA}
\affiliation{Department of Physics and Astronomy, Pomona College, Claremont, CA 91711, USA}

\author{Olivier Simon}\email[]{osimon@princeton.edu}
\affiliation{Princeton Center for Theoretical Science, Princeton University, Princeton, New Jersey, 08544, USA}
\affiliation{Department of Physics, Princeton University, Princeton, New Jersey, 08544, USA}

\author{Zachary J. Weiner}\email[]{zweiner@perimeterinstitute.ca}
\affiliation{Perimeter Institute for Theoretical Physics, Waterloo, Ontario N2L 2Y5, Canada}

\date{\today}
\begin{abstract}
No experimental test precludes the possibility that the dark matter experiences forces beyond
general relativity---in fact, a variety of cosmic microwave background observations suggest greater
late-time structure than predicted in the standard $\Lambda$ cold dark matter model.
We show that minimal models of scalar-mediated forces between dark matter particles do not enhance
the growth of unbiased tracers of structure: weak lensing observables depend on the total density
perturbation, for which the enhanced growth of the density contrast in the matter era is cancelled
by the more rapid dilution of the background dark matter density.
Moreover, the same background-level effects imply that scenarios compatible with CMB temperature and
polarization anisotropies in fact \textit{suppress} structure growth, as fixing the distance to last
scattering requires a substantially increased density of dark energy.
Though massive mediators undo these effects upon oscillating, they suppress structure even further
because their gravitational impact as nonclustering subcomponents of matter outweighs the enhanced
clustering strength of dark matter.
We support these findings with analytic insight that clarifies the physical impact of dark forces
and explains how primary CMB measurements calibrate the model's predictions for low-redshift
observables.
We discuss implications for neutrino mass limits and other cosmological anomalies, and we also
consider how nonminimal extensions of the model might be engineered to enhance structure.
\end{abstract}

\maketitle
\makeatletter
\def\l@subsubsection#1#2{}
\makeatother
\tableofcontents

\section{Introduction}
\label{sec:introduction}

In the absence of direct evidence for nongravitational interactions of dark matter with the Standard
Model, insight into its fundamental nature may derive from searches for dynamics beyond the
predictions of cold dark matter.
Evidence for nonminimal dynamics from cosmological and astrophysical observations---those that
motivate dark matter's existence~\cite{Bertone:2004pz, Bertone:2016nfn, Buckley:2017ijx,
Bertone:2018krk, Cirelli:2024ssz}---would suggest an underlying particle nature and inform the
details of its microphysical description, such as the degrees of freedom involved and their masses,
spins, and initial conditions.
Such evidence must be interpreted in the context of the broader cosmological model, including both
the hallmark successes of the $\Lambda$ cold dark matter (\LCDM{}) paradigm and any of its
deficiencies in explaining contemporary observations.

From a phenomenological point of view, models featuring nonstandard dynamics after photon-baryon
decoupling are of contemporary interest due to an influx of precision observations of the geometry
and structure of the Universe at low redshift.
Moreover, an increasing number of such probes exhibit tension with the inference of \LCDM{}
cosmology from cosmic microwave background (CMB) temperature and polarization anisotropies.
The lensing of the CMB, for instance, indicates a higher amplitude of structure at moderate
redshifts, both as inferred via its effect on two-point statistics~\cite{Planck:2018vyg,
ACT:2025fju, SPT-3G:2025bzu} and from direct reconstruction from higher-point
statistics~\cite{Planck:2018lbu, Carron:2022eyg, ACT:2023kun, SPT-3G:2024atg, ACT:2025qjh}.
Baryon acoustic oscillation (BAO) data additionally infer that the Universe was less dense late in
the matter-dominated epoch than what \LCDM{} extrapolates from the recombination
epoch~\cite{DESI:2025zgx, Loverde:2024nfi, Lynch:2025ine}.

Though these tensions are more moderate (and more recent) than the Hubble
tension~\cite{Riess:2016jrr, Riess:2019cxk, Riess:2021jrx, Freedman:2019jwv, Freedman:2020dne,
Freedman:2021ahq, Freedman:2023jcz}, they derive from measurements that are thought to be less
susceptible to uncertainty in empirically calibrated astrophysical modeling; independent surveys of,
e.g., CMB lensing are also mutually concordant~\cite{SPT-3G:2024atg}.
Intriguingly, a larger lensing amplitude would also reduce the matter density deficit between CMB
and BAO data via correlated shifts in the CMB's inferred density in cold dark
matter~\cite{Loverde:2024nfi, Lynch:2025ine}.
Calibrating the amplitude of the primordial power spectrum with the primary CMB, however, does
require measuring the optical depth of photons in the late Universe due to reionization, which is
currently achieved via \Planck{} observations of large-scale polarization.
Recent work has pointed out that a larger optical depth could explain the lensing excess, but the
requisite values would amount to a $5 \sigma$ upward fluctuation from current
measurements~\cite{Craig:2024tky, Loverde:2024nfi, Sailer:2025lxj, Jhaveri:2025neg}.
Though predictions of the signal generated during reionization require a model of the ionization
history, inference of the optical depth itself appears to be highly insensitive to such
details~\cite{Heinrich:2021ufa, Ilic:2025idl, Cheng:2025cmb}; moreover, the \Planck{} measurement is
consistent with independent astrophysical ones~\cite{Mason:2019oeg, Paoletti:2024lji, Cain:2025usc,
Sims:2025hfm, Elbers:2025xvk}.

Given the observational motivation for increasing the inferred degree of CMB lensing, consistently
assessing candidate microphysical models is imperative.
Conversely, a lack of such models might further motivate consideration of whether or how the optical
depth could be so severely miscalibrated, even in spite of intensive prior
scrutiny~\cite{Planck:2019nip,Planck:2018vyg}, or of other potential systematics.
In particular, the common consistency test of a phenomenological rescaling of the lensing
amplitude~\cite{Calabrese:2008rt, Craig:2024tky, Green:2024xbb} is unlikely to offer a reasonable
proxy for actual models (other than a miscalibrated optical depth, perhaps), given that there is no
reason \emph{a priori} that nonnegligible modifications to structure growth would not be
accompanied by commensurate changes to the expansion history.
Moreover, current data measure low-redshift distances more precisely than the amplitude of late-time
structure.

The possibility that dark matter is subject to an additional long-range force (LRF or ``dark
force''), as postulated by Ref.~\cite{Green:2024xbb} to explain the lensing excess, indeed
nonnegligibly affects the expansion history~\cite{Archidiacono:2022iuu, Bottaro:2023wkd,
Bottaro:2024pcb}.
Refs.~\cite{Archidiacono:2022iuu,Bottaro:2024pcb}, which considered a minimal model of a light (or
massless) scalar mediating a Yukawa force between dark matter particles, observed that background
effects offset the impact of enhanced clustering on the CMB lensing spectrum.
In this work, we demonstrate that in the minimal case this cancellation is in fact exact (at leading
order in the force strength).
Using model-agnostic analytic solutions, we explain why dark forces do not modify the growth rate of
lensing potentials in this case in \cref{sec:structure-growth}.

Furthermore, we show in \cref{sec:suppression} that for model parameters that respect the
best-measured features in the primary CMB anisotropies---namely, the angular extent of the sound
horizon and the dark matter density around recombination---the same modifications to the expansion
history in fact suppress CMB lensing.
Along the way, we demonstrate, both analytically and numerically, that the generation of small-scale
anisotropies at last scattering is insensitive to the modified dynamics of dark matter density
perturbations, building on prior results for cold~\cite{Weinberg:2002kg, Weinberg:2008zzc} and
warm~\cite{Voruz:2013vqa} dark matter.
The primary CMB itself is thus also most sensitive to the modified background evolution of dark
matter.

Building on these theoretical developments, \cref{sec:extrapolation} compares the predictions of the
model, as calibrated to primary CMB data, to CMB lensing measurements and low-redshift BAO
distances.
We consider the impact of nonzero mediator masses and more nonminimal extensions in
\cref{sec:nonminimal}.
\Cref{sec:mnu,sec:cosmic-shear} discuss the implications of our results for cosmological
measurements of neutrino masses and other extant tensions, and \cref{sec:nonlinear} comments on
future directions beyond the regime of linear perturbation theory.
We summarize our results and conclude in \cref{sec:conclusions}.
Finally, \cref{app:eoms} summarizes our notation and formalism, \cref{app:supplementary-results}
presents supplementary results, \cref{app:relic-abundance} analytically computes the relic abundance
of hyperlight scalars linearly coupled to dark matter that begin to oscillate in the matter era, and
\cref{app:numerics} enumerates technical details of our numerical implementation.

\section{Structure growth with long-range forces}
\label{sec:structure-growth}

We begin by outlining the general theory of nonrelativistic particles $\chi$ that experience a
long-range force mediated by a real scalar $\varphi$.
\Cref{sec:formalism} lays out the general formalism, with details relegated to \cref{app:eoms}, and
\cref{sec:subhorizon-growth} discusses the appropriate limit to analytically understand structure
growth in the matter-dominated era.
In \cref{sec:linear-coupling} we then show that for the minimal case of a linearly coupled, massless
mediator, absolute (rather than relative) fluctuations in energy density grow no faster than in
\LCDM{}, meaning the long-range force on its own does not enhance the growth of lensing potentials
(nor source any integrated Sachs-Wolfe [ISW] effect).
We identify precisely which physical effects cancel, guiding the engineering of nonminimal
extensions to modify this result as explored in \cref{sec:nonminimal}.

\subsection{Scalar-mediated forces in kinetic theory}
\label{sec:formalism}

We describe dark matter as a collection of point particles in kinetic theory, which provides an
appropriate description for a cold species whether it is bosonic or
fermionic~\cite{Archidiacono:2022iuu} (or even if an imperfect fluid with pressure and shear
stress).
Field-theoretic effects, such as $\chi$ having a macroscopic de Broglie wave length or itself
mediating forces for $\varphi$, are not captured by this description, but they could be treated with
a dedicated effective theory of fluids derived from the underlying Klein-Gordon
equations~\cite{twoscalars}.
On the other hand, coupling $\varphi$ to the dark matter's kinetic term yields an equivalent system
for dark matter as treated in this work (see \cref{sec:kinetic-couplings}).
In any case, an underlying description in terms of fields is of course necessary to interpret the
model as a quantum theory, e.g., to assess radiative
stability~\cite{DAmico:2016jbm,Archidiacono:2022iuu}.

The action describing a scalar mediator and its interaction with dark matter particles $\chi$ is
obtained by promoting the particle mass to a functional of $\varphi$ in the single-particle
action~\cite{Nordstrom1912, Damour:1990tw, Damour:1994zq, Farrar:2003uw, Uzan:2023dsk}:
\begin{align}
	S
	&= 2 \Mpl^2 \int \ud^4 x \, \sqrt{-g} \left[
            - \frac{1}{2} \partial_\mu \varphi \partial^\mu \varphi
            - V_\varphi(\varphi)
        \right]
        - \sum_p \int \ud \uptau_p \, m_\chi\big( \varphi[x_p(\uptau_p)] \big),
    \label{eqn:kinetic-theory-action}
\end{align}
where $\varphi$ is dimensionless and related to the canonical field $\phi$ via
$\phi \equiv \varphi / \sqrt{4 \pi G} = \sqrt{2} \Mpl \varphi$, and where $\uptau_p$ and
$x_p^\alpha$ are the proper time and worldline of particle $p$.
Leaving more detailed exposition to \cref{app:parametrization}, we work with a perturbed,
conformal-time Friedmann-Lema\^itre-Robertson-Walker (FLRW) metric of the form
$g_{\mu \nu} \equiv a(\tau)^2 \left( \eta_{\mu \nu} + h_{\mu \nu} \right)$, with
$\eta_{\mu \nu}$ the Minkowski metric in the mostly positive signature and $h_{\mu \nu}$ a small
perturbation.
We use primes to denote derivatives with respect to conformal time $\tau$ and overbars to denote
spatial averages.
As derived in \cref{app:dark-matter-dynamics,app:mediator-dynamics}, variation of the action yields
\begin{subequations}
\begin{align}
    \nabla_\mu T^{\mu \nu}_\chi
    &= - \pd{\ln m_\chi}{x_\nu} \left( \rho_\chi - 3 P_\chi \right)
    \label{eqn:chi-conservation}
    \\
    \nabla^\mu \nabla_\mu \varphi
    &= \dd{V_\varphi}{\varphi}
        + \pd{\ln m_\chi}{\varphi} \frac{\rho_\chi - 3 P_\chi}{2 \Mpl^2}
    \equiv \frac{\partial V / \partial \varphi}{2 \Mpl^2},
    \label{eqn:covariant-klein-gordon}
\end{align}
\end{subequations}
where $\rho_\chi$ and $P_\chi$ are the (spacetime-dependent) energy density and pressure of $\chi$.
The mediator's kinetic and potential energies are
$\Mpl^2 H^2 \left( \partial \varphi / \partial \ln a \right)^2$ and $2 \Mpl^2 V_\varphi(\varphi)$
(where $H = a' / a^2$ is the standard Hubble rate), and its coupling to dark matter is fully
specified by the function $\partial \ln m_\chi / \partial \varphi$.

Though dark matter exchanges energy and momentum with the mediator, its particle number $n_\chi$
still satisfies the same conservation law as CDM since the mediator only modifies geodesics.
Taking $\chi$ to have vanishing pressure sets $\bar{n}_\chi' + 3 \mathcal{H} \bar{n}_\chi = 0$ at
the background level, where $\mathcal{H} = a' / a = a H$.
The energy density itself evolves as
\begin{align}
    \bar{\rho}_\chi'
    + 3 \mathcal{H} \bar{\rho}_\chi
    &= \dd{\ln m_\chi}{\tau} \bar{\rho}_\chi
    \label{eqn:bg-continuity-rho}
    ,
\end{align}
which integrates to
\begin{align}
	\bar{\rho}_\chi(a)
	&= \frac{m_\chi(a)}{m_\chi(a_i)}
		\frac{\bar{\rho}_\chi(a_i)}{(a / a_i)^3}
    \label{eqn:rho-chi-solution}
\end{align}
with $a_i$ an arbitrary reference scale factor.
For practical reasons explained in
\cref{app:mediator-dynamics}, we work in synchronous gauges with $h_{0 0} = 0$, but the results in
\cref{app:eoms} are written in a general gauge; \cref{app:parametrization} notes the relationship
between our parametrization and more conventional parametrizations of Newtonian and synchronous
gauges.
As reviewed in \cref{app:dark-matter-dynamics}, in synchronous gauges the perturbations to the dark
matter fluid evolve according to
\begin{subequations}\label{eqn:chi-conservation-equations}
\begin{align}
    \delta_{n_\chi}'
    + \partial_i \partial_i \delta u_\chi
    + \psi
    &= 0
    \\
    \delta u_\chi'
    + \left( \mathcal{H} + \dd{\ln m_\chi}{\tau} \right)
    \delta u_\chi
    &= - \delta \ln m_\chi
    = - \pd{\ln m_\chi}{\varphi} \delta \varphi
    .
    \label{eqn:chi-conservation-equations-momentum}
\end{align}
\end{subequations}
Here $\delta_{n_\chi} = \delta n_\chi / \bar{n}_\chi$ is the number density contrast,
$\delta u_\chi$ is the scalar component of the fluid velocity $\upsilon_{\chi, i}$ divided by $a$
(i.e., $\upsilon_{\chi, i} = a \partial_i \delta u_\chi$), and
$\psi = \delta^{i j} h_{i j}' / 2 - \partial_i h_{i 0}$ is the only combination of metric
perturbations (other than $h_{0 0}$) that appears in the energy-momentum and Klein-Gordon equations.

With the above definitions and also decomposing the mediator into a background and small
perturbation as
$\varphi(\tau, \mathbf{x}) = \bar{\varphi}(\tau) + \delta \varphi(\tau, \mathbf{x})$,
\cref{eqn:covariant-klein-gordon} reads
\begin{subequations}
\begin{align}
    0
    &= \bar{\varphi}''
        + 2 \mathcal{H} \bar{\varphi}'
        + \frac{a^2}{2 \Mpl^2} \pd{V}{\varphi}
    \label{eqn:bg-kg}
    \\
    0
    &= \delta \varphi''
        + 2 \mathcal{H} \delta \varphi'
        + \left( - \partial_i \partial_i + a^2 m_\mathrm{eff}^2 \right) \delta \varphi
        + \frac{a^2 \bar{\rho}_\chi}{2 \Mpl^2} \pd{\ln m_\chi}{\varphi} \delta_\chi
        + \bar{\varphi}' \psi
    \label{eqn:perturbed-kg}
    ,
\end{align}
\end{subequations}
defining the effective mass
\begin{align}
    m_\mathrm{eff}^2
    &\equiv \ddd{V_\varphi}{\varphi}
        + \frac{\bar{\rho}_\chi}{2 \Mpl^2} \pdd{\ln m_\chi}{\varphi}
    .
\end{align}
\Cref{eqn:perturbed-kg} is written in terms of the energy (rather than number) density contrast
$\delta_\chi \equiv \delta \rho_\chi / \bar{\rho}_\chi$, like the Einstein equations; we account for
the distinction between the two below, but we discuss the form of \cref{eqn:perturbed-kg} in terms
of $\delta_{n_\chi}$ in \cref{app:eoms}.

\subsection{Subhorizon growth of structure}\label{sec:subhorizon-growth}

Large-scale structure observables are primarily sensitive to comoving scales that reentered the
horizon in the radiation era (or very early in the matter era); these modes were therefore
subhorizon for the entire matter era.
In the subhorizon limit (discussed in more detail in \cref{app:subhorizon-limit}), the Klein-Gordon
equation permits a quasistatic approximation of the form
\begin{align}\label{eqn:mediator-poisson}
    \left( \partial_i \partial_i - a^2 m_\mathrm{eff}^2 \right) \delta \varphi
    &\simeq \frac{a^2 \bar{\rho}_\chi}{2 \Mpl^2} \pd{\ln m_\chi}{\varphi} \delta_\chi
        + \bar{\varphi}' \psi
    ,
\end{align}
which bears a close resemblance to Newtonian gravity.
To be precise, this limit only takes
$\delta \varphi'' + 2 \mathcal{H} \delta \varphi'$ to be negligible compared to
$\partial_i \partial_i \delta \varphi$; the coupling to the metric $\bar{\varphi}' \psi$ is not
negligible \emph{a priori} nor in practice.\footnote{
    Neglecting this term led prior work to miss contributions of the scalar-mediated force to the
    effective friction acting on dark matter density perturbations; we discuss this and related
    subtleties in \cref{app:subhorizon-limit}.
}
\Cref{app:subhorizon-limit} explains how power counting in $k / a H$ justifies the neglect of
fast-varying modes in the subhorizon limit.
We may therefore combine the energy and momentum equations for $\chi$
[\cref{eqn:chi-conservation-equations}] into a second-order equation for the density contrast, in
addition substituting the Einstein equation \cref{eqn:svt-field-equations-trace-psi} for $\psi$ in
terms of $\delta \rho + 3 \delta P$:
\begin{align}
    \delta_\chi''
    + \left(
        \mathcal{H}
        + \frac{
            \ud \ln m_\chi / \ud \tau
        }{
            1 + \left( a m_\mathrm{eff} / k \right)^2
        }
    \right)
    \delta_\chi'
    &\simeq
        \frac{a^2 \bar{\rho}_\chi}{2 \Mpl^2}
        \left(
            1
            + \frac{
                \left( \partial \ln m_\chi / \partial \varphi \right)^2
            }{
                1 + \left( a m_\mathrm{eff} / k \right)^2
            }
        \right)
        \delta_\chi
        + \frac{a^2}{2 \Mpl^2}
        \sum_{I \neq \chi} \left(
            \delta \rho_I
            + 3 \delta P_I
        \right)
    \label{eqn:delta-chi-second-order-pressureless-subhorizon-synchronous}
    ,
\end{align}
with $\delta \rho_I$ and $\delta P_I$ the energy-density and pressure perturbations for species $I$.
In \cref{eqn:delta-chi-second-order-pressureless-subhorizon-synchronous}, we identified that the
same terms neglected in the quasistatic approximation to the Klein-Gordon equation are those which
differentiate time derivatives of the energy and number density contrasts $\delta_\chi$ and
$\delta_{n_\chi}$, permitting their exchange.\footnote{
    The only additional approximation made in
    \cref{eqn:delta-chi-second-order-pressureless-subhorizon-synchronous} is to neglect a remaining
    term involving the velocity $\delta u_\chi$; \cref{app:subhorizon-limit} explains that this term
    is suppressed in both the $k \gg a m_\mathrm{eff}$ and $k \ll a m_\mathrm{eff}$ limits and also
    enters at next-to-leading order in the mediator coupling.
}
Note that \cref{eqn:delta-chi-second-order-pressureless-subhorizon-synchronous} holds for all
$k / a m_\mathrm{eff}$ and is restricted only to subhorizon scales.

As must be for scalar-mediated self-interactions~\cite{Peskin:1995ev}, the dark force's impact on
clustering is manifestly attractive, whatever the form of its microphysical coupling, since it
arises via $\left( \partial \ln m_\chi / \partial \varphi \right)^2 \delta_\chi$ in
\cref{eqn:delta-chi-second-order-pressureless-subhorizon-synchronous}.
\Cref{eqn:delta-chi-second-order-pressureless-subhorizon-synchronous}, moreover, demonstrates that
the long-range force modifies not only the clustering strength of dark matter particles but also the
friction it experiences~\cite{Archidiacono:2022iuu}.
Both effects are cut off at length scales longer the comoving range of the scalar-mediated force,
$1 / a m_\mathrm{eff}$, but they are differentiated by the friction term's dependence on the
background evolution of the mediator $\bar{\varphi}'$.
So long as $\bar{\varphi}'$ (and all other coefficients in the equation) varies relatively slowly,
constant-coefficient solutions to
\cref{eqn:delta-chi-second-order-pressureless-subhorizon-synchronous} provide useful analytic
insight into the general dependence of structure growth on
$\partial \ln m_\chi / \partial \varphi$ and $\ud \ln m_\chi / \ud \tau$ even without specifying the
mediator's background dynamics.

\subsubsection{Dynamics after photon-baryon decoupling}\label{sec:structure-growth-after-decoupling}

For the observable modes that reenter the horizon in the decade or two of expansion before equality,
most of the evolution proceeds after photon-baryon decoupling when stress-energy perturbations are
dominated by baryons and dark matter, i.e.,
$\sum_{I \neq \chi} \left( \delta \rho_I + 3 \delta P_I \right) = \bar{\rho}_b \delta_b$
in \cref{eqn:delta-chi-second-order-pressureless-subhorizon-synchronous}.
Following Ref.~\cite{Archidiacono:2022iuu}, we change variables from $\delta_\chi$ and $\delta_b$ to
total and relative density contrasts,
$\delta_{\chi b} \equiv f_\chi(a) \delta_\chi + \left[ 1 - f_\chi(a) \right] \delta_b$
and $\delta_r \equiv \delta_\chi - \delta_b$, where the fraction in $\chi$ is
$f_\chi(a) \equiv \bar{\rho}_\chi(a) / \left[ \bar{\rho}_\chi(a) + \bar{\rho}_b(a) \right]$.
We assume that $\chi$ and baryons are the only clustering species (i.e., metric potentials are
sourced by $\delta \rho_\chi + \delta \rho_b$ alone) but not necessarily the only matterlike
contributors to expansion.
Ref.~\cite{Archidiacono:2022iuu} showed that the relative density contrast is generated only by the
dark force and appears in the equation of motion for the total density contrast multiplied
by $\ud \ln m_\chi / \ud \tau$ or $\left( \partial \ln m_\chi / \partial \varphi \right)^2$.
As such, the effect of relative density contrasts on the growth of structure is subleading in the
mediator coupling; for brevity, we omit these terms.
The total density contrast thus evolves as
\begin{align}
\begin{split}
    \ddd{\delta_{\chi b}}{\ln a}
    &\simeq - \left(
            2 + \dd{\ln H}{\ln a}
            + \frac{
                f_\chi \ud \ln m_\chi / \ud \ln a
            }{
                1 + \left( a m_\mathrm{eff} / k \right)^2
            }
        \right)
        \dd{\delta_{\chi b}}{\ln a}
        + \frac{3 \Omega_{\chi b}(a)}{2}
        \left(
            1
            + \frac{
                f_\chi^2 \left( \partial \ln m_\chi / \partial \varphi \right)^2
            }{
                1 + \left( a m_\mathrm{eff} / k \right)^2
            }
        \right)
        \delta_{\chi b}
    \label{eqn:delta-chi-b-eom}
    ,
\end{split}
\end{align}
differing from that for $\delta_\chi$ only by an additional factor of $f_\chi$ multiplying terms
generated by the dark force.
Here $\Omega_{\chi b}(a) \equiv [\bar{\rho}_\chi(a) + \bar{\rho}_b(a)] / \bar{\rho}(a)$ is
the fractional density in dark matter and baryons.

From here on out, we focus on scales within the range of the LRF and therefore take
$k \gg a m_\mathrm{eff}$ for simplicity.
It proves convenient to parametrize the deviations of \cref{eqn:delta-chi-b-eom} from the result
for a CDM-dominated Universe via
\begin{align}
    \ddd{\delta_{\chi b}}{\ln a}
    &= - \left[
            2 - \frac{3}{2} \left( 1 + \Delta_\gamma \right)
        \right]
        \dd{\delta_{\chi b}}{\ln a}
        + \frac{3}{2} \left( 1 + \Delta_\omega \right)
        \delta_{\chi b}
\end{align}
where
\begin{subequations}
\begin{align}
    \Delta_\gamma
    &= \Omega_m(a) - 1
        + \Omega_m(a) w_\varphi(a) f_\varphi(a)
        - \frac{2}{3} f_\chi \dd{\ln m_\chi}{\ln a}
    \\
    \Delta_\omega
    &= \Omega_{\chi b}(a) - 1 + \Omega_{\chi b}(a) \left( f_\chi \pd{\ln m_\chi}{\varphi} \right)^2
    ,
\end{align}
\end{subequations}
with $\Omega_m(a) \equiv \bar{\rho}_m(a) / \bar{\rho}(a)$ the fractional abundance of all matterlike
components including those that may not participate in clustering.
That is, $\Omega_m(a)$ is larger than $\Omega_{\chi b}(a)$ by a fractional amount
$f_\mathrm{ncl} \equiv \bar{\rho}_\mathrm{ncl} / \left( \bar{\rho}_{\chi b} + \bar{\rho}_\mathrm{ncl} \right)$.
Nonrelativistic neutrinos are one such example, as can be the mediator itself (as its density
perturbations are quite subdominant to dark matter's).
Since the mediator may have nonzero pressure (even when redshifting like $a^{-3}$, since it is not
an uncoupled fluid), we define its equation of state
$w_\varphi(a) \equiv \bar{P}_\varphi(a) / \bar{\rho}_\varphi(a)$ and fractional contribution to the
total matter density $f_\varphi = \bar{\rho}_\varphi(a) / \bar{\rho}_m(a)$.
We drop the scale dependence of the couplings for simplicity, since we focus on scenarios where
observable structure is uniformly modified by the long-range force (i.e., taking $a m_\mathrm{eff}$
smaller than the horizon scale at matter-radiation equality, $k_\mathrm{eq}$).

The growth rate, approximated as $\dot{y} / y$ for solutions to the equation
$\ddot{y} + \gamma \dot{y} - \omega y = 0$ with time-independent coefficients, is
\begin{align}
    \dd{\ln \delta_{\chi b}}{\ln a}
    &= \frac{- \gamma \pm \sqrt{\gamma^2 + 4 \omega}}{2}
    \approx \begin{dcases}
            1 + \frac{3}{5} \Delta_\gamma + \frac{3}{5} \Delta_\omega, & +,
            \\
            - \frac{3}{2} + \frac{9}{10} \Delta_\gamma - \frac{3}{5} \Delta_\omega, & -.
        \end{dcases}
\end{align}
Focusing on the matter era, with $\Omega_m(a) = 1$ the growth rate of the total density contrast is
\begin{align}
    \dd{\ln \delta_{\chi b}}{\ln a}
    \approx 1
        - \underbrace{\frac{2}{5} f_\chi \dd{\ln m_\chi}{\ln a}}_{\substack{\text{modified}\\ \text{friction}}}
        + \underbrace{\frac{3}{5} \left( f_\chi \pd{\ln m_\chi}{\varphi} \right)^2}_\text{enhanced clustering}
        + \underbrace{\frac{3}{5} w_\varphi f_\varphi}_{\substack{\text{modified}\\ \text{expansion}}}
        - \underbrace{\frac{3}{5} f_\mathrm{ncl},}_{\substack{\text{nonclustering}\\ \text{matter}}}
    \label{eqn:delta-growth-rate}
\end{align}
identifying the contributions due to the modified friction term, the enhanced clustering strength,
the modified expansion rate, and the presence of nonclustering components of matter.
In the decoupling limit (with $f_\varphi \to 0$), \cref{eqn:delta-growth-rate} recovers the
solution $\delta_{cb} \propto a^{1 - 3 f_\mathrm{ncl} / 5}$ applicable for CDM with, e.g.,
nonclustering massive neutrinos.

However, physical observables that derive from metric potentials, like gravitational lensing and the
ISW effect, depend on the total density perturbation $\delta \rho$, not the density contrast.
The growth rate of the Bardeen potential(s)~\cite{Bardeen:1980kt} is the same as that of
$a^2 \delta \rho_{\chi b} = a^3 \bar{\rho}_{\chi b} \delta_{\chi b} / a$, i.e., is reduced by the
degree to which $a^3 \bar{\rho}_{\chi b} \propto m_\chi[\varphi(a)]$ decays.
Using \cref{eqn:bg-continuity-rho} to write
$\ud \ln a^3 \bar{\rho}_{\chi b} / \ud \ln a = f_\chi \ud \ln m_\chi / \ud \ln a$,
\begin{align}
    \dd{\ln \Phi_\mathrm{B}}{\ln a}
    = \dd{\ln a^3 \bar{\rho}_{\chi b}}{\ln a}
        + \dd{\ln \delta_{\chi b} / a}{\ln a}
    &\approx
        \underbrace{\frac{3}{5} f_\chi \dd{\ln m_\chi}{\ln a}}_{\text{mass evolution}}
        + \underbrace{\frac{3}{5} \left( f_\chi \pd{\ln m_\chi}{\varphi} \right)^2}_\text{enhanced clustering}
        + \underbrace{\frac{3}{5} w_\varphi f_\varphi}_{\substack{\text{modified}\\ \text{expansion}}}
        - \underbrace{\frac{3}{5} f_\mathrm{ncl}.}_{\substack{\text{nonclustering}\\ \text{matter}}}
    \label{eqn:Phi-growth-rate}
\end{align}
When the mediator's coupling to dark matter dominates its effective potential, we generically expect
$\ud \ln m_\chi / \ud \ln a < 0$, since $\chi$'s particle number is conserved and decreasing the
dark matter mass is energetically favorable.
\Cref{eqn:delta-growth-rate} indicates that this effective reduction in total friction (i.e., on
top of Hubble friction) enhances the growth rate of the density \emph{contrast} relative to that
from the enhancement to clustering alone~\cite{Archidiacono:2022iuu}.
However, \cref{eqn:Phi-growth-rate} demonstrates that in the growth of metric potentials (and
therefore the total density perturbation $\delta \rho_{\chi b}$) this effect is overcompensated by
the faster dilution of the background density which has the same physical origin, putting
modifications to the dynamics of the background and (relative) perturbations at odds.

The mediator, whose pressure modifies expansion and whose energy density acts as nonclustering
matter, does not directly modify the growth rate unless it has a bare potential: when it carries
only kinetic energy, its equation of state $w_\varphi = 1$.
Because a scalar's pressure is no larger than its energy density (unless its potential were negative
in a physically meaningful way), supplying the mediator with a bare potential only suppresses
the growth rate insofar as its contribution to the stress-energy tensor is effectively spatially
homogeneous.
(Aside from its impact on the Einstein equations, the bare potential also alters the growth
rate via the evolution of $m_\chi$.)
We explore how this effect might be outweighed by modifications to the growth of dark matter
perturbations in \cref{sec:nonminimal}, but we first study the minimal scenario with no bare
potential.

\subsection{Linearly coupled, massless mediator}\label{sec:linear-coupling}

When the dark matter mass depends linearly on the mediator, the mediator's background dynamics
permit an analytic solution~\cite{Archidiacono:2022iuu}.
Here by a linear (or quadratic, etc.) coupling we mean that entering $\ln m_\chi(\varphi)$ rather
than the Lagrangian in terms of the underlying field, since derivatives of $\ln m_\chi$ are
what appear in the equations of motion.
In other words, we write $m_\chi(\varphi) = m_{\chi, 0} \exp g_{m_\chi}(\varphi)$ in terms of the
coupling function $g_{m_\chi}$~\cite{Baryakhtar:2025uxs}.\footnote{
    A Yukawa coupling to fermions of the form $d_{m_\chi}^{(1)} \varphi \bar{\chi} \chi$
    corresponds to $g_{m_\chi}(\varphi) = \ln (1 + d_{m_\chi}^{(1)} \varphi)$ and one to bosons
    $d_{m_\chi}^{(1)} m_{\chi, 0} \varphi \chi^2$ to
    $g_{m_\chi}(\varphi) = \ln \sqrt{1 + 2 d_{m_\chi}^{(1)} \varphi}$.
    \label{footnote:yukawa}
}
Specifically, we Taylor expand the coupling function about its initial condition and denote
the linear coefficient as $d_{m_\chi}^{(1)}$.
As our results for the growth rate depend only on the instantaneous gradient of the coupling
function (given that we took a constant-coefficient approximation), we later simply express
our results in terms of $\partial \ln m_\chi / \partial \varphi$ that is understood to indicate the
linear coefficient evaluated at the current field value.

We take initial conditions $\bar{\varphi} \to \bar{\varphi}_i$ and
$\ud \bar{\varphi} / \ud \ln a \to 0$ as $a \to 0$, as appropriate given that Hubble friction deep
in the radiation era far exceeds the effective potential sourced by dark matter.
Note that, with a linear coupling and no bare potential for the mediator, the action
\cref{eqn:kinetic-theory-action} is invariant under constant shifts of $\bar{\varphi}$ and a
redefinition of the ``bare'' value of $m_\chi$, i.e., the initial misalignment $\bar{\varphi}_i$ is
irrelevant to the dynamics.
At leading order in $f_\chi d_{m_\chi}^{(1)}$, the homogeneous Klein-Gordon equation
[\cref{eqn:bg-kg}] is of the form
\begin{align}
	\ddd{\bar{\varphi}}{\ln a}
	+ \frac{1}{2} \left( 3 - \frac{1}{1 + a / a_\mathrm{eq}} \right) \dd{\bar{\varphi}}{\ln a}
	&= - \frac{3}{2}
		\frac{d_{m_\chi}^{(1)} f_\chi}{1 + a_\mathrm{eq} / a}
\end{align}
with $a_\mathrm{eq}$ the scale factor at matter-radiation equality, which is solved by
$\bar{\varphi}(a) = \bar{\varphi}_i - f_\chi d_{m_\chi}^{(1)} (1 - 1 / y + 2 \ln y)$ where
$2 y = 1 + \sqrt{1 + a / a_\mathrm{eq}}$.
In the matter era ($a \gg a_\mathrm{eq}$),
\begin{align}
    \bar{\varphi}(a)
    &= \bar{\varphi}_i - f_\chi d_{m_\chi}^{(1)} \ln \frac{a}{4 a_\mathrm{eq} / e}
        + \mathcal{O}(a_\mathrm{eq} / a)
    \label{eqn:massless-mediator-matter-era-bg-soln}
\end{align}
meaning the mass evolves as
$\ud \ln m_\chi / \ud \ln a \approx - f_\chi \left( \partial \ln m_\chi / \partial \varphi \right)^2$
and the mediator's kinetic energy comprises a time-independent fraction of the total matter
density\footnote{
    Recall that the energy density of a perfect fluid with equation of state $w$ only redshifts like
    $a^{- 3 (1 + w)}$ if its stress-energy tensor is independently conserved.
}
$f_\varphi \approx \left( f_\chi \partial \ln m_\chi / \partial \varphi \right)^2 / 3$.
Well after equality, the density in dark matter and baryons therefore evolves as
$\bar{\rho}_{\chi b} \propto a^{- 3 - f_\chi^2 \left( \partial \ln m_\chi / \partial \varphi \right)^2}$.
All physical observables are sensitive to the LRF in this same combination $\beta f_\chi^2$ of the
strength of the LRF relative to gravity,
$\beta \equiv \left( \partial \ln m_\chi / \partial \varphi \right)^2$,
and the fraction of matter in $\chi$, $f_\chi$~\cite{Archidiacono:2022iuu}.
That is, $d_{m_\chi}^{(1)} f_\chi$ is effectively a vertex factor that appears twice in all
gravitational effects---one factor weights the dark-matter source in the Klein-Gordon equation,
which either is squared insofar as the mediator directly sources gravity or is multiplied by another
factor of its interaction strength and of the fraction $f_\chi$ of matter whose dynamics it
modifies.

\begin{figure}[t!]
\begin{centering}
    \includegraphics[width=\textwidth]{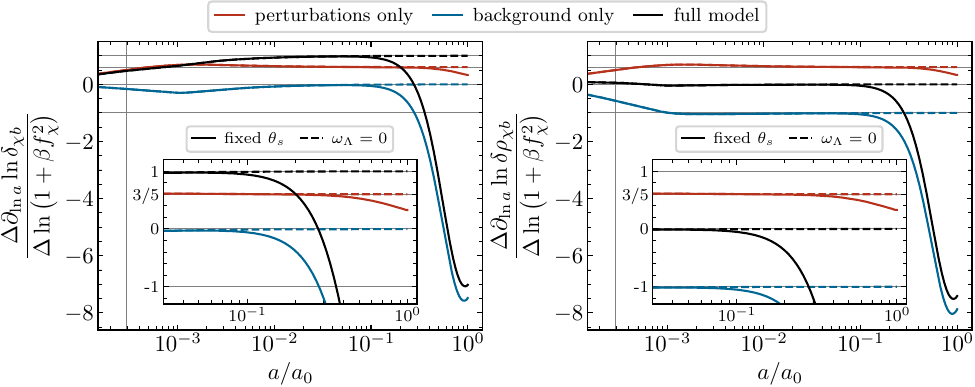}
    \caption{
        Sensitivity of the growth rate $\ud \ln X / \ud \ln a$ to the strength of a
        long-range force acting on dark matter, with $X$ the density contrast $\delta_{\chi b}$
        (left) or the total density perturbation $\delta \rho_{\chi b}$ (right) in baryons and dark
        matter.
        The sensitivity is quoted as the response coefficient multiplying $\beta f_\chi^2$,
        the combination of the LRF strength relative to gravity,
        $\beta = \left( \partial \ln m_\chi / \partial \varphi \right)^2$, and the dark matter
        fraction, $f_\chi$, that parametrizes the effects; it is computed via the
        difference $\Delta \partial_{\ln a} \ln X$ for $k = 10~\mathrm{Mpc}^{-1}$ modes between
        a cosmology with a force strength relative to gravity of $\beta = 10^{-2}$ and \LCDM{}
        divided by $\beta f_\chi^2$.
        Horizontal lines mark the analytic, matter-era results \cref{eqn:delta-growth-rate,eqn:Phi-growth-rate}; the vertical line marks matter-radiation equality.
        The results labeled ``background/perturbations only'' isolate the LRF effects in the
        equations of motion for the background and linear perturbations as described in the main
        body; all effects combine to yield zero net sensitivity of $\delta \rho_{\chi b}$ (and
        therefore of metric potentials) to the LRF in matter domination.
        Inset panels enlarge the matter-era dynamics.
        Dashed and solid lines respectively fix the dark energy density to zero (to better
        illustrate the effects in a pure-matter Universe) and the angular extent of the photon
        sound horizon at last scattering $\theta_s$; the latter incurs a strong suppression of the
        growth rate that is explained in \cref{sec:suppression}.
    }
    \label{fig:growth-deviation}
\end{centering}
\end{figure}

Inserting these results into \cref{eqn:delta-growth-rate} yields an enhanced growth rate of the
density contrast:\footnote{
    Our result differs slightly from that of Ref.~\cite{Archidiacono:2022iuu}, who instead obtain
    $6/5 \left( f_\chi \partial \ln m_\chi / \partial \varphi \right)$; the decrement of $1/5$ in
    the coefficient derives from accounting for the mediator as a nonclustering component, i.e.,
    that $\Omega_{\chi b}$ in \cref{eqn:delta-chi-b-eom} is
    $1 - f_\varphi \approx 1 - \beta f_\chi^2 / 3$ rather than unity in the matter-dominated era.
}
\begin{align}
    \dd{\ln \delta_{\chi b}}{\ln a}
    &\approx
        1
        - \underbrace{
            \frac{2}{5} f_\chi \times \left[
                - f_\chi \left( \pd{\ln m_\chi}{\varphi} \right)^2
            \right]
        }_{\substack{\text{modified friction}}}
        + \underbrace{\frac{3}{5} \left( f_\chi \pd{\ln m_\chi}{\varphi} \right)^2}_\text{enhanced clustering}
    = 1 + \left( f_\chi \pd{\ln m_\chi}{\varphi} \right)^2
    \label{eqn:delta-growth-rate-linear-massless}
    .
\end{align}
(As mentioned previously, the gravitational impact of the mediator itself only derives from its bare
potential, which we currently take to be negligible.)
On the other hand, the various contributions to the growth rate of the Bardeen potentials
[\cref{eqn:Phi-growth-rate}] cancel exactly (i.e., at leading order in $\beta f_\chi^2$):
\begin{align}
    \dd{\ln \Phi_\mathrm{B}}{\ln a}
    &\approx
        \underbrace{
            \frac{3}{5} f_\chi \times \left[
                - f_\chi \left( \pd{\ln m_\chi}{\varphi} \right)^2
            \right]
        }_{\substack{\text{mass evolution}}}
        + \underbrace{\frac{3}{5} \left( f_\chi \pd{\ln m_\chi}{\varphi} \right)^2}_\text{enhanced clustering}
    = 0.
    \label{eqn:Phi-growth-rate-linear-massless}
\end{align}
In other words, the growth rate of absolute density perturbations $\delta \rho_{\chi b}$ is
unaffected by a long-range force mediated by a linearly coupled, massless scalar.
We corroborate these analytic results with full solutions to the Einstein-Boltzmann equations in
\cref{fig:growth-deviation} [which match well with numerical solutions to
\cref{eqn:delta-chi-second-order-pressureless-subhorizon-synchronous} that include radiation and
dark energy].
In order to measure the response coefficient (or sensitivity) of the growth rate to the dark force,
we evaluate the difference in growth rate between an LRF model and \LCDM{} and divide by
$\beta f_\chi^2$.

In addition to the result that consistently implements the model, which shows that indeed
$\ud \ln \delta_{\chi b} / \ud \ln a \approx 1 + \beta f_\chi^2$ and
$\ud \ln \delta \rho_{\chi b} / \ud \ln a \approx 0$ in matter domination, we consider cases that
account for the mediator coupling in only the background or perturbation equations.
The background-only case [which includes only the background mass evolution and not mediator
friction, which are collectively denoted ``mass evolution'' in
\cref{eqn:Phi-growth-rate,eqn:Phi-growth-rate-linear-massless}], shows two effects:
a sensitivity $-1$ in matter domination for $\delta \rho_{\chi b}$ from the mass evolution and a
slightly negative sensitivity for $\delta_{\chi b}$ around matter-radiation equality.
The latter derives from the slower dilution of radiation, whose presence slows the growth of
structure, compared to the total density [i.e., $\Omega_r(a)$ decreasing more slowly], which is not
encoded in the above analytic results, as they fix matter domination.
The case that only accounts for the enhanced clustering rate shows the expected sensitivity of
$3/5$ in matter domination.

\Cref{fig:growth-deviation} thus confirms the analytic results derived in this section for a
matter-dominated Universe, which describes a majority of the growth of structure.
However, \cref{fig:growth-deviation} also shows a strongly negative sensitivity at low redshift when
adjusting the dark energy density to fix the angular extent of the sound horizon measured by the
CMB, $\theta_s$.
We explain the origin of this suppression of structure growth in \cref{sec:suppression} by
identifying how the information in CMB anisotropies calibrates the predictions of dark forces for
low-redshift dynamics.

\section{Suppression of growth predicted by the primary CMB}
\label{sec:suppression}

Cosmic microwave background temperature and polarization anisotropies provide rich, scale-dependent
information on the dynamics around recombination, while weak lensing and distance measurements offer
relatively greater sensitivity to deviations of the matter content from CDM-like behavior at late
times.
In this section, we characterize and quantify the constraining power on dark long-range forces
deriving from the CMB's sensitivity to early-time physics and the acoustic scale.
We show in \cref{sec:calibration} that the generation of small-scale anisotropies at last scattering
is almost exclusively impacted by the modifications to the background evolution of the dark matter
density and not the enhanced growth of its overdensities.
In \cref{sec:extrapolation} we then study the predictions of the model, as calibrated by this
early-time information, for late-Universe observables in the minimal scenario of a linearly coupled,
massless mediator (\cref{sec:linear-coupling}).
In particular, we quantify the distortion of low-redshift distances, as measured by supernovae and
acoustic scale measurements from galaxy surveys, and show that weak lensing observables are in fact
\emph{suppressed} relative to \LCDM{} predictions.

The CMB is only (directly) observable in its lensed form, for which reason one cannot truly
compartmentalize its sensitivity to early- and late-time information.
In \LCDM{}-like models, however, power on larger angular scales ($\ell \lesssim 1000$, say) is
relatively less affected by lensing than on smaller ones and are therefore less directly sensitive
to late-time structure than higher-resolution observations (from, e.g., ACT and SPT) and lensing
reconstruction from higher-point statistics.\footnote{
    One can use lensing reconstruction to marginalize over the impact of late-time structure on
    primary anisotropies, which for \Planck{} only marginally weakens measurements of
    \LCDM{} parameters~\cite{Lemos:2023xhs}.
}
Likewise, the inference of the acoustic scale is insensitive to the effects of lensing and only
provides information on the integrated late-time expansion history, whereas low-redshift distances
from BAO or SNe data directly trace it.
Organizing observables in this manner thus facilitates interpreting the physical origin of
constraints on cosmological models, especially those deviating from \LCDM{} at late times.

To study the calibration of (i.e., predictions for) low-redshift observables by the primary CMB, we
perform parameter inference with various combinations of CMB temperature and polarization data (as
well as BAO data in \cref{sec:distances}).
In particular, we employ a subset of \Planck{} PR3 observations~\cite{Planck:2019nip} cut to
multipoles $\ell \leq 1000$ in temperature and $\leq 600$ in polarization and temperature-polarization
cross correlation---for convenience, the same subset used in combination with ACT CMB data, so
chosen to effectively remove overlap between the two surveys~\cite{ACT:2025fju}.
\label{text:data-description}
This subset includes the $\ell < 30$ temperature and polarization likelihoods from PR3.
At times, we also use the full PR3 dataset and also the aforementioned subset combined with ACT
DR6~\cite{ACT:2025fju} and SPT-3G D1~\cite{SPT-3G:2025bzu} data.
For theoretical predictions for the latter datasets, we derive sufficient precision settings that
are substantially reduced compared to those recommended in Ref.~\cite{AtacamaCosmologyTelescope:2025nti, Hill:2021yec,
Bolliet:2023sst, Jense:2024llt}; we enumerate these and discuss other implementation details in
\cref{app:numerics}.

We use a modified version of \textsf{CLASS}~\cite{Blas:2011rf, Lesgourgues:2011re} that implements a
long-range force mediated by a scalar with arbitrary coupling and potential functions.
We briefly comment on technical aspects of the implementation in \cref{app:numerics}.
Models of nonlinear structure growth that account for additional long-range forces are not readily
available, the development of which we defer to future work.
For this reason, we never include lensing reconstruction observations in parameter inference.
While those from \Planck{}~\cite{Planck:2018lbu, Carron:2022eyg} are not especially sensitive to
nonlinear structure growth, they offer little information~\cite{Bottaro:2024pcb} (due in part to
their precision as well as the suppressed effect of dark forces on lensing).
Measurements from ACT and SPT~\cite{ACT:2023kun, SPT-3G:2024atg, ACT:2025qjh} are substantially
more precise and include smaller scales, but are much more sensitive to nonlinear structure growth.
We therefore include no lensing reconstruction data in our analyses and note that doing so can lead
to spuriously strong evidence for nonzero LRF strength (see \cref{sec:nonlinear} for further
discussion).

We parametrize the LRF mediated by a massless scalar via the early-time comoving energy density in
$\chi$,
$\tilde{\omega}_\chi \equiv \lim_{a \to 0} a^3 \bar{\rho}_\chi(a) / 3 \Mpl^2 H_{100}^2$,
and the long-range force strength relative to gravity $\beta$, equal to $(d_{m_\chi}^{(1)})^2$ for
our baseline coupling (i.e., linear in $\partial \ln m_\chi / \partial \varphi$ but exponential in
the Lagrangian).
Here $H_{100} \equiv H_0 / h = 100~\mathrm{km}/\mathrm{s}/\mathrm{Mpc}$; we often parametrize energy
densities of species $X$ with units
$\omega_X(a) \equiv \bar{\rho}_X(a) / 3 \Mpl^2 H_{100}^2$, with
$\omega_X \equiv \omega_X(a_0) = \Omega_X(a_0) h^2$ the conventional present-day density parameter.
Using $\mathcal{U}(a, b)$ to denote a uniform prior between $a$ and $b$,
we sample $\tilde{\omega}_\chi \sim \mathcal{U}(0.01, 0.25)$, i.e., the same prior as for
$\omega_c$ in \LCDM{}.
Since the physical effects we seek to study are linear in $\beta$ rather than $d_{m_\chi}^{(1)}$,
we take a uniform prior over the former, $\beta \sim \mathcal{U}(10^{-6}, 10^{-1})$.
This choice does shift marginal posterior distributions over both $\beta$ and $d_{m_\chi}^{(1)}$
to slightly larger values, since it weights $d_{m_\chi}^{(1)}$ in proportion to $d_{m_\chi}^{(1)}$,
but not enough to alter qualitative conclusions.

We take standard priors for the remaining \LCDM{} parameters: the present baryon density
$\omega_b \sim \mathcal{U}(0.005, 0.035)$, the angular extent of the sound horizon
$100 \theta_s \sim \mathcal{U}(0.9, 1.1)$ (which, being less degenerate with $\beta$ than the Hubble
rate $h$ or the dark energy density $\omega_\Lambda$, can be sampled over more efficiently), the
tilt $n_s \sim \mathcal{U}(0.8, 1.2)$ of the scalar power spectrum, its amplitude $A_s$ via
$\ln(10^{10} A_s) \sim \mathcal{U}(1.61, 3.91)$, and the optical depth to reionization
$\tau_\mathrm{reion} \sim \mathcal{U}(0.02, 0.2)$.
In some analyses we sample the neutrino mass sum
$\summnu / \mathrm{eV} \equiv \sum_i m_{\nu_i} / \mathrm{eV} \sim \mathcal{U}(0, 1.5)$, taking a
degenerate mass hierarchy.
We perform parameter sampling with \textsf{emcee}~\cite{Foreman-Mackey:2012any, Hogg:2017akh,
Foreman-Mackey:2019}.

\subsection{Information in primary anisotropies}\label{sec:calibration}

The CMB is sensitive to the nature of dark matter through several distinct physical processes that
take place at both high and low redshift.
We discuss the impact of dark matter on the dynamics of plasma perturbations when photons last
scattered (\cref{sec:last-scattering}) and on their propagation to late times
(\cref{sec:propagation-effects}) in turn (but defer discussing lensing until
\cref{sec:extrapolation}).

\subsubsection{Generation of anisotropies at last scattering}\label{sec:last-scattering}

Before photon-baryon decoupling, dark matter modulates the propagation of acoustic waves in the
plasma as a contribution to the expansion rate that decays more slowly than radiation and to density
perturbations that, unlike baryons, is unsupported by pressure~\cite{Hu:1994jd, Hu:1995en,
Zaldarriaga:1995gi, Hu:1996mn}.
As matter-radiation equality occurs shortly before recombination, the acoustic peaks in the CMB
reflect the growing importance of dark matter to the Einstein equations, manifesting as an ISW
effect around the first acoustic peak and by diminishing the so-called radiation driving effect at
horizon crossing.
In \LCDM{}, these features contribute to the CMB's constraining power on the CDM density $\omega_c$,
which is intrinsically sensitive to ratios of the (dimensionful) densities in various components
(since CMB anisotropies are a dimensionless observable~\cite{Baryakhtar:2024rky}).
The primary temperature and polarization anisotropies best measure the density ratio of matter to
radiation (and of baryons to photons) around peak visibility, i.e., around hydrogen
recombination~\cite{Hu:2004kn, Weinberg:2008zzc, Baryakhtar:2024rky}.
Since baryons and cold dark matter redshift in a fixed manner ($\propto a^{-3}$), these ratios
evaluated at recombination fully parametrize their physical effects; combined with the precise
measurement of the present-day CMB temperature and the Standard Model's prediction for temperature
at last scattering and the density in relativistic neutrinos, the CMB uniquely measures the
present-day densities of baryons and CDM.

In fact, the effects of dark matter on the generation of small-scale CMB anisotropies at last
scattering are largely through its impact on the expansion history rather than via the detailed
dynamics of its perturbations, as has been shown for cold dark matter~\cite{Hu:1995en,
Weinberg:2002kg, Weinberg:2008zzc} and warm dark matter~\cite{Voruz:2013vqa}.
As referenced in \cref{sec:subhorizon-growth} (and elaborated on in \cref{app:subhorizon-limit}),
solutions to the (linear) Einstein-Boltzmann system may be decomposed into modes that evolve on
comoving timescales of order $k$ and $a H$; power counting in $k / a H$ shows that, on scales
smaller than the comoving horizon at equality and from a few $e$-folds prior to equality until last
scattering, the fast and slow modes are dominated by plasma and dark-matter perturbations,
respectively~\cite{Weinberg:2002kg, Weinberg:2008zzc}.
The plasma perturbations that source the primary CMB and the dark-matter perturbations that govern
late-time structure are thus effectively gravitationally decoupled.

\begin{figure}[t!]
\begin{centering}
    \includegraphics[width=\textwidth]{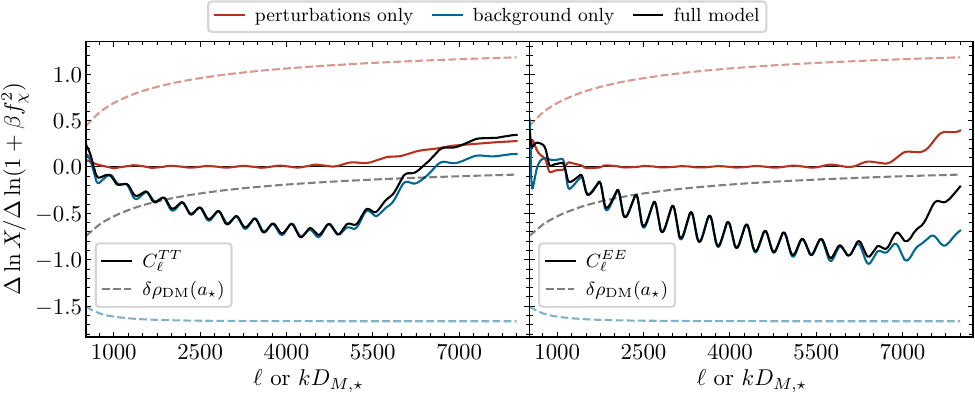}
    \caption{
        Sensitivity of unlensed CMB temperature (left) and polarization (right) anisotropies to a
        scalar-mediated long-range force acting on dark matter, fixing the early-time dark matter
        density $\omega_\chi(a \to 0)$ and angular size of the sound horizon $\theta_s$.
        Each panel depicts the relative residual between a cosmology with a force strength relative
        to gravity of $\beta = 10^{-2}$ and \LCDM{} divided by $\beta f_\chi^2$, i.e., an
        approximation to $\partial \ln C_\ell / \partial \ln (1 + \beta f_\chi^2)$.
        For comparison, transparent, dashed lines depict the sensitivity of the amplitude of
        dark matter density perturbations, evaluated at peak visibility ($a_\star$) and at the
        comoving scales $k = \ell / D_{M, \star}$ that predominantly contribute to each $\ell$.
        Each panel depicts results that consistently implement the model (black) and artificially
        disable the mediator's impact on the evolution of the dark matter's background (red) or
        perturbations (blue), following \cref{fig:growth-deviation}.
        As elaborated in the main text, these results evidence the gravitational decoupling of
        plasma and dark-matter perturbations on small scales $500 \lesssim \ell \lesssim 5000$;
        the dark force therefore only impacts the generation of small-scale anisotropies at
        last scattering via the expansion history, largely by modifying diffusion damping
        (see \cref{fig:cmb-sensitivity-undamped}).
    }
    \label{fig:cmb-sensitivity}
\end{centering}
\end{figure}

We argue in \cref{app:subhorizon-limit} that dark, long-range forces do not undo the gravitational
decoupling of the plasma and dark matter, which is corroborated by full solutions to the
Einstein-Boltzmann equations in \cref{fig:cmb-sensitivity}.\footnote{
    The logarithmic sensitivity displayed in \cref{fig:cmb-sensitivity}, i.e.,
    $\partial \ln C_\ell / \partial \ln \theta$ for a parameter $\theta$, measures the response
    $\Delta C_\ell / C_\ell$ to small variations $\Delta \theta / \theta$ and therefore
    indicates the relative precision $\sigma(\theta) / \theta$ with which $C_\ell$ can measure
    $\theta$.
    That is, the Fisher information on $\theta$ from idealized, cosmic-variance--limited measurements
    of a single map with angular spectrum $C_\ell$ is
    $\theta^2 F_{\theta \theta} = \sum_\ell \left( \partial \ln C_\ell / \partial \ln \theta \right)^2 \left( 2 \ell + 1 \right) / 2$~\cite{Tegmark:1996bz}.
    With all other parameters fixed, $\theta$ is measured by $C_\ell$ with relative precision
    $\sigma(\theta) / \theta = 1 / \sqrt{F_{\theta \theta} \theta^2}$.
}
We study the dark force's separate impacts on the dynamics of the homogeneous dark matter
density and its spatial perturbations by artificially disabling one or the other.
\Cref{fig:cmb-sensitivity} shows that the modified evolution of $\delta_\chi$ alone (due solely to
enhanced clustering) indeed has no impact on the primary anisotropies at $\ell \gtrsim 500$, despite
the order-unity sensitivity of $\delta \rho_\mathrm{DM}(a_\star)$ itself.
Diffusion and cancellation damping eventually suppress the amplitude of fast modes below slow
modes~\cite{Hu:1995en, Weinberg:2002kg, Weinberg:2008zzc} taking effect for $\ell \gtrsim 5000$ and
$6000$ for temperature and polarization, respectively; the observable CMB on these scales, on the
other hand, is overwhelmed by the impact of gravitational lensing (not to mention foregrounds).
The results that instead disable the impact of the force on $\delta_\chi$ are nearly identical to
those that neglect neither effect, despite the substantial difference in the sensitivity of
$\delta \rho_\mathrm{DM}(a_\star)$.

The dark force therefore only affects the generation of small-scale anisotropies via the expansion
history---namely, because the scalar-mediated force introduces freedom to the background evolution
of dark matter.
The sensitivity evident in \cref{fig:cmb-sensitivity} mostly derives from modifications to the
diffusion damping rate with time, which is the origin of the secular drift between multipoles of
$1000$ and $5000$ in \cref{fig:cmb-sensitivity} (see \cref{fig:cmb-sensitivity-undamped} and further
discussion in \cref{app:supplementary-results}).
At larger scales, however, the evolution of $a^3 \bar{\rho}_\chi$ modulates the rate with which the
radiation-driving and ISW effects abate.
The plasma and dark-matter perturbations are not gravitationally decoupled on these scales, but the
actual subhorizon evolution of $\delta \rho_\chi$ is less affected than that of $\delta_\chi$ or
$\bar{\rho}_\chi$ themselves, even before matter domination.

In sum, the signatures of dark matter in the primary CMB depend on its dynamics around
recombination, mostly at the background level.
Phenomenological quantifications of the CMB's sensitivity to the instantaneous CDM
abundance~\cite{Ilic:2020onu} (or that of other exotic components~\cite{Samsing:2012qx}) affirm that
information peaks at recombination, with additional support in the decade of expansion prior.
We therefore expect the primary CMB to most precisely measure the combination of coupling parameters
and the early-time comoving density in dark matter, $\lim_{a \to 0} a^3 \bar{\rho}_\chi(a)$, that
determines its abundance at or just before last scattering; we find the scale factor of peak
sensitivity to be $a_\mathrm{CMB} \approx 0.5 a_\star \approx 1.6 a_\mathrm{eq}$.
For the linear coupling discussed in \cref{sec:linear-coupling}, the best-measured combination
happens to be approximately
$a_\mathrm{CMB}^3 \omega_\chi(a_\mathrm{CMB}) \simeq  \left( 1 - \beta f_\chi \right) \lim_{a \to 0} a^3 \omega_\chi(a)$, reducing
the parameter space describing dark matter from two dimensions to one.
The dynamics of $a^3 \bar{\rho}_\chi(a)$ relative to this fixed value remains a distinguishing
signature of the model that depends on $\beta$, e.g., generating a scale-dependent modulation of
diffusion.

\begin{figure}[t!]
\begin{centering}
    \includegraphics[width=4.5in]{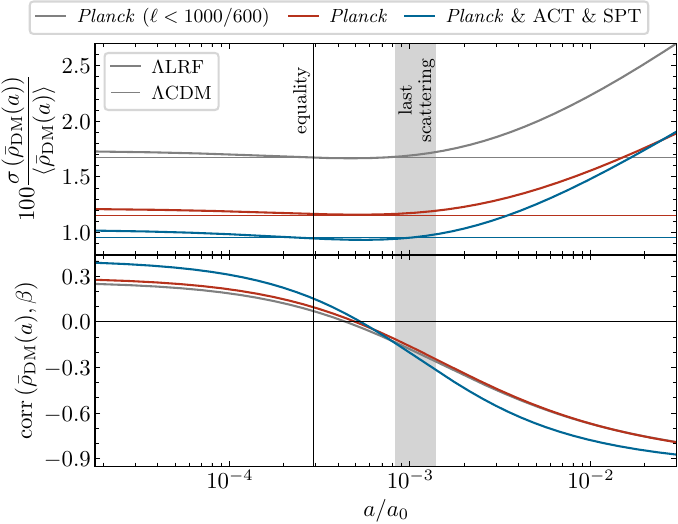}
    \caption{
        Calibration of the dark matter density near last scattering by CMB temperature and
        polarization data.
        Under a massless, linearly coupled mediator, dark matter begins redshifting faster than CDM
        around matter-radiation equality.
        Since the shape of the temperature and polarization spectra are most sensitive to the
        background evolution of dark matter in the epoch leading up to when CMB photons last scatter
        (vertical grey band), CMB data most strongly constrain the dark matter abundance at this
        time (top panel).
        Indeed, the dark matter abundance shortly before recombination is constrained
        with the same precision as is standard CDM (thin horizontal lines).
        Moreover, its correlation coefficient with the long-range force strength $\beta$ (bottom
        panel) vanishes at the same moment.
    }
    \label{fig:early-calibration}
\end{centering}
\end{figure}

\Cref{fig:early-calibration} confirms this expectation, displaying the posterior uncertainty in
$a^3 \bar{\rho}_\chi(a)$ as a function of redshift for this model.
While the specific shape of the uncertainty with redshift is strongly constrained by the dynamics
allowed by the model, it still shows a minimum just before last scattering where, moreover, the
precision (and central value, not show in \cref{fig:early-calibration}) matches that for the
abundance of CDM.
\Cref{fig:early-calibration} further demonstrates that the calibration of the early-time density is
uncorrelated with the coupling $\beta$ to the mediator, highlighting its insensitivity to the
dynamics of dark matter after last scattering.
In particular, the subset of \Planck{} data cut to multipoles below $1000$ in temperature and $600$
in polarization in \cref{fig:early-calibration} is effectively insensitive to late-time lensing,
providing measurements that derive almost exclusively from dynamics before last scattering.
The full dataset from \Planck{} and the combination with ACT and SPT provide more information from
the damping tail but are also increasingly sensitive to lensing; nevertheless, they lead to the same
qualitative conclusions.

\subsubsection{Propagation effects after last scattering}\label{sec:propagation-effects}

After last scattering, observed CMB photons are influenced by (cold) dark matter only
gravitationally, i.e., via their geodesic motion along the line of sight.
At the background level, the distance photons propagate between last scattering and today is
governed by the expansion history in the matter- and dark-energy--dominated epochs.
Metric perturbations source the ISW effect on large angular scales if dynamical and distort
temperature and polarization anisotropies via gravitational lensing (discussed in
\cref{sec:extrapolation}).

The angular extent of the photon sound horizon on the sky, $\theta_s$, is the best-measured summary
statistic from CMB anisotropies (with precision just above the $10^{-4}$ level~\cite{Planck:2018vyg,
ACT:2025fju, SPT-3G:2025bzu}) and remains so quite robustly in extensions of \LCDM{}.
In standard cosmology, the shape of the primary anisotropies provide a high-redshift anchor for the
densities in dark matter and baryons that tightly constrains the sound horizon at last scattering,
$r_{s, \star}$, which maps to $\theta_s \equiv r_{s, \star} / D_{M, \star}$ where $D_{M, \star}$ is
the transverse distance to last scattering (defined below).
In \LCDM{}, this combined information fixes the density $\omega_\Lambda$ of the cosmological
constant, as it provides the only remaining parameter freedom to fit $\theta_s$ via $D_{M, \star}$
(and is much more weakly constrained by its impact on the growth of structure insofar as it affects
the CMB).
The evolution of the dark-matter mass due to the LRF opens up a geometric degeneracy in the CMB,
since the DM density at late times is no longer uniquely determined by what the shape of the spectra
measure at early times.
Since the CMB best measures the dark matter density near last scattering (see
\cref{fig:early-calibration}), the slight evolution of $m_\chi$ beforehand in fact
has no discernible impact on the inferred sound horizon $r_{s, \star}$ (being measured by PR3 CMB
data as $144.5 \pm 0.3~\mathrm{Mpc}$ both without and with the LRF).

The geometric information in the CMB thus effectively remains a constraint on the distance to last
scattering, which is sensitive to the integrated evolution of dark matter insofar as it determines
the line-of-sight comoving distance~\cite{Hogg:1999ad} (not to be confused with the DM field
$\chi$),
\begin{align}
    \chi_\mathrm{C}(a)
    &= \int_{a}^{a_0} \frac{\ud \tilde{a}}{\tilde{a}} \,
        \frac{1}{\tilde{a} H(\tilde{a})}
    ,
    \label{eqn:comoving-distance}
\end{align}
between recombination and the present (here fixing spatially flat cosmologies, such that the
transverse distance $D_M = \chi_\mathrm{C}$).
Since photons travel the furthest when the horizon is the largest, \cref{eqn:comoving-distance} is
dominated by low redshifts, at which point the dark matter mass has evolved much more substantially
than at last scattering.
Recalling \cref{eqn:massless-mediator-matter-era-bg-soln} and that
$a_\star^3 \bar{\rho}_\chi(a_\star) \approx \left( 1 - \beta f_\chi \right) \lim_{a \to 0} a^3 \bar{\rho}_\chi(a)$
(neglecting the small evolution between $a_\mathrm{CMB}$ and $a_\star$)
and accounting for the mediator's fractional contribution $\beta f_\chi^2 / 3$ to the matter
density at $a \gg a_\mathrm{eq}$,
\begin{align}
    \frac{a^3 \bar{\rho}_m(a)}{a_\star^3 \bar{\rho}_m(a_\star)} - 1
    &\approx - \beta f_\chi^2 \left( \ln \frac{a}{a_\mathrm{eq}} - 1.7 \right)
    \label{eqn:rho-m-apx-evolution}
\end{align}
provides an excellent approximation during matter domination.
Equality of matter and dark energy occurs around $a / a_0 = 0.77$ in \LCDM{}, at which point the
dark matter density is smaller than what would be in CDM by a factor of $1 - 6.2 \beta f_\chi^2$.

Deviations in the expansion history are thus nominally six times greater at late times than around
recombination.
Because dark energy becomes important so near the present, however, its density must be increased
disproportionately to compensate for the reduced matter density in fixing $D_{M, \star}$, in
close analogy to the reduction required with massive neutrinos~\cite{Planck:2018vyg,
Baryakhtar:2024rky, Loverde:2024nfi} or the increase required if a fraction of dark matter decays
after last scattering~\cite{Lynch:2025ine}.
Refs.~\cite{Baryakhtar:2024rky, Loverde:2024nfi} computed the geometric degeneracies that hold at
fixed $\omega_\mathrm{DM}(a_\star)$, $\omega_b$, and $\theta_s$ when allowing the matter density to
differ before and after recombination.\footnote{
    The canonical such example is massive neutrinos, which contribute to the matter density when they become nonrelativistic after recombination. The comoving matter densities at late and early times differ by a factor $1+f_\nu$, where $f_\nu$ is the neutrino density fraction today.
    In the LRF scenario, the comoving dark matter density varies by the redshifting of the dark matter mass, which is approximately captured by translating the results in Ref.~\cite{Baryakhtar:2024rky, Loverde:2024nfi} as in \cref{eqn:Wm-approx-degeneracy}.
}
The present-day matter fraction varies as
\begin{align}
    \Omega_m
    &\equiv \frac{\bar{\rho}_m(a_0)}{\bar{\rho}(a_0)}
    \propto \left( \frac{a_0^3 \bar{\rho}_m(a_0)}{a_\star^3 \bar{\rho}_m(a_\star)} \right)^{5}
    \label{eqn:Wm-approx-degeneracy}
\end{align}
which implies that the Hubble constant
$h \propto \left[ a_0^3 \bar{\rho}_m(a_0) / a_\star^3 \bar{\rho}_m(a_\star) \right]^{-2}$.
The latter scaling points to the potential of LRF models to alleviate the Hubble
tension~\cite{Archidiacono:2022iuu, Pitrou:2023swx, Uzan:2023dsk}.
Inserting \cref{eqn:rho-m-apx-evolution} into \cref{eqn:Wm-approx-degeneracy} suggests that the
present matter fraction decreases by as much as \emph{thirty} times $\beta f_\chi^2$, resulting in
an increase to the Hubble constant $h \propto (1 + \beta f_\chi^2)^{12}$.
Because $\chi$ does not redshift with $a^{-3}$ as the above analytic results assume [in particular,
since the comoving distance \cref{eqn:comoving-distance} has substantial support at moderately
higher redshifts when $a^3 \bar{\rho}_\chi$ is higher than its value at the present], the actual
scaling is slightly shallower:
\begin{subequations}\label{eqn:Wm-h-scaling}
\begin{align}
    \Omega_m
    &\propto (1 + \beta f_\chi^2)^{-24}
    \intertext{and}
    h
    &\propto (1 + \beta f_\chi^2)^{9}.
\end{align}
\end{subequations}
(For the same reason, the present matter fraction $\Omega_m$ is not as useful a summary statistic as
in \LCDM{}, since it does not uniquely parametrize the low-redshift expansion history.)

Finally, the modified evolution of the dark matter background transiently impacts the dynamics of
metric potentials after decoupling, as radiation becomes subdominant at a different rate.
The early-time ISW effect is therefore enhanced at $20 \lesssim \ell \lesssim 200$, as displayed in
the inset panel in \cref{fig:cmb-isw-lensing-sensitivity}, but only with marginal sensitivity of
order $2 \beta f_\chi^2$ because it is only sourced in the first $e$-folds after equality.
The cancellation between modifications to the background and to perturbations discussed in
\cref{sec:linear-coupling} greatly diminish the would-be effect of the enhanced growth of
$\delta_\chi$, for which reason the early ISW effect is only appreciably modified when radiation is
nonnegligibly abundant rather than throughout the matter era.\footnote{
    Ref.~\cite{Copeland:2006wr} relatedly noted the absence of an ISW effect in coupled dark energy
    models during matter domination.
}
We discuss the late-time ISW effect in the following section in parallel with CMB lensing and
low-redshift distances, as the main impact of the LRF on all three derives from the aforementioned
modifications to the onset of dark-energy domination incurred by fixing $\theta_s$.

\subsection{Predictions for low-redshift observables}\label{sec:extrapolation}

When the mediator's early-time dynamics are sourced only by its coupling to dark matter, as in
\cref{sec:linear-coupling}, the evolution of the dark matter density only differs appreciably from
CDM in the single $e$-fold of expansion between matter-radiation equality and last scattering.
Likewise, the perturbations that imprint on small-scale CMB anisotropies only entered the horizon a
handful of $e$-folds before recombination.
Observables that depend on the seven $e$-folds between recombination and the present offer greater
leverage to constrain deviations of the dark matter's dynamics from
CDM's~\cite{Archidiacono:2022iuu}.
In this section we discuss the predictions for late-time observables (CMB lensing in
\cref{sec:lensing} and low-redshift distances in \cref{sec:distances}) within the LRF model as
calibrated to the primary CMB (as a constraint on early-time dynamics).

\subsubsection{Gravitational lensing}\label{sec:lensing}

The consequences for low-redshift observables of the severe modification to late-time expansion
history incurred by fixing the distance to last scattering are best encoded by the early onset of
dark-energy domination~\cite{Baryakhtar:2024rky, Loverde:2024nfi}.
Equality between matter and dark energy occurs at
$\amL / a_0 \propto (1 + \beta f_\chi^2)^{-11.7}$.
While CMB lensing is sensitive to structure at slightly larger redshifts than other observables from
galaxy surveys, i.e., $1 \lesssim z \lesssim 5$~\cite{Lewis:2006fu}, an appreciable suppression of
scale-independent growth due to dark energy starts substantially earlier because of the steep
sensitivity of $\amL$ to the coupling strength.
Since the dark force does not enhance the growth of lensing potentials in matter domination
(\cref{sec:structure-growth-after-decoupling}), its net impact is to suppress growth after last
scattering through the correlated increase in $\omega_\Lambda$ imposed by fixing
$\theta_s$---namely, the growth rate decreases rapidly with $\beta$ at $z \lesssim 8$, as evident in
\cref{fig:growth-deviation}, with sensitivity saturating at $-7$ or so by the present.

\begin{figure}[t!]
\begin{centering}
    \includegraphics[width=\textwidth]{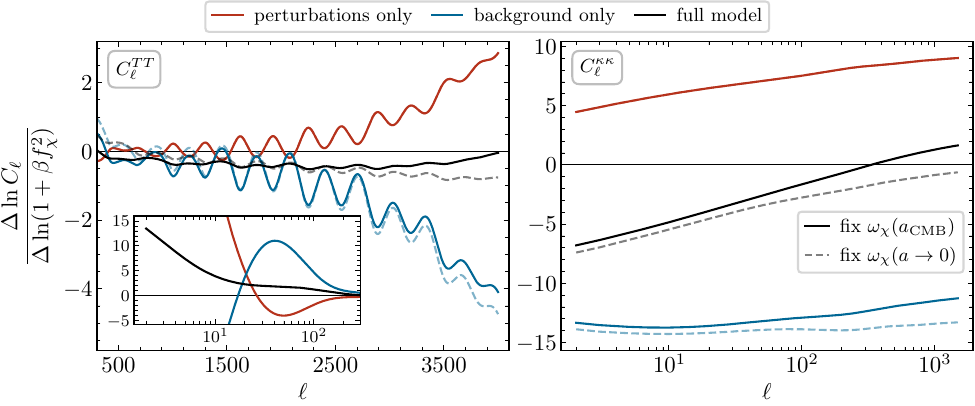}
    \caption{
        Sensitivity of lensed CMB temperature anisotropies (left) and the CMB lensing spectrum
        (right) to a scalar-mediated long-range force acting on dark matter, measured as described
        in \cref{fig:cmb-sensitivity}.
        The inset panel displays the impact of the ISW effect for lower multipoles on different axes
        scales.
        All results fix the angular size of the sound horizon $\theta_s$ and either the dark matter
        density at last scattering (opaque lines), as motivated by \cref{fig:early-calibration}, or
        the early-time dark matter density $\omega_\chi(a \to 0)$ (transparent lines), for comparison
        with \cref{fig:cmb-sensitivity}.
        Lensing is suppressed on all observationally relevant scales in the latter case; the CMB's
        calibration of $\omega_\mathrm{DM}(a_\mathrm{CMB})$ alters the shape of the transfer
        function such that small-scale lensing is marginally enhanced by $< 2 \beta f_\chi^2$,
        deriving from changes to the background evolution.
        These scales are not those for which current data exhibit an excess and are also
        nonnegligibly impacted by nonlinear structure formation (not accounted for here given its
        lack of study under additional long-range forces); see \cref{fig:lensing-spaghetti}.
        Like \cref{fig:cmb-sensitivity}, red, blue, and black lines respectively depict results when
        modifying the dynamics of perturbations, the background, or both, demonstrating the
        substantial cancellation between the two effects.
    }
    \label{fig:cmb-isw-lensing-sensitivity}
\end{centering}
\end{figure}

\Cref{fig:cmb-isw-lensing-sensitivity} demonstrates the substantial cancellation between the
modified dynamics of the dark matter background and perturbations as they impact CMB lensing.
The case that leaves the equation of motion for $\delta_\chi$ unchanged shows a substantial
suppression of lensing, both in the lensing potential itself and its impact on CMB temperature
anisotropies at $\ell \gg 2000$, due to the earlier onset of dark-energy domination.
(CMB polarization anisotropies exhibit similar features.)
In the consistent implementation, the enhanced growth of the dark matter density contrast due to the
long-range force does not overcompensate---in particular, the sensitivity coefficient of the CMB
temperature spectrum is negative at all $\ell$ between $500$ and $4000$, i.e., including scales
where lensing is important.
The sensitivity of the lensing spectrum $C_\ell^{\kappa\kappa}$ itself is negative
at low and moderate multipoles (as a consequence of the earlier onset of dark-energy domination) and
only marginally exceeds zero at high $\ell$ when holding $\omega_\chi(a_\mathrm{CMB})$ fixed.
The difference between the cases that fix $\omega_\chi(a_\mathrm{CMB})$ and $\omega_\chi(a \to 0)$
points to an origin in the modified dynamics between horizon crossing and last scattering due to the
background evolution of $a^3 \bar{\rho}_\chi$, i.e., effects on the transfer function rather than
the growth of structure in the postdecoupling Universe.
The relative change in $a^3 \bar{\rho}_m$ is smaller in magnitude than $\beta f_\chi^2$ by
recombination, which in practice is smaller than the effects from the intrinsic variation in the
CMB's measurement of $\omega_\chi(a_\mathrm{CMB})$ alone (as evident in
\cref{fig:early-calibration}).

\begin{figure}[t!]
\begin{centering}
    \includegraphics[width=\textwidth]{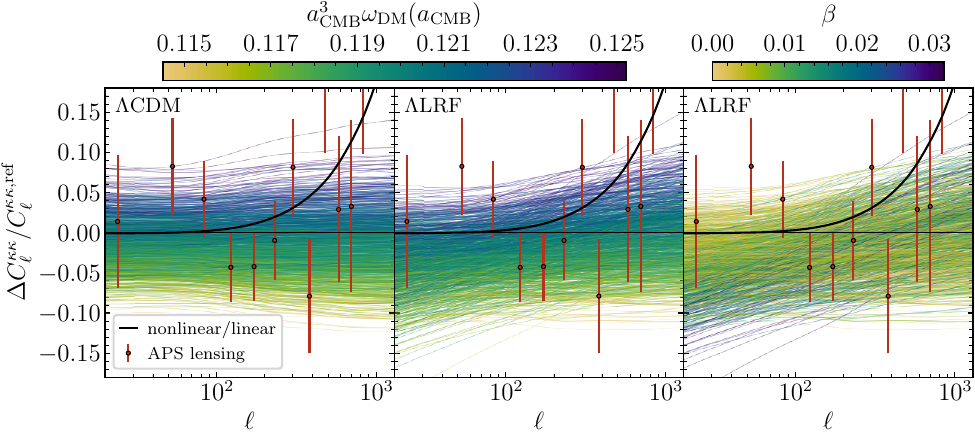}
    \caption{
        Residuals of the CMB lensing convergence relative to the \LCDM{} best fit (to all \Planck{}
        PR3 temperature and polarization data) for samples from posteriors for \LCDM{} (left) and for
        the dark force model (middle and right), each calibrated to PR3 temperature data at
        $\ell \leq 1000$ and polarization and the cross spectrum at $\ell \leq 600$.
        This subset is essentially insensitive to lensing; the comparison to the best fit to all PR3
        data indicates the degree to which each model, calibrated to the unlensed CMB, underpredicts
        the lensing amplitude preferred by the full dataset.
        Curves in the left and middle panels are colored by the comoving dark matter density at
        recombination, $a_\mathrm{CMB}^3 \omega_\mathrm{DM}(a_\mathrm{CMB})$, which reduces to
        $\omega_c$ in \LCDM{}; the right panel colors by the LRF strength relative to gravity,
        $\beta$.
        Joint CMB lensing reconstruction data (which are not included in the fits) are displayed in
        red~\cite{Carron:2022eyg, ACT:2023dou, ACT:2023dou, SPT-3G:2024atg, ACT:2025qjh}.
        Solid black lines show the residual for the reference \LCDM{} cosmology when modeling
        nonlinear structure growth, which indicates the error made by neglecting it (i.e., because
        we presently lack a nonlinear model that accounts for the dark force).
    }
    \label{fig:lensing-spaghetti}
\end{centering}
\end{figure}

To contextualize the lensing signatures displayed in \cref{fig:cmb-isw-lensing-sensitivity},
\cref{fig:lensing-spaghetti} presents the relative change in $C_\ell^{\kappa\kappa}$ for a sample of
a posterior calibrated to CMB temperature and polarization data from the lensing-insensitive subset
of \Planck{} PR3 described on page~\pageref{text:data-description}.
Comparing with the \LCDM{} results in \cref{fig:lensing-spaghetti} indicates that most of the
variability derives from \LCDM{} physics, i.e., the shape of the transfer function as determined by
the matter density at early times (and, naturally, the primordial power spectrum).
Just as expected from \cref{fig:cmb-isw-lensing-sensitivity}, the LRF itself suppresses power at
large scales and modestly enhances it at small ones, which opposes the relatively prominent trend in
the residuals of observed data at $\ell \lesssim 300$.
These larger scales are those for which current lensing reconstruction observations are signal
dominated~\cite{ACT:2025qjh} and also that are most responsible for the smoothing of the acoustic
peaks (which also exhibit an excess).
\Cref{fig:lensing-spaghetti-extra} in \cref{app:supplementary-results} presents results analogous to
\cref{fig:lensing-spaghetti} for posteriors calibrated to more CMB data (all of \Planck{} and its
combination with ACT DR6 and SPT-3G D1).

Furthermore, \cref{fig:lensing-spaghetti} shows that the impact of nonlinear structure formation on
CMB lensing reconstruction exceeds the typical posterior variation at the higher multipoles that
experience any enhancement.\footnote{
    CMB temperature and polarization are also sensitive to nonlinear structure growth, but only on
    very small scales---the impact exceeds a percent only at $\ell \gtrsim 3000$, at which point
    foregrounds and resolution degrade the precision of current measurements.
}
Without an extension of existing models of nonlinear structure growth~\cite{Mead:2020vgs} that
consistently accounts for the dark force and the mediator's dynamics, we can draw no
strong conclusions from CMB lensing on these scales (nor use any measurements thereof in parameter
inference) except that the impact of the dark force specifically on linear dynamics is unlikely to
be especially important.
Since CMB lensing reconstruction measurements mostly skew high at low multipoles where the impact of
the LRF definitively suppresses structure and nonlinear effects are negligible (the same scales that
would be predominantly responsible for the excess smoothing of the primary CMB), a dark force due to
a linearly coupled, massless mediator is a poor candidate explanation for the $6$ to $8\%$ lensing
excess measured via phenomenological rescalings of CMB lensing potentials~\cite{SPT-3G:2024atg}.

Finally, the weak sensitivity of the small-scale CMB (evident in
\cref{fig:cmb-isw-lensing-sensitivity}) positions it as a poor probe of dark forces.
The CMB is most sensitive in temperature at very low multipoles, owing to the enhancement of the
late-time ISW effect from the correlated decrease of $\amL$, with coefficient varying from $2$ to
$15$ below $\ell = 30$.
Such low multipoles are subject to the largest sample variance ($\sim 1 / \sqrt{2 \ell + 1}$);
however, existing measurements thereof prefer lower power than predicted in
\LCDM{}~\cite{Planck:2019nip}, which modestly disfavors larger dark energy densities (i.e., lower
$\amL$) in cosmologies with geometric degeneracies that can otherwise accommodate
them~\cite{Planck:2018vyg, Ivanov:2020mfr, Baryakhtar:2024rky}.
On the other hand, the results presented in \cref{fig:cmb-isw-lensing-sensitivity} that artificially
disable modifications to the background or to perturbations exhibit far more severe sensitivity
through the ISW effect, reaching $-30$ and $55$, respectively.
Such extreme effects again suggest that phenomenological modifications to the redshifting or
clustering of dark matter would be unrealistic or at the least challenging to interpret in relation
to consistent microphysical theories.

\subsubsection{Low-redshift distances}\label{sec:distances}

Low-redshift distance measurements, in combination with CMB data, stand to be much more informative
than CMB lensing because of the steep sensitivity of the late-time expansion history to the
dark force at fixed $\theta_s$.
At the present, the most precise such observable is the BAO feature extracted from spectroscopic
galaxy surveys.\footnote{
    Uncalibrated supernova luminosities, as a probe of relative distances, also break the
    low-redshift degeneracy.
    We do not present these results for the sake of brevity, but we note that, given their
    preference for substantially larger matter fractions (in \LCDM{}) than BAO
    data~\cite{DES:2024jxu}, current supernova distance datasets combined with CMB data would likely
    place tighter upper limits.
}
Cosmological inference from the acoustic scale, including $\theta_s$ from CMB data, are fully
specified in \LCDM{} by the present-day matter fraction $\Omega_m$ and density relative to the drag
horizon squared, $\omega_m r_\mathrm{d}^2$~\cite{Eisenstein:2004an, Loverde:2024nfi}; so
parametrized, the inference is effectively independent of physics before decoupling (which remains
true when augmented with, e.g., spatial curvature or dark energy dynamics).

The dark force has the greatest impact on the distance to last scattering, which integrates
over nearly the entire matter era.
As established in \cref{sec:calibration}, the primary CMB is only weakly sensitive to the LRF
coupling itself, instead measuring best the density in dark matter around last scattering; the
calibration of the drag horizon $r_\mathrm{d}$ by the CMB is thus only weakly modified.
Since dark matter's mass evolves by $\sim \beta f_\chi^2$ per $e$-fold
[\cref{eqn:massless-mediator-matter-era-bg-soln}], the late-time expansion history relevant to BAO
distances resembles a \LCDM{} cosmology with a comoving dark matter density (i.e.,
$a^3 \bar{\rho}_\chi$) anchored to its average value in the observational interval
($0 < z \lesssim 4$).
This qualitative picture suggests that scalar-mediated forces could reconcile the matter-density
deficit that is partly responsible for CMB and BAO data's incompatibility with the neutrino
masses~\cite{Loverde:2024nfi, Lynch:2025ine} expected from neutrino oscillation
experiments~\cite{Esteban:2020cvm, deSalas:2020pgw}; we discuss this connection in \cref{sec:mnu}.

\begin{figure}[t!]
\begin{centering}
    \includegraphics[width=3.2in]{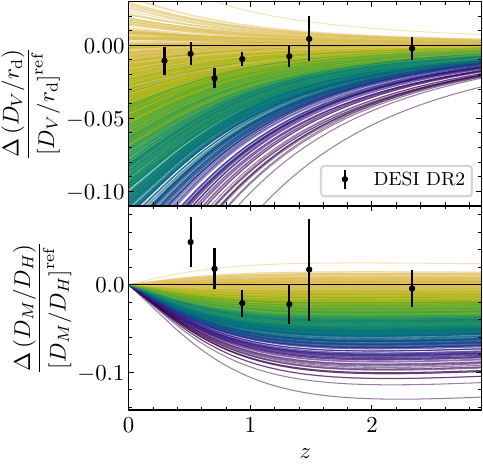}
    \includegraphics[width=3.2in]{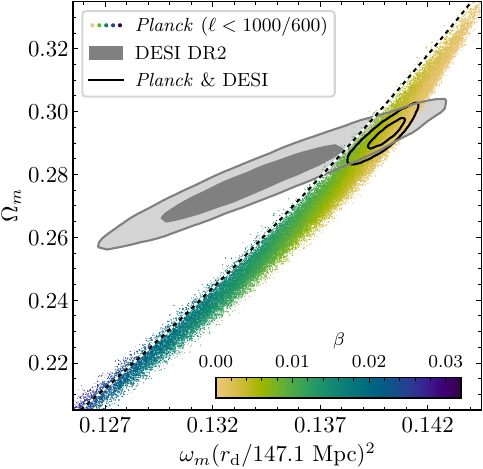}
    \caption{
        Predictions for baryon acoustic oscillation measurements calibrated by PR3 temperature data
        at $\ell \leq 1000$ and polarization and the cross spectrum at $\ell \leq 600$ (as in
        \cref{fig:lensing-spaghetti}).
        In both panels, colors label the long-range force strength $\beta$ (with the same scale
        as in \cref{fig:lensing-spaghetti}).
        The left panels display the isotropic BAO scale $D_V / r_\mathrm{d}$ and the ratio of the
        transverse and longitudinal BAO scales $D_M / D_H$ relative to the \LCDM{} best fit (to all
        \Planck{} PR3 temperature and polarization data), with measurements from DESI DR2
        superimposed in black.
        The right panel depicts posterior distributions from the aforementioned subset of \Planck{}
        data (colored scatter) and DESI DR2 BAO data (grey) alone and combined (black, unfilled),
        the contours marking $1$ and $2 \sigma$ regions (i.e., $39.3\%$ and $86.5\%$ mass levels).
        Posteriors are presented in the plane that fully parametrizes the acoustic scale in \LCDM{},
        independent of physics before photon/baryon decoupling---namely, the present-day matter
        fraction and density relative to the drag horizon squared, $\Omega_m$ and $\omega_m
        r_\mathrm{d}^2$~\cite{Eisenstein:2004an, Loverde:2024nfi}.
        The black and white dashed line indicates the degeneracy direction
        \cref{eqn:Wm-h-scaling} from fixing $\theta_s$.
    }
    \label{fig:bao}
\end{centering}
\end{figure}

\Cref{fig:bao} depicts the substantial geometric degeneracy realized by the dark force when
calibrated to primary CMB data.\footnote{
    Reference~\cite{Archidiacono:2022iuu} notes that the relative density perturbation between
    baryons and dark matter generated by the LRF could in principle modify the inferred BAO
    position but argues that such effects are likely negligible for observationally relevant
    parameters.
}
The CMB's calibration aligns precisely with the expectation from \cref{eqn:Wm-h-scaling} that
$\Omega_m \propto (1 + \beta f_\chi^2)^{-24}$ and
$\omega_m r_\mathrm{d}^2 \propto (1 + \beta f_\chi^2)^{-6.6}$.
The posterior samples of BAO observables versus redshift in \cref{fig:bao} illustrate how the
dark force accommodates features in DESI data that are not well explained in
\LCDM{}---foremost a reduction in the isotropic BAO scale $D_V(z) = \sqrt[3]{z D_M(z)^2 D_H(z)}$
[where $D_H(z) \equiv 1 / H(z)$] with decreasing $z$, since the Universe has a higher density in the
dark energy era at fixed $\theta_s$.
DESI measurements of the anisotropic factor $D_M(z) / D_H(z)$ also exhibit a monotonic trend at low
redshifts relative to \Planck{}'s preferred \LCDM{} cosmology, which roughly aligns with the
predictions of more strongly coupled scenarios; however, the positive residuals cannot be explained
because, in the trade off between the decaying dark matter mass and the increased dark energy
density from fixing $\theta_s$, transverse distances shrink more than $H \propto \sqrt{\bar{\rho}}$
increases at any redshift.

The net result of these expansion-history effects is a mild preference for a nonzero coupling to a
light mediator with strength relative to gravity $10^3 \beta = 2.7^{+1.9}_{-1.6}$, in line with the
results from DESI's first data release presented in Ref.~\cite{Bottaro:2024pcb} and from DR2 in
scenarios where the mediator has an exponential potential and makes up the dark
energy~\cite{Bedroya:2025fwh}.
These measurements are not insensitive to priors: a uniform prior on $d_{m_\chi}^{(1)}$ penalizes
larger $\beta$, shifting posteriors to, e.g., $10^3 \beta = 1.8^{+2.0}_{-1.5}$ for \Planck{}
combined with DESI DR2.
On the other hand, marginalizing over the neutrino mass sum (rather than fixing it to zero)
favors larger $\beta$, as discussed in \cref{sec:mnu}.

\section{Discussion}
\label{sec:discussion}

A massless scalar mediator linearly coupled to dark matter is the simplest prototype of a number of
proposed dark sector models~\cite{Amendola:1999er}, including for instance those motivated to mimic
``phantom'' dark energy~\cite{Das:2005yj, Khoury:2025txd}, inspired by string theory and conjectures
on quantum gravity~\cite{Agrawal:2018own, Agrawal:2019dlm, McDonough:2021pdg, Bedroya:2025fwh,
Anchordoqui:2025fgz, Andriot:2025los, Smith:2024ibv, Smith:2024ayu}, or to explain coincidence
problems in theories of early dark energy~\cite{McDonough:2021pdg, Karwal:2021vpk, Lin:2022phm,
Smith:2025grk}.
The analysis of \cref{sec:structure-growth,sec:suppression} clarifies the manifestation of
long-range forces in genuine cosmological observables---in particular highlighting the outsized
importance of background dynamics on not just low-redshift distances and
structure~\cite{Archidiacono:2022iuu} but also the generation of anisotropies at last scattering
(\cref{sec:calibration}).
In this section, we discuss the broader implications of our results for contemporary observations,
in particular those that inform neutrino masses (\cref{sec:mnu}) and inference of structure from
cosmic shear (\cref{sec:cosmic-shear}).
Using the analytic results of \cref{sec:structure-growth} as a guide, in \cref{sec:nonminimal} we
explain why mediators with bare masses only further suppress structure growth and explore how one
might engineer a nonminimal model that genuinely enhances it.
Finally, in \cref{sec:nonlinear} we discuss modeling developments necessary to test dark force
models against future (and even current) measurements of CMB lensing and small-scale temperature and
polarization anisotropies, and we also comment on possible implications of our results for the
interpretation of galaxy clustering.

\subsection{Neutrino masses, positive and ``negative''}\label{sec:mnu}

Long-range forces acting on dark matter are of particular contemporary interest because of their
purported potential to explain the incompatibility of present cosmological data with massive
neutrinos~\cite{Planck:2018vyg, DESI:2024mwx, DESI:2025zgx}.
Reference~\cite{Craig:2024tky} rephrased the phenomenological lensing rescaling parameter
$A_\mathrm{lens}$, long known to be measured greater than unity by multiple generations of CMB
data~\cite{Calabrese:2008rt, Planck:2013pxb, Planck:2015fie, Planck:2019nip, SPT-3G:2024atg,
ACT:2025fju, SPT-3G:2025bzu}, in terms of a signed neutrino mass.\footnote{
    See Ref.~\cite{SPT-3G:2024atg} for exposition on the interpretation of constraints
    on $A_\mathrm{lens}$ parameters that derive separately and jointly from CMB lensing
    reconstruction and from temperature and polarization anisotropies.
}
Using a toy model in which dark matter couples to itself with a larger gravitational constant,
Ref.~\cite{Craig:2024tky} claimed that long-range dark sector forces could mediate the CMB lensing
excess.
Reference~\cite{Archidiacono:2022iuu}, however, had previously observed in scalar-mediated models
that CMB lensing is not enhanced to the degree expected from analytic results for the matter power
spectrum, which was further supported by parameter inference presented in
Ref.~\cite{Bottaro:2024pcb}.
As established in \cref{sec:extrapolation}, the net effect of a LRF mediated by a massless scalar is
a suppression of CMB lensing; the main impact for neutrino masses is therefore to mediate the
geometric tension between CMB and BAO measurements, which Ref.~\cite{Loverde:2024nfi} showed is at
least as important as the CMB lensing excess in contemporary neutrino mass limits.

More recently, Ref.~\cite{Graham:2025dqn} extended the study of an effective, signed neutrino mass
impacting the lensing amplitude to include a second such parameter that modulates BAO observables,
seeking to quantify the relationship between the two tensions.
Quoting results from Refs.~\cite{Archidiacono:2022iuu, Bottaro:2023wkd, Bottaro:2024pcb} for the
evolution of the dark matter density
[that $a^3 \bar{\rho}_\chi(a) \propto 1 - \beta f_\chi \ln (a / a_\mathrm{eq})$\footnote{
    Ref.~\cite{Graham:2025dqn} discusses the impact of new physics on the expansion
    history via $\Omega_m$, which typically denotes the matter fraction rather than the matter
    density $\omega_m \equiv \Omega_m h^2$, and writes
    $\Omega_m \propto 1 - 6 \beta f_\chi^2$; the distinction between the two variables is
    significant~\cite{Loverde:2024nfi} [see \cref{eqn:Wm-h-scaling}], and this particular
    scaling applies to the matter density.
}] and density perturbations
[that $\delta_m / a \propto 1 + 6 / 5 \cdot \beta f_\chi^2 \ln (a / a_\mathrm{eq})$],
Ref.~\cite{Graham:2025dqn} maps the long-range force strength $\beta$ into these two effective
neutrino masses.

Our results demonstrate two key issues with this approach.
First, \cref{sec:subhorizon-growth} shows that the matter power spectrum is a poor proxy for weak
lensing observables and that lensing potentials grow no faster than in \LCDM{}.
Second, the phenomenological model of Ref.~\cite{Graham:2025dqn} only accounts for the impact of
expansion history modifications on BAO observables and not on the CMB or the growth of structure
itself.
The adjustment to the dark energy density required to fix $\theta_s$ leads to the net suppression of
lensing by scalar-mediated forces demonstrated in \cref{sec:extrapolation}, just as it offsets some
of the suppression of structure growth due to massive neutrinos~\cite{Pan:2015bgi, Loverde:2024nfi}.
Reference~\cite{Graham:2025dqn}, which did not test a consistent implementation of the dark force,
neglects the correlated impact of expansion-history modifications on the growth of structure; their
model therefore assumes the LRF enhances structure and mediates the matter density deficit in
tandem, whereas in reality the improved fit to BAO data with nonzero $\beta$ incurs a suppression of
structure.

\begin{figure}[t!]
\begin{centering}
    \includegraphics[width=\textwidth]{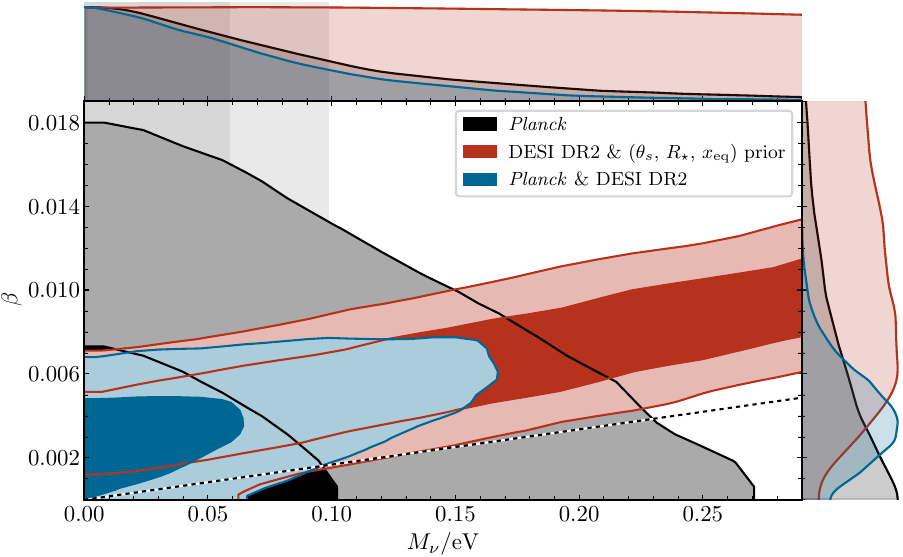}
    \caption{
        Joint posterior distribution over the neutrino mass sum $\summnu$ and the long-range
        force strength $\beta$ deriving from \Planck{} PR3 CMB data alone (black), excluding
        lensing reconstruction; DESI DR2 BAO data with a PR3 prior on $\theta_s$, the
        baryon-to-photon and radiation-to-matter ratios at recombination, $R_\star$ and
        $x_\mathrm{eq}$ (red); and PR3 and DESI DR2 combined (blue).
        The acoustic scale constraints (red) exhibit the geometric degeneracy that essentially fixes
        the late-time matter density, while the CMB limits the net suppression of large-scale
        lensing (via its effect on two-point statistics) incurred by both massive neutrinos and the
        dark force.
        Marginalization over $\beta$ thus lifts the impact of BAO data but not lensing information
        on neutrino mass limits.
        Vertical grey shading indicates neutrino mass sums incompatible with the normal and
        inverted hierarchies, and the dashed line shows the degeneracy that fixes the matter
        density at redshift $z \approx 2$, using the analytic result \cref{eqn:rho-m-apx-evolution}.
        The lower left panel displays the 1 and $2 \sigma$ contours (i.e., the $39.3\%$ and $86.5\%$
        mass levels) of the two-dimensional marginal posterior density, and the outer panels depict
        kernel density estimates of the one-dimensional marginal posteriors normalized relative to
        the peak density.
    }
    \label{fig:mnu-beta}
\end{centering}
\end{figure}

Scalar-mediated dark forces provide a particularly striking example of the interplay between the
expansion history and growth of structure as probed by cosmological observations, highlighting the
importance of consistently testing concrete models of new physics.
In general, it is challenging to interpret the microphysical implications of phenomenological models
that compartmentalize effects by observable rather than physical inputs (like the expansion history
and the growth rate of structure), in particular when these physical effects impact multiple
observables nontrivially.
We illustrate the actual impact of the dark force on the inference of neutrino masses in
\cref{fig:mnu-beta}.
Background effects---the combination of DESI DR2 BAO data with a prior on $\theta_s$ and the
baryon-to-photon and radiation-to-matter ratios at recombination, $R_\star$ and $x_\mathrm{eq}$ (see
Refs.~\cite{Baryakhtar:2024rky,Loverde:2024nfi})---exhibit a strong degeneracy, reflecting that the
dark matter mass evolution allows for larger contributions to the late-time matter density from
neutrinos.
The posterior degeneracy matches the analytic result \cref{eqn:rho-m-apx-evolution}, i.e.,
attributing the drop in the dark matter density by redshifts $\sim 2$ to neutrinos with mass sum
$\Delta \omega_\chi \cdot 93.1~\mathrm{eV} \approx 60 \beta~\mathrm{eV}$, up to the impact of the
faster redshifting of dark matter relative to the reduced value at this time.

\begin{figure}[t!]
\begin{centering}
    \includegraphics[width=\textwidth]{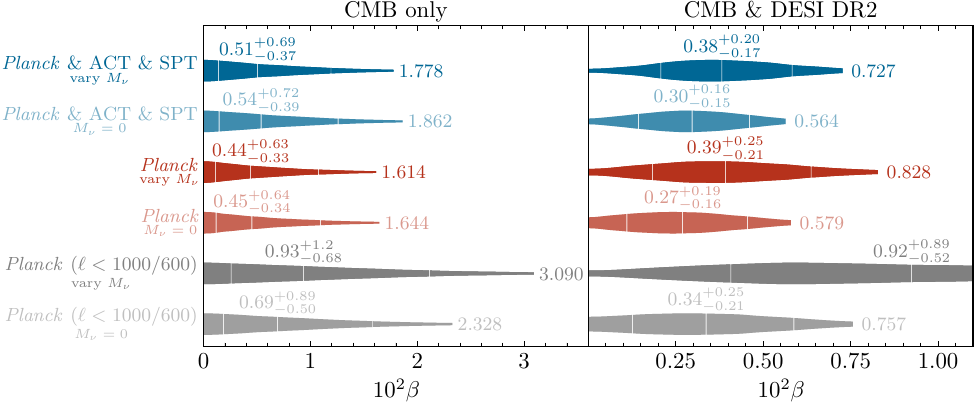}
    \caption{
        Marginal posterior distributions over the long-range force strength relative to gravity,
        $\beta$, deriving from CMB temperature and polarization anisotropies alone (left) and
        in combination with DESI DR2 BAO data (right).
        Results that vary the neutrino mass sum or fix it to zero are solid and transparent,
        respectively, and the CMB datasets are the same as appear in \cref{fig:early-calibration}.
        Posteriors are truncated at the $95$th percentile (whose value is labeled); vertical
        white lines indicate the median and $\pm 1 \sigma$ quantiles (also indicated above each
        distribution).
    }
    \label{fig:beta-violin}
\end{centering}
\end{figure}

As argued, dark forces from a massless scalar mediator thus remove much of the impact of BAO data on
neutrino mass limits when combined with CMB data without ameliorating the penalty from the CMB
lensing excess (rather, exacerbating it).
Indeed, the unmarginalized upper limit on $M_\nu$ from CMB data decreases monotonically with
increasing values of $\beta$.
The result resembles the relaxation of neutrino mass limits in scenarios with extended geometric
degeneracies in the CMB whose late-time expansion history still resembles \LCDM{}'s, such as early
recombination~\cite{Baryakhtar:2024rky, Loverde:2024nfi}, decaying subcomponents of dark
matter~\cite{Lynch:2025ine}, or varying spatial curvature~\cite{Chen:2025mlf} (although the latter
effectively only modulates the distance to last scattering rather than all distances).
The doubling of the geometric degeneracy introduced by neutrino masses, moreover, shifts marginal
preferences from CMB data and DESI DR2 toward larger LRF strengths, as evident in
\cref{fig:beta-violin}.
The 95th percentile of the $M_\nu$ posteriors deriving from \Planck{} and DESI DR2, for instance,
are relaxed from $0.08~\mathrm{eV}$ to $0.19~\mathrm{eV}$ by marginalizing over $\beta$, whereas
both are about $0.26~\mathrm{eV}$ without BAO data.\footnote{
    The \Planck{}-only upper limit in \cref{fig:beta-violin} is slightly relaxed from the analogous
    result in Refs.~\cite{Archidiacono:2022iuu, Bottaro:2023wkd, Bottaro:2024pcb}, deriving in part
    from our broader prior on $\beta$ (and use of an exponential rather than Yukawa coupling).
    We also either fix the neutrino mass sum to zero or marginalize over it, whereas
    Refs.~\cite{Archidiacono:2022iuu, Bottaro:2023wkd, Bottaro:2024pcb} fixed a single-eigenstate
    neutrino mass sum of $0.06~\mathrm{eV}$ (rather than degenerate hierarchy), which leads to a
    slightly tighter upper limit as indicated by \cref{fig:mnu-beta}.
}

\subsection{Cosmic shear}\label{sec:cosmic-shear}

The implications of our findings for CMB lensing carry over to the weak lensing of galaxies, an
observable that of late has indicated an amplitude of matter clustering lower than that calibrated
by the CMB in \LCDM{}~\cite{KiDS:2020suj, DES:2022qpf, Wright:2025xka}, though only relatively
mildly and with varying significance.
This tension is typically summarized in terms of $\sigma_8$, the present-day, root-mean-squared
matter overdensity in spheres of radius $8/h~\mathrm{Mpc}$, or a rescaling thereof
$S_8 = \sqrt{\Omega_m / 0.3} \, \sigma_8$ which is less correlated with $\Omega_m$ in \LCDM{}.
The features of scalar-mediated dark force models, however, invalidate the effectiveness of
every aspect of these summary statistics.
First, $\sigma_8$ is a weighted integral of the power in the relative density perturbation
$\delta_m$, which is enhanced by the LRF, but lensing actually traces the absolute density
perturbation $\delta \rho$ which grows no more than in \LCDM{}.
Second, the extrapolation to the present from the actual redshift of a given galaxy sample
introduces parameter dependence irrelevant to the observable (particularly relevant given the strong
sensitivity of the late-time expansion history to the LRF as calibrated by CMB data).
Likewise, $S_8$ is unlikely to remain a better-measured combination, even were $\sigma_8$ a relevant
measure.
Finally, even when evaluated at the redshift of the galaxy sample instead of today, $\sigma_8$
coarse grains over length scales $R / h$ that correspond to different angular scales $R / h
\chi(z)$; $\sigma_8$ therefore summarizes the power spectrum at different apparent scales because $h
\chi(z)$ is not cosmology independent (except at lowest order in small $z$, valid at much smaller
redshifts than the actually observed galaxies).

\begin{figure}[t!]
\begin{centering}
    \includegraphics[width=\textwidth]{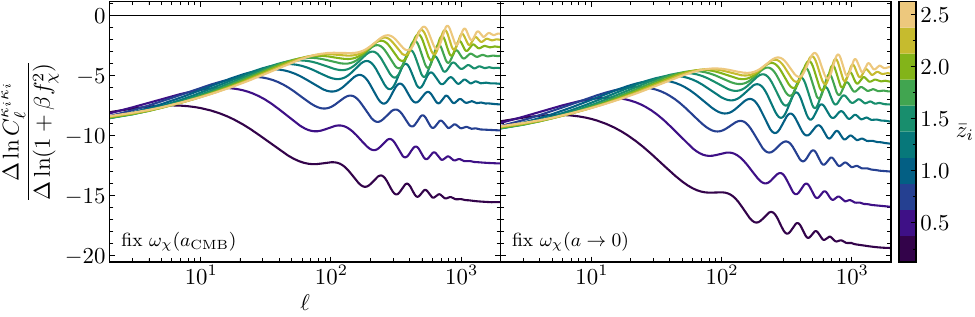}
    \caption{
        Sensitivity of the cosmic shear spectra to the dark force, taking source galaxy
        distributions centered at various redshifts (by color).
        Results fix the angular extent of the sound horizon $\theta_s$ and the dark matter
        abundance either at last scattering (left) or in the far past (right).
        Observable lensing power is strongly suppressed at lower redshifts due to the increased
        density in dark energy required to fix $\theta_s$.
        Note that nonlinear structure growth, for which we presently lack a model that accounts for
        the dark force, has a $\gtrsim 10\%$ effect in \LCDM{} at multipoles beyond $300$ to
        $600$ depending on redshift; these linear-theory results are nonetheless indicative of the
        directionality and rough degree of the effect, in particular as compared to $\sigma_8$ and
        $S_8$ which are themselves defined in terms of the linear matter power spectrum.
    }
    \label{fig:low-z-lensing}
\end{centering}
\end{figure}

Short of a complete reanalysis of cosmic shear data, we may estimate the impact of the LRF with the
sensitivity of shear power spectra over redshift, computed in the Limber approximation as
\begin{align}
    C_\ell^{\kappa_i \kappa_i}
    &= \frac{2 \pi^2 \ell^2 (\ell+1)^2}{L(\ell)^3}
        \int_0^\infty \frac{\ud \chi}{\chi} \,
        \left( 1 - \chi \int_{z(\chi)}^\infty \ud z \, \frac{n_{\kappa_i}(z)}{\chi(z)} \right)^2
        \Delta_{\Phi + \Psi}^2(\eta(\chi), L(\ell) / \chi)
    \label{eqn:C-ell-XY-limber}
\end{align}
where $L(\ell) = \sqrt{\ell (\ell + 1)}$~\cite{Limber:1954zz, LoVerde:2008re}.
For illustrative purposes, we take toy redshift distributions of source galaxies $n_{\kappa_i}$
modeled as Gaussians separated by intervals of $0.25$ each with standard deviation $0.1$; these
choices roughly correspond to, e.g., the binning employed in forecasts for the Vera C. Rubin
Observatory~\cite{LSSTScience:2009jmu} or Euclid~\cite{EuclidTheoryWorkingGroup:2012gxx}.
The shear spectra in \cref{fig:low-z-lensing} exhibit a substantially negative sensitivity to
$\beta f_\chi^2$ due to the increased dark energy density, which both suppresses the growth rate and
decreases the comoving distance to a fixed redshift, thereby projecting smaller length scales
$\chi(z) / \ell$ where the Weyl potential has less (dimensionless) power.\footnote{
    The impact of geometric projection is less relevant for CMB lensing
    (\cref{fig:cmb-isw-lensing-sensitivity}) because, in contrast to galaxy lensing, the
    line-of-sight distance to the source is also precisely measured (via $\theta_s$), rather than
    just its redshift.
    The power spectrum of the Weyl potential at fixed wave number rather than fixed $\ell$ (as in
    \cref{fig:low-z-lensing}) exhibits smaller sensitivity.
}
The results of \cref{fig:low-z-lensing} sharply contrast the increase in $\sigma_8$ due to the
long-range force, with sensitivity $\sigma_8 \propto (1 + \beta f_\chi^2)^{8}$ or so at fixed
$\theta_s$ and $\omega_\chi(a_\mathrm{CMB})$;\footnote{
    The matter power spectrum also shows substantial enhancement rather than suppression,
    since the LRF indeed enhances the growth of the density contrast.
    The conventional dimensionful matter power spectrum versus $k / h$ shows especially exaggerated
    sensitivity coefficients, as large as 40 or 50, which is an artifact of its choice of units
    that differentiate it from actual lensing observables.
} on the other hand, $S_8 \propto (1 + \beta f_\chi^2)^{-4}$, which though correct in sign still
severely underestimates the actual sensitivity of low-redshift lensing.
(With $\Omega_m$ and $\sigma_8$ evaluated at nonzero redshift, $S_8$ has positive sensitivity above
$z \simeq 0.2$-$0.3$, i.e., at observationally relevant redshifts.)
Ironically, scalar-mediated forces acting on dark matter could potentially explain the putative
deficit of structure inferred from galaxy lensing (though the preference is diminished in recent
results~\cite{Wright:2025xka} and such a suppression is nominally at odds with the excess in CMB
lensing).

\subsection{Nonminimal models}\label{sec:nonminimal}

The cancellation in the growth rate of structure during matter domination and the suppression of
structure growth incurred at fixed distance to last scattering are both particular to the dynamics
generated by the mediator's coupling to dark matter alone.
In this section we assess whether these findings extend to next-to-minimal models, using the general
result \cref{eqn:Phi-growth-rate} for the growth rate in the matter era as a guide.
The growth of structure is sensitive to scalar-mediated forces via the intrinsic enhancement of
clustering and the mediator's own gravitational effects to a comparable degree.
Nonlinear coupling functions, which we consider in \cref{sec:nonlinear-couplings}, modify both force
mediation and dynamics, while nonzero bare potentials (\cref{sec:potentials}) impact dynamics alone
(though they limit the force to a finite range).
In \cref{sec:cde} we then discuss the connection between our results and previously studied models
in which the mediator is identified as coupled dark energy (or early dark energy).

\subsubsection{Beyond linear couplings}\label{sec:nonlinear-couplings}

The solution for the growth rate of the Bardeen potentials in matter domination
\cref{eqn:Phi-growth-rate}, which makes no assumption other than that the background value of the
mediator does not evolve rapidly compared to $H$, shows that the enhancement of clustering is
sensitive to the slope of the coupling function about the mediator's instantaneous value.
If the dark matter mass only changes by a perturbatively small amount, then the system only probes
the coupling function's local gradient near its initial condition and not its global structure,
i.e., the model is effectively linear.
That said, if the Lagrangian coupling is strictly monomial rather than exponential (see
Footnote~\ref{footnote:yukawa}), then a large initial misalignment suppresses the interaction
strength~\cite{Bottaro:2023wkd}.

If one instead allows for substantial evolution, the mediator is driven to minimize its effective
potential.
Since the gradient of the coupling function sets the strength of the dark force, the mediator thus
evolves toward values where it vanishes, akin to the suppression of equivalence principle violation
from quadratically coupled scalars~\cite{Damour:1994zq, Olive:2007aj, Hees:2018fpg,
Sibiryakov:2020eir, Bouley:2022eer, Banerjee:2022sqg, Baryakhtar:2025uxs}.
In this regime, the mediator also begins evolving nonnegligibly around matter-radiation equality,
since $\bar{\rho}_\chi / \bar{\rho}$ weights the importance of the interaction in the Klein-Gordon
equation.
The dark matter mass then evolves maximally before recombination, an effect to which primary CMB
anisotropies are quite sensitive (see \cref{sec:last-scattering}).

Without a minimum, the mediator evolves monotonically and the dark matter mass (squared, if bosonic)
may cross zero.
One could imagine constructing coupling functions where the onset of evolution is parametrically
delayed after equality, i.e., if the mediator is initialized in a relatively flat part of the
coupling function (where the dark force is weak) and starts to roll later, but enhanced clustering
still has to compete against mass evolution in \cref{eqn:Phi-growth-rate}.
In any case, this regime by definition probes the nonperturbative structure of the coupling, a case
for which effective field theory (i.e., a perturbative expansion of the coupling function) is
nominally unsuitable.
In summary, while nonminimal couplings might be of interest in their own right, they do not offer
any particularly obvious means to viably enhance structure growth at late times.

\subsubsection{Beyond massless mediators}\label{sec:potentials}

Without a compelling reason to consider nonlinear couplings, we next explore the engineering of
mediator dynamics with bare potentials.
\Cref{eqn:Phi-growth-rate} shows that a necessary (though not a sufficient) condition for enhanced
growth of $\delta \rho_{\chi b}$ is that the rate of mass evolution
$\partial \ln m_\chi / \partial \varphi \cdot \ud \varphi / \ud \ln a$ be smaller than in the
massless case, as can be arranged if the scalar undergoes decaying oscillations about the minimum of
its potential.
Nonzero effective masses also limit the force range to comoving length scales smaller than
$1 / a m_\mathrm{eff}(a)$; taking $a m_\mathrm{eff}(a) \lesssim k_\mathrm{eq}$ ensures the dark
force is mediated on observable scales.
Massive, misaligned scalars, however, do not cluster below their Jeans scale
$k_J \sim a \sqrt{H m_\mathrm{eff}}$~\cite{Hu:2000ke, Amendola:2005ad}, which is a larger length
scale than the force range when the scalar oscillates, i.e., when $H < m_\mathrm{eff}$.

The natural first step---quadratic potentials---was studied in Ref.~\cite{Bottaro:2024pcb}:
crucially, the mediator's dynamics in the early, near-massless regime generate a nonzero
misalignment of order $\sqrt{\beta} f_\chi \ln(a_\mathrm{osc} / a_\mathrm{eq})$ by the time it
begins oscillating (at scale factor $a_\mathrm{osc}$); as derived in \cref{app:relic-abundance},
$f_\varphi \approx \beta f_\chi^2 \ln(m_\varphi / H_\mathrm{eq})^2 / 3$ for
$H_0 \lesssim m_\varphi \lesssim H_\mathrm{eq}$.
Once oscillating, though the growth factor retains the enhanced clustering contribution
$3/5 \cdot \beta f_\chi^2$ and $m_\chi$ regresses to its vacuum value and ceases to evolve, the
mediator's gravitational contribution in \cref{eqn:Phi-growth-rate} as a nonclustering matter
component no longer vanishes (because $w_\varphi \approx 0$ rather than $1$).
Once $m_\chi$ stops evolving appreciably, the growth indices of $\Phi_B$ and $\delta_{\chi b}$
[\cref{eqn:Phi-growth-rate,eqn:delta-growth-rate}] coincide and
for $a > a_\mathrm{osc} > a_\mathrm{eq}$ are
\begin{align}
    \dd{\ln \Phi_\mathrm{B}}{\ln a}
    \approx \dd{\ln \delta_{\chi b}}{\ln a}
    &\approx
        \underbrace{\frac{3}{5} \beta f_\chi^2}_{\substack{\text{enhanced}\\ \text{clustering}}}
        - \underbrace{\frac{3}{5} \frac{\beta f_\chi^2}{3} \ln(m_\varphi / H_\mathrm{eq})^2.}_{\substack{\text{nonclustering}\\ \text{matter}}}
    \label{eqn:growth-rate-massive}
\end{align}
The mediator's gravitational effects thus outweigh the impact of the dark force unless the scalar
begins oscillating around or before equality.
In this marginal regime of mass, $m_\varphi \simeq H(a_\mathrm{osc}) \approx H_\mathrm{eq}$, the
force range (which decreases as $1 / a m_\varphi$) soon drops below observable scales, and the
evolution of the dark matter mass and the scalar's own nontrivial gravitational effects also yield
direct signatures in the primary CMB, like in some cases discussed in
\cref{sec:nonlinear-couplings}.

\begin{figure}[t!]
\begin{centering}
    \includegraphics[width=\textwidth]{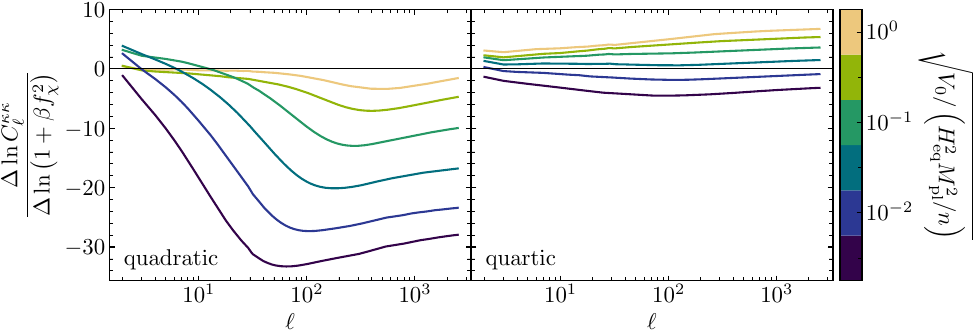}
    \caption{
        Sensitivity of CMB lensing to finite-range dark forces with quadratic (left) and quartic
        (right) potentials, depicted as in \cref{fig:cmb-isw-lensing-sensitivity} and holding
        $\theta_s$ and $\omega_\chi(a \to 0)$ fixed.
        In each case, the potential is parametrized as
        $V(\phi) = V_0 (\phi / \Mpl)^n / n$, such that a slowly rolling (and uncoupled) scalar
        with order-unity misalignment would begin oscillating around
        $3 H_\mathrm{osc}^2 = V'(\phi) / \phi$; the colors thus label approximate values of
        $H_\mathrm{osc} / H_\mathrm{eq} \sim \sqrt{n V_0 / H_\mathrm{eq}^2 \Mpl^2}$,
        although for $n \neq 2$ this relationship holds only approximately.
    }
    \label{fig:bare-potentials}
\end{centering}
\end{figure}

\Cref{fig:bare-potentials} demonstrates the severe suppression of CMB lensing incurred by
increasingly light mediators: the squared logarithmic enhancement of their abundance
$\propto \ln(m_\varphi / H_\mathrm{eq})^2 \simeq 9 \ln(a_\mathrm{osc} / a_\mathrm{eq})^2 / 4$
outweighs the shorter interval $\ln(a / a_\mathrm{osc})$ over which their suppression of structure
growth accumulates.\footnote{
    The increase in the matter density after recombination is also disfavored by current
    BAO data, which would prefer a decrement~\cite{Loverde:2024nfi, Lynch:2025ine};
    replacing BAO with supernova distance distances, however, would lead to the opposite
    conclusion for the same reason that they prefer nonzero neutrino masses~\cite{Loverde:2024nfi}.
}
Their irreducible abundance cannot even be eliminated by tuning the mediator's initial misalignment
so that it happens to reach zero by the time of oscillations: the mediator fraction $f_\varphi$ can
be no smaller than $\beta f_\chi^2 / 3 \cdot \pi^2 / 4$ [per the analytic result
\cref{eqn:hyperlight-relic}].
While such a tuning can eliminate the logarithmic enhancement for lighter mediators, the growth rate
is still no larger than
$3 \beta f_\chi^2 / 5 \cdot \left( 1 - \pi^2 / 12 \right) \approx 3 \beta f_\chi^2 / 5 \cdot 0.18$.

The massive case motivates constructions that eliminate the mediator's own gravitational effects
while preserving a nonnegligible long-range force---that is, potentials steep enough that the
mediator redshifts faster than matter.
Scalars whose dynamics are dominated by quartic self-interactions, for instance, redshift like
radiation and thus nominally permit a regime in which all effects on the growth of structure
[\cref{eqn:Phi-growth-rate}] drop out except for the enhancement of clustering, such that
$\Phi_B \propto a^{1 + 3 \beta / 5}$ or so (though the mediator's oscillations remain apparent in
the growth rate for some time).
Unlike the quadratic case, the comoving force range is now time independent~\cite{Domenech:2023afs}.
\Cref{fig:bare-potentials} shows that indeed the CMB lensing spectrum is enhanced by mediators with
sufficiently large quartic potentials.
The interval over which growth is enhanced decreases with decreasing effective mass, while the
mediator's peak abundance (at the onset of oscillations) increases just as for quadratic potentials.
For effective masses of $10^{-2} H_\mathrm{eq}$ or lower, the mediator's transient suppression of
growth when it starts to oscillate outweighs the subsequent period of enhanced growth.

The quartic examples in \cref{fig:bare-potentials}, or steep potentials more generally, thus provide
an existence proof of models that realize a regime of enhanced clustering with no side effects (at
sufficiently late times).
Yet more parameter freedom arises with bare potentials as they grant dynamical relevance to the
mediator's initial misalignment.
We leave to future work a dedicated analysis to assess whether the CMB can accommodate (or actually
prefers) such nontrivial dynamics around recombination for parameters that address the CMB lensing
excess.
A robust determination thereof requires a model of nonlinear structure growth that accounts for the
dark force in order to make reliable predictions for CMB lensing reconstruction, as discussed
in \cref{sec:nonlinear}.

\subsubsection{The mediator as dark energy}\label{sec:cde}

Coupled scalars with low-scale (but not negligible) potentials are interesting dark energy
candidates~\cite{Amendola:1999er, Burgess:2025vxs}.
Reference~\cite{Archidiacono:2022iuu} considered coupled dark energy with Yukawa interactions, for
which the large misalignments required to match the dark energy density suppress the force strength
(see Footnote~\ref{footnote:yukawa}), but strictly exponential couplings are unsuppressed.
More recently, Ref.~\cite{Bedroya:2025fwh} showed that augmenting an exponentially coupled scalar
with an exponential potential achieves a fit to cosmological data (including DESI DR2 and supernova
distances) comparable to uncoupled, evolving dark energy models with phenomenologically parametrized
equations of state~\cite{DESI:2025zgx}.
Reference~\cite{Khoury:2025txd} also proposed a model of apparently phantom dark energy featuring
axionlike dark energy with sinusoidal coupling and potential with minima offset by a half period.
In this scenario, the dark matter mass increases at late times, which enhances absolute density
perturbations and therefore lensing observables.
Reference~\cite{Khoury:2025txd} also posited that matter-radiation equality would be delayed such
that structure grows less; however, the analysis of \cref{sec:calibration} suggests that the CMB
would in general prefer mass evolution for which the scale factor of matter-radiation equality is
unchanged.
Axiodilaton dark sectors provide another example model~\cite{Smith:2024ibv, Smith:2024ayu,
Smith:2025grk}: in the limit of negligible axion pressure, dilaton potential, and couplings to the
Standard Model, the axion and dilaton in these theories reduce to the dark matter and mediator
considered here (see \cref{sec:kinetic-couplings}).

Separately, Ref.~\cite{Chen:2025ywv} suggested that the limits on $\beta$ derived in
Ref.~\cite{Archidiacono:2022iuu} for coupled dark energy models strongly constrain models with bare
potentials differing from that considered in Ref.~\cite{Archidiacono:2022iuu}.
Given that Ref.~\cite{Chen:2025ywv} performed parameter inference with similar datasets---the same
CMB data and more recent BAO data---it is unclear what unique information the results of
Ref.~\cite{Archidiacono:2022iuu} add to their analysis.
Presumably Ref.~\cite{Chen:2025ywv} neglected any modifications to the dynamics of perturbations
(which \cref{sec:calibration} shows is in fact a good approximation for some observables and scales
but an extremely poor one for others), but, as emphasized in Ref.~\cite{Archidiacono:2022iuu} and
reiterated here, the modifications to the background are the most significant signature of dark
matter couplings.
Since the constraining power of CMB lensing on modified clustering is subdominant (especially for
the \Planck{} lensing data used in Ref.~\cite{Archidiacono:2022iuu}), it is certainly not the case
that significant background effects of dark matter couplings are excluded by their impact on
structure formation.

Coupled scalars with steep potentials resembling those entertained in \cref{sec:potentials} were
previously invoked as early dark energy candidates~\cite{Karwal:2021vpk, McDonough:2021pdg,
Lin:2022phm, Smith:2025grk} to address the Hubble tension.
Reference~\cite{McDonough:2021pdg} determined that such models (specifically considering bare
potentials with sextic minima) are not simultaneously compatible with cosmic shear data because they
increase the $S_8$ parameter (however, \cref{sec:cosmic-shear} shows that $S_8$ is not in general an
appropriate summary statistic for these models).
These works, given their focus on possible interactions of an early dark energy field, did not
recognize that a time-varying dark matter mass by itself provides a means to increase the
CMB-inferred Hubble constant without the need for the mediator to modify the sound
horizon~\cite{Archidiacono:2022iuu} (see also Refs.~\cite{Uzan:2023dsk, Pitrou:2023swx}) and,
moreover, that this simplified setup could simultaneously decrease the amplitude of structure
inferred from cosmic shear (see \cref{sec:cosmic-shear}).
More recent analyses~\cite{Wright:2025xka}, however, yield weaker evidence for a lower amplitude of
structure than predicted by the CMB-calibrated \LCDM{} model; one might also consider the excess
from CMB lensing, whose measurements are more precise and less sensitive to nonlinear scales and
baryonic feedback, stronger motivation to instead seek to enhance structure.

\subsubsection{Kinetic couplings}\label{sec:kinetic-couplings}

Thus far we have explicitly studied couplings that manifest purely as spacetime variation of the
dark matter particle mass.
In field theory, the free Lagrangian (of a scalar dark matter field, for concreteness) comprises a
kinetic and potential (mass) term, both of which may be promoted to depend on the mediator.
We now derive the conditions under which both couplings reduce to a common description in terms of a
$\varphi$-dependent mass and show that they hold for dark matter as treated in this work.

We start with an action with general couplings of the form
\begin{align}
    S_\psi
    &= \int \ud^4 x \, \sqrt{-g} \left[
            - \frac{1}{2} X(\varphi)^2 g^{\alpha \beta} \nabla_\alpha \psi \nabla_\beta \psi
            - Y(\varphi)^2 V_\psi(\psi)
        \right]
    .
    \label{eqn:psi-action}
\end{align}
The field $\chi \equiv X(\varphi) \psi$ has a canonical kinetic term:
\begin{align}
    S_\chi
    &= \int \ud^4 x \, \sqrt{-g} \left[
            - \frac{1}{2} \nabla^\alpha \chi \nabla_\alpha \chi
            - Y(\varphi)^2 V_\psi(\chi / X)
            + \chi \nabla_\alpha \chi \nabla^\alpha \ln X
            - \frac{1}{2} \chi^2 \nabla_\alpha \ln X \nabla^\alpha \ln X
        \right].
    \label{eqn:chi-canonical-action}
\end{align}
A quadratic potential for $\psi$ translates to
$Y(\varphi)^2 V_\psi(\chi / X(\varphi)) = Y(\varphi)^2 m_\psi^2 \chi^2 / 2 X(\varphi)^2$---that is,
the mediator-dependent mass of $\chi$ is
$m_\chi(\varphi) = Y(\varphi) / X(\varphi) \cdot m_\psi$.
Pure couplings to the kinetic or potential terms are distinguished by the derivative couplings in
\cref{eqn:chi-canonical-action}, which in general can lead to tachyonic instabilities
(see, e.g., Ref.~\cite{Cyncynates:2023zwj, Cyncynates:2024yxm} for related discussion with slightly
different notation).
However, given a sufficient hierarchy in $m_\chi(\varphi)$ and gradients of $\ln X(\varphi)$,
derivative interactions are subdominant and $\chi$ evolves adiabatically (see also
Refs.~\cite{Smith:2024ibv, Smith:2024ayu, Smith:2025grk}).
In this case, the relevant dynamics are still captured by the kinetic theory description we employ.

To make the preceding argument more rigorous and to better qualify the distinction between the two
classes of couplings, we derive the equation of motion for $\varphi$, replacing its coupling to dark
matter in \cref{eqn:kinetic-theory-action} with those in \cref{eqn:psi-action}.
[Working with the canonical field in \cref{eqn:chi-canonical-action} is more cumbersome.]
Variation gives
\begin{align}
    \nabla_\mu \nabla^\mu \varphi
    &= \dd{V_\varphi}{\varphi}
        + \frac{1}{2 \Mpl^2} \pd{\ln X}{\varphi} X(\varphi)^2 \nabla^\mu \psi \nabla_\mu \psi
        + \frac{1}{\Mpl^2} \pd{\ln Y}{\varphi} Y(\varphi)^2 V_\psi(\psi)
    .
\end{align}
To cast this in a form similar to \cref{eqn:covariant-klein-gordon}, note that
\begin{align}
    g^{\mu \nu} T_{\mu \nu}^{\psi}
    &= - 2 g^{\mu \nu} \pd{\mathcal{L}_\psi}{g^{\mu \nu}}
        + 4 \mathcal{L}_\psi
    = - X(\varphi)^2 \nabla^\mu \psi \nabla_\mu \psi
        - 4 Y(\varphi)^2 V_\psi(\psi)
    ,
\end{align}
where $\mathcal{L}_\psi$ is the Lagrangian density corresponding to \cref{eqn:psi-action};
shuffling factors of the coupling functions yields
\begin{align}
    \nabla_\mu \nabla^\mu \varphi
    &= \dd{V_\varphi}{\varphi}
        - \frac{1}{2 \Mpl^2} \pd{\ln m_\chi}{\varphi} g^{\mu \nu} T_{\mu \nu}^{\psi}
        + \frac{1}{\Mpl^2} \pd{\ln \left( Y / X^2 \right)}{\varphi} \mathcal{L}_\psi
    ,
    \label{eqn:kg-general-couplings}
\end{align}
with $m_\chi(\varphi)$ given in terms of $X$ and $Y$ as above.

\Cref{eqn:kg-general-couplings} shows that the kinetic and potential couplings coincide when the
Lagrangian (evaluated along the equations of motion) is negligible compared to the trace of the
stress-energy tensor.
The Lagrangian vanishes for free plane waves in flat spacetime, and is therefore suppressed relative
to $g^{\mu \nu} T_{\mu \nu}^{\psi}$ by $\mathcal{O}([H / m_\psi]^2)$ or
$\mathcal{O}(m_\psi' / a m_\psi^2)$---exactly the ratios of scales we assume to be small when
treating dark matter as a gas of nonrelativistic particles, i.e., when field-theoretic/wave effects
are unimportant.
The coupling to the Lagrangian alone also vanishes if $Y = X^2$, the precise condition for which
\cref{eqn:psi-action} is equivalent to a scalar-tensor--type coupling that derives from replacing
$g_{\mu \nu}$ with $X(\varphi)^2 g_{\mu \nu}$ in the free-theory action.
In other words, the two scalar couplings in \cref{eqn:psi-action} compartmentalize into
``conformal'' and ``nonconformal'' combinations that respectively couple to the trace of $\chi$'s
stress-energy tensor and to $\mathcal{L}_\chi$.
Energy-momentum conservation requires that the dark matter dynamics are identical to the kinetic
theory description employed in this work if the mediator couples conformally or if
$\mathcal{L}_\psi \ll g^{\mu \nu} T_{\mu \nu}^{\psi}$ (and $X \neq Y$).

\subsection{The nonlinear frontier}\label{sec:nonlinear}

The analysis of dark force models in this work is limited to linear scales and has roughly saturated
the cosmological information from CMB anisotropies that can be reliably modeled in linear theory.
We opt not to include CMB lensing data, for instance, because those scales that are justifiably
linear provide little information on the coupling strength of massless mediators and inference from
smaller-scale data would be biased without modeling nonlinear effects.\footnote{
    Namely, the blue tilt of the transfer function induced by the modified background evolution of
    dark matter (see \cref{fig:cmb-isw-lensing-sensitivity,fig:lensing-spaghetti}) can mimic the
    enhancement of small-scale power by nonlinear structure formation.
    We find that fitting to the latest CMB lensing reconstruction data without a nonlinear model
    contributes a spurious preference for nonzero long-range force strengths upwards of $4$ to $5
    \sigma$ (compared to our finding of just over $2 \sigma$ in \cref{fig:beta-violin}).
    This modeling error very likely underlies the much stronger preference for a coupling between
    dark matter and dark energy claimed by Refs.~\cite{Bedroya:2025fwh,Li:2026xaz}.
}
Even the lensed temperature and polarization anisotropies are marginally sensitive to nonlinear
effects at the scales measured by ACT~\cite{ACT:2025fju} and SPT-3G~\cite{SPT-3G:2025bzu}; though we
deemed the potential biases small enough to justify the analysis (since constraints from ACT and SPT
in \LCDM{} are quite unaffected by modeling nonlinear structure), our results are caveated on the
neglect of nonlinear growth.
Given how little the posteriors over $\beta$ in \cref{fig:beta-violin} differ between the
\Planck{}-only and joint \Planck{}, ACT, and SPT results, any effect is likely to be negligible,
especially considering the paramount importance of the expansion history.
To test dark force models with all available (and future) cosmological data, however, requires
nonlinear modeling, which could be accomplished by extending either the halo
model~\cite{Peacock:2000qk, Seljak:2000gq, Cooray:2002dia, Smith:2002dz, Takahashi:2012em} or
effective field theory methods~\cite{Chudaykin:2020aoj, Linde:2024uzr}.
Augmenting the results of \cref{sec:cosmic-shear} with a nonlinear model is also necessary to test
long-range forces against cosmic shear observations.

More broadly, the cancellation of dark-force--dependent contributions to the evolution of
$\delta \rho_{\chi b}$ in linear theory (\cref{sec:linear-coupling}) raises intriguing questions for
its relevance to other probes of large-scale structure.
The gravitational impact of dark matter overdensities on baryons (just as on photons) depends on the
total rather than relative density perturbation, suggesting that at linear level a (massless) dark
force might only impact galaxy clustering via background dynamics.
Biased tracers of structure are conventionally defined relative to the matter overdensity field
$\delta_{c b}$, however, whose growth is modified~\cite{Bottaro:2023wkd}.
The relevance of the distinction between the two might motivate an alternative bias expansion, or at
least offer physical insight into the relationship between biases and the long-range force.

Furthermore, redshift-space contributions to the observable galaxy power spectrum incur a
substantial suppression due to the dark force, as evident in \cref{fig:growth-deviation} which shows
a substantially negative sensitivity of the growth rate $f \equiv \ud \ln \delta_{\chi b} / \ud \ln
a$ at observable redshifts.
Such effects may bear on the suppressed growth rate, as interpreted in a model otherwise described
by \LCDM{}, found in Ref.~\cite{Nguyen:2023fip} from a combination of the CMB, galaxy surveys,
peculiar velocities, and redshift-space distortions; it would be interesting to assess whether this
result could be explained in models (like scalar-mediated dark forces) where the relative evolution
of the total and relative density perturbation in time is not simply a factor of $a^2$.

It is implausible, moreover, that a cancellation persists to nonlinear order in perturbation theory,
as required to model (for instance) biased tracers in the effective field theory of large-scale
structure.
The BAO scale is also typically extracted from galaxy surveys after
reconstruction~\cite{Eisenstein:2006nk}, whose application remains valid in
equivalence-principle--violating theories due to a specific cancellation of bulk flow contributions
at $\mathcal{O}(\beta)$~\cite{Bottaro:2023wkd}; the same contributions could shift the BAO position.
This cancellation hinges on a specific relation between the friction term and the enhanced
clustering term, which holds for a massless, linearly coupled mediator as discussed in
\cref{sec:linear-coupling} but not in general.

\section{Summary and conclusions}\label{sec:conclusions}

Dark matter's dominance over the energy budget of the Universe offers a unique window into possible
new forces, whether they arise via gravitational degrees of freedom beyond general
relativity~\cite{Brans:1961sx, Scherk:1974ca, Cho:1987xy, Damour:1995kt, Fujii:2003pa} or simply via
other particles in the dark sector containing the dark matter~\cite{Taylor:1988nw, Damour:1994ya,
Dimopoulos:1996kp, Kaplan:2000hh, Gasperini:2001pc, Damour:2002mi, Farrar:2003uw, Strassler:2006im,
Green:2012pqa}.
In this work we studied the simplest (but still microphysically grounded) example in which the
dynamics of dark matter are modified at cosmologically long distances, focusing on clearly
identifying how the various physical aspects of the model manifest in cosmological observables.

Building from Refs.~\cite{Archidiacono:2022iuu, Bottaro:2023wkd, Bottaro:2024pcb}, in
\cref{sec:structure-growth} we outlined a general description of dark matter dynamics with
scalar-mediated dark forces.
In \cref{sec:subhorizon-growth}, we derived the subhorizon limit of the equations of motion that
describes structure growth, clarifying the physical impacts of the mediator on dark matter dynamics
that arise not just directly but also as mediated by general relativity.
The compartmentalization of solution modes that evolve on fast and slow time scales, which
rigorously justifies the limit taken~\cite{Weinberg:2002kg, Weinberg:2008zzc}, underscores a perhaps
underappreciated aspect of the radiation-era dynamics of the Standard Model plasma: that its
perturbations are effectively decoupled from dark matter perturbations, so long as the latter
introduces no additional fast timescale to the problem~\cite{Weinberg:2002kg, Weinberg:2008zzc,
Voruz:2013vqa}.
\Cref{app:subhorizon-limit} shows that the analytic argument of Ref.~\cite{Weinberg:2002kg,
Weinberg:2008zzc} carries through in dark force models (with sufficiently light mediators),
and \cref{fig:cmb-sensitivity} illustrates its striking realization in full solutions to the
Einstein-Boltzmann equations.
The paramount role of background dynamics emphasized in Refs.~\cite{Archidiacono:2022iuu,
Bottaro:2023wkd, Bottaro:2024pcb} thus extends even to the radiation era, as the small-scale CMB
anisotropies generated at last scattering are most sensitive to the modulation of photon diffusion
due to the faster redshifting of dark matter (see \cref{sec:last-scattering} and
\cref{fig:cmb-sensitivity-undamped}).

Specializing to a massless, linearly coupled mediator, \cref{sec:linear-coupling} demonstrated that
its various physical effects---enhanced clustering, background mass evolution, and its own
contributions to the Einstein equations---precisely cancel in the growth rate of structure during
matter domination at leading order.
This cancellation appears not in the density contrast, which does grow faster, but rather the total
density perturbation as sources gravity and therefore weak lensing and the integrated Sachs-Wolfe
effect.
Holding fixed the relative densities of matter and radiation at recombination and of matter and dark
energy near the present, dark matter's mass evolution would manifest in these observables only
during transitions into and out of matter domination: the faster dilution of dark matter allows
radiation to persist longer and dark energy to take over faster, in both cases slightly slowing
structure growth.
The phenomenologically motivated parameter direction to consider, however, is that which fixes not
the relative amount of dark matter and dark energy but rather the angular extent of the photon sound
horizon.
As explained in \cref{sec:propagation-effects,sec:extrapolation}, the substantially larger
cosmological constant required to do so reduces both the distances measured via the acoustic scale
in the galaxy distribution and, crucially, the rate of structure growth at late times.

As such, the only bias-free tracers of large-scale structure---i.e., those that map trivially to the
density field $\delta \rho$ or to gravitational potentials---are suppressed in amplitude relative to
the predictions of the \LCDM{} model, each calibrated to fit primary CMB data.
The consequences of this result for model preferences are nontrivial, as measurements of the weak
lensing of the CMB and of galaxies of late have driven contradictory inferences of the amplitude of
late-time structure.
Counterintuitively, the mild preference for a lower structure amplitude by cosmic shear could be
accommodated by dark forces (see \cref{sec:cosmic-shear}).
On the other hand, the stronger and more persistent evidence for excess lensing of the CMB,
responsible in part for the incompatibility of current cosmological data with the neutrino masses
expected from neutrino oscillations, cannot be explained by dark forces, though the geometric
tension~\cite{Loverde:2024nfi, Lynch:2025ine} can be (see \cref{sec:mnu}).
As explained in Ref.~\cite{Bottaro:2024pcb} and \cref{sec:nonminimal}, massive mediators exacerbate
both issues because they contribute to expansion without clustering on observationally relevant
scales, just like massive neutrinos.
\Cref{sec:potentials} provided an existence proof of a model that genuinely enhances structure,
affixing the mediator with a steep potential; testing the proposal merits dedicated study and
modeling developments to account for nonlinear structure growth (\cref{sec:nonlinear}), which we
will undertake in future work.

It has not escaped our notice that a cancellation between modified background and perturbation
dynamics suggests a deeper physical explanation or that a reformulation of the problem would make
the physics more transparent.
The result echoes the constancy of the Bardeen potentials in a pure-CDM
Universe~\cite{Bardeen:1980kt}.
The simplest notion of forces as deflecting particle trajectories is wholly insufficient in
relativity, where the microscopic description of a scalar-mediated force is perhaps better phrased
as a variation of the theory's ``fundamental constants''---namely, particle
masses~\cite{Terazawa:1981ga, Bekenstein:1982eu, Damour:2010rm, Damour:2010rp}.
The cosmological implications of this distinction are tantamount to the expansion of the Universe in
general relativity.
From this perspective, perhaps it is \emph{not} surprising that absolute density perturbations
(i.e., those that enter the field equations) evolve no faster in extensions of \LCDM{} with dark
forces mediated by massless scalars.

The analogy with standard gravity may be more deeply grounded, as the broad class of scalar
couplings we consider are closely related to scalar-tensor theories of gravity~\cite{Brans:1961sx,
Jordan1955, Fujii:2003pa}---namely, dark matter effectively self-gravitates under a metric related
to that of general relativity by a conformal factor.
Indeed, the equations of motion of dark matter are identical to those in GR in the ``dark Jordan
frame''~\cite{Kase:2020hst, Karwal:2021vpk, Uzan:2023dsk} in terms of the components of that frame's
metric [as could be guessed from the form of \cref{eqn:pressureless-momentum-equation}].
The lack of a cancellation at the end of the radiation era and the onset of dark-energy domination
observed in \cref{fig:growth-deviation} is unsurprising since the dark force we study is
nonuniversal.
On the other hand, the cancellation persists in matter domination despite the presence of uncoupled
baryons.
Moreover, photon geodesics are invariant under conformal transformations of the metric, which may
explain the absence of direct modifications to CMB lensing and the ISW effect in matter domination.
We will investigate these ideas and their possible implications for cosmological observables more
broadly in future work.

\begin{acknowledgments}
We thank Junwu Huang, Lloyd Knox, Marilena Loverde, Gabe Lynch, Jessie Muir, Sergey Sibiryakov, and
especially Diego Redigolo and Ennio Salvioni for numerous useful discussions.
We also thank Maria Archidiacono and Emanuele Castorina for sharing the code developed for
Ref.~\cite{Archidiacono:2022iuu} to cross-check our implementation as well as the authors of
Ref.~\cite{ACT:2025qjh} for sharing the joint CMB lensing bandpowers produced in that work.
We thank the Galileo Galilei Institute for Theoretical Physics for its hospitality and the INFN
for partial support during the completion of this work.
O.S. is supported by the Princeton Center for Theoretical Science, the Princeton University
Department of Physics, their patrons and Trustees.
C.C.S. is supported by a grant from the Simons Foundation (SFI-MPS-T-Institutes-00012000, ML).
Research at Perimeter Institute is supported in part by the Government of Canada through the
Department of Innovation, Science and Economic Development and by the Province of Ontario through
the Ministry of Colleges and Universities.
This work made use of the software packages
\textsf{emcee}~\cite{Foreman-Mackey:2012any,Hogg:2017akh,Foreman-Mackey:2019},
\textsf{corner.py}~\cite{corner}, \textsf{NumPy}~\cite{Harris:2020xlr},
\textsf{SciPy}~\cite{Virtanen:2019joe}, \textsf{matplotlib}~\cite{Hunter:2007ouj},
\textsf{xarray}~\cite{hoyer2017xarray}, \textsf{ArviZ}~\cite{arviz_2019},
\textsf{SymPy}~\cite{Meurer:2017yhf}, and \textsf{CMasher}~\cite{cmasher}.
\end{acknowledgments}

\appendix

\section{Equations of motion}\label{app:eoms}

In this appendix, we outline the general formalism we employ for cosmological perturbation theory,
both to identify its relation to conventional notation and to comment on some gauge-dependent
subtleties for the systems we consider.
We describe our parametrization in \cref{app:parametrization}, the dynamics of dark matter and the
mediator in \cref{app:dark-matter-dynamics,app:mediator-dynamics}, and the subhorizon limit that
describes structure growth in \cref{app:subhorizon-limit}.
Our treatment is similar to that of Ref.~\cite{Weinberg:2008zzc}, differing notably by working in
conformal rather than cosmic time and in the definition of pressure perturbations and scalar
anisotropic stress.

\subsection{Parametrization}\label{app:parametrization}

We take perturbed, conformal-time FLRW metrics of the form
$g_{\mu \nu} \equiv a(\tau)^2 \left( \eta_{\mu\nu} + h_{\mu\nu} \right)$,
where $\eta_{\mu\nu}$ is the Minkowski metric with the mostly positive signature and $h_{\mu \nu}$ a
small perturbation.
Rather than fixing a gauge, we employ a general decomposition of the scalar degrees of freedom in
$h_{\mu \nu}$ as
\begin{subequations}\label{eqn:metric-svt-decomposition}
\begin{align}
    \label{eqn:metric-svt-decomposition-h00}
    h_{00}
    &= - E
    \\
    \label{eqn:metric-svt-decomposition-h0i}
    h_{i 0}
    &= \partial_i F
    \\
    \label{eqn:metric-svt-decomposition-hij}
    h_{i j}
    &= A \delta_{i j}
        + \partial_i \partial_j B
    .
\end{align}
\end{subequations}
We similarly decompose the scalar perturbations to the stress-energy tensor in terms of density and
pressure perturbations $\delta \rho$ and $\delta P$, velocity $\partial_i \delta u$, and anisotropic
stress $\pi^S$:
\begin{subequations}
\label{eqn:stress-tensor-decomp-mixed}
\begin{align}
    \label{eqn:stress-tensor-decomp-mixed-T00}
    \delta T^{0}_{\hphantom{0}0}
    &= - \delta \rho
    \\
    \label{eqn:stress-tensor-decomp-mixed-T0i}
    \delta T^{0}_{\hphantom{0}i}
    &= \left( \bar{\rho} + \bar{P} \right)
        \partial_i \delta u
    \\
    \label{eqn:stress-tensor-decomp-mixed-Tij}
    \delta T^{i}_{\hphantom{i}j}
    &= \delta_{ij} \delta P
        + \left( \partial_i \partial_j - \frac{1}{3} \delta_{ij} \partial_k \partial_k \right) \pi^S
    .
\end{align}
\end{subequations}
The homogeneous stress-energy tensor has components
$\bar{T}^0_{\hphantom{0}0} = - \bar{\rho}$,
$\bar{T}^0_{\hphantom{0}i} = 0$,
and $\bar{T}^i_{\hphantom{i} j} = \bar{P} \delta^i_{\hphantom{i} j}$ as usual, with bars denoting
spatially averaged quantities.
The stress-energy tensor for any individual constituent's contribution to $T_{\mu \nu}$
takes the same form.
With this parametrization, the scalar metric perturbations $B$ and $F$ only enter the
Einstein and energy-momentum equations in the combination~\cite{Weinberg:2008zzc}
\begin{align}
    \label{eqn:weinberg-psi-general}
    \psi
    &\equiv \frac{1}{2} \left(
            3 A' + \partial_i \partial_i \left[ B' - 2 F \right]
        \right).
\end{align}
For reference, the conformal Newtonian gauge has $E = 2 \Psi$, $A = - 2 \Phi$, and both $F$ and $B$
zero, while the synchronous gauge used in Ref.~\cite{Ma:1995ey} sets $A = - 2 \eta$ and $\partial_i
\partial_i B = h + 6 \eta$ with $E$ and $F$ zero.
In these two gauges, $\psi$ equals $-3 \Phi'$ and $h' / 2$, respectively.

In terms of $\psi$, $E$, and $A$, the perturbations to the Einstein equation are
\begin{subequations}\label{eqn:svt-field-equations-psi}
\begin{align}
    \label{eqn:svt-field-equations-delta-rho-psi}
        \frac{a^2}{\Mpl^2} \delta \rho
        &= - 3 \mathcal{H}^2 E
            - \partial_i \partial_i A
            + 2 \mathcal{H} \psi
    \\
    \label{eqn:svt-field-equations-delta-u-psi}
        \frac{a^2}{\Mpl^2} \left( \bar{\rho} + \bar{P} \right) \delta u
        &= A' - \mathcal{H} E
    \\
    \label{eqn:svt-field-equations-delta-P-psi}
        \frac{a^2}{\Mpl^2} \delta P
        &= \mathcal{H} E'
            + \left( 2 \mathcal{H}' + \mathcal{H}^{2} \right) E
            + \frac{1}{3} \partial_i \partial_i \left( A + E \right)
            - \frac{2}{3} \psi'
            - \frac{4}{3} \mathcal{H} \psi
    \\
    \label{eqn:svt-field-equations-pi-S-psi}
        \frac{a^2}{\Mpl^2} \partial_i \partial_i \pi^S
        &= \mathcal{H} \left( 2 \psi - 3 A' \right)
            + \frac{1}{2} \left( 2 \psi' - 3 A'' \right)
            - \frac{1}{2} \partial_i \partial_i \left( A + E \right)
    .
\end{align}
\end{subequations}
For our purposes, the most pertinent combination is the sum of the diagonal entries of the Einstein equation,
\begin{align}
    \label{eqn:svt-field-equations-trace-psi}
    - \frac{a^2}{2 \Mpl^2} \left(
        \delta \rho
        + 3 \delta P
    \right)
    &= \psi'
        + \mathcal{H} \psi
        - \frac{1}{2} \partial_i \partial_i E
        - 3 \mathcal{H}' E
        - \frac{3}{2} \mathcal{H} E'
    .
\end{align}
Finally, the divergence of the stress-energy tensor is
\begin{subequations}
\label{eqn:energy-momentum-scalar-psi}
\begin{align}
    \label{eqn:scalar-energy-conservation-psi}
    - \nabla_\mu T^{\mu}_{\hphantom{\mu} 0}
    &= \bar{\rho}'
        + 3 \mathcal{H} \left( \bar{\rho} + \bar{P} \right)
        + {\delta \rho}'
        + 3 \mathcal{H} \left( \delta \rho + \delta P \right)
        + \left( \bar{\rho} + \bar{P} \right)
        \left(
            \psi
            + \partial_j \partial_j \delta u
        \right)
    \\
    \label{eqn:scalar-momentum-conservation-pi-S-psi}
    \nabla_\mu T^{\mu}_{\hphantom{\mu} i}
    &= \partial_i \left[
            \partial_\tau \left[
                \left( \bar{\rho} + \bar{P} \right) \delta u
            \right]
            + \left( \bar{\rho} + \bar{P} \right)
            \left( 4 \mathcal{H} \delta u + E / 2 \right)
            + \delta P
            + \frac{2}{3} \partial_j \partial_j \pi^S
        \right]
    ,
\end{align}
\end{subequations}
which equals zero when describing the full stress-energy tensor but not when describing that of a
species that exchanges energy and/or momentum with another (as is the case for $\chi$ and its
mediator $\varphi$).

\subsection{Dark matter dynamics}\label{app:dark-matter-dynamics}

Variation of the action \cref{eqn:kinetic-theory-action} with respect to the trajectory $x^\mu_p$ of
particle $p$ yields
\begin{align}
    \ddd{x_p^\mu}{\uptau_p}
    + \Gamma^{\mu}_{\hphantom{\mu} \alpha \beta}
        \dd{x_p^\alpha}{\uptau_p} \dd{x_p^\beta}{\uptau_p}
    &= - \left(
            g^{\mu \nu}
           + \dd{x_p^\mu}{\uptau_p} \dd{x_p^\nu}{\uptau_p}
        \right)
        \pd{\ln m_\chi}{x_p^\nu}
    ,
    \label{eqn:geodesic-equation}
\end{align}
with $\Gamma^{\mu}_{\hphantom{\mu} \alpha \beta}$ the usual connection symbol, i.e.,
the standard geodesic equation augmented by a scalar force represented by a particle mass that
depends on spacetime (which we later identify to derive from dependence on the mediator $\varphi$).
The energy-momentum equations may be computed at the level of kinetic theory, with the stress-energy
tensor for $\chi$ particles in \cref{eqn:kinetic-theory-action} being
\begin{align}
    T^{\mu \nu}_\chi(x)
    &= \frac{1}{\sqrt{-g(x)}} \sum_p
        \int \ud \uptau_p \,
        m_\chi[x_p(\uptau_p)]
        \dd{x_p^\mu}{\uptau_p}
        \dd{x_p^\nu}{\uptau_p}
        \delta^4( x - x_p(\uptau_p))
    \label{eqn:kinetic-theory-stress-energy}
\end{align}
and the conservation equation reading
$\nabla_\mu T^{\mu \nu}_\chi = \partial_\mu \left( \sqrt{-g} T^{\mu \nu}_\chi \right) / \sqrt{-g} + \Gamma^{\nu}_{\hphantom{\nu} \mu \beta} T^{\mu \beta}_\chi$.
Evaluating the first term by using the delta function to swap $\partial / \partial x^\mu$ for
$- \partial / \partial x_p^\mu$, combining that derivative with $\ud x_p^\mu / \ud \uptau_p$,
and integrating by parts (and assuming that $x$ coincides with none of the $x_p$ at the endpoints
of integration) yields
\begin{align}
    \pd{\left( \sqrt{-g} T^{\mu \nu}_\chi \right)}{x^\mu}
    &= \sum_p \int \ud \uptau_p \,
        \delta^4(x - x_p(\uptau_p))
        m_\chi[x_p(\uptau_p)]
        \left[
            \pd{\ln m_\chi}{x_p^\beta}
            \dd{x_p^\beta}{\uptau_p}
            \dd{x_p^\nu}{\uptau_p}
            + \ddd{x_p^\nu}{\uptau_p}
        \right]
    .
\end{align}
Inserting the geodesic equation \cref{eqn:geodesic-equation} cancels the first term in
brackets and the second term in the above expression for $\nabla_\mu T^{\mu \nu}_\chi$.
Identifying the trace of \cref{eqn:kinetic-theory-stress-energy} in the remaining coefficient of
$\partial_\nu \ln m_\chi$ gives
\begin{align}
    \nabla_\mu T^{\mu \nu}_\chi
    &= \pd{\ln m_\chi}{x_\nu} g_{\alpha \beta} T^{\alpha \beta}_\chi
    \label{eqn:chi-conservation-app}
    ,
\end{align}
demonstrating that the scalar couples to the trace of the stress-energy tensor.

With the form of the interaction so motivated, we proceed by taking a fluid ansatz of the
form \cref{eqn:stress-tensor-decomp-mixed} for $T^{\mu \nu}_\chi$, such that the right-hand side of
\cref{eqn:chi-conservation-app} reduces to that of \cref{eqn:chi-conservation}.
To emphasize the physics pertinent to our main results, we assume $\chi$ has neither pressure nor
anisotropic stress, leaving two scalar degrees of freedom such that the energy equation and the
scalar part of the momentum equations fully specify the dynamics of dark matter perturbations.
One could equivalently derive the Boltzmann equation for the phase-space distribution of $\chi$ and
compute its moments~\cite{Archidiacono:2022iuu}.
The energy-momentum equations reduce to
\begin{align}
    \label{eqn:background-continuity-sourced-chi}
    \bar{\rho}_\chi'
        + 3 \mathcal{H} \bar{\rho}_\chi
    &= \bar{\rho}_\chi \dd{\ln m_\chi}{\tau}
    = \bar{\rho}_\chi \pd{\ln m_\chi}{\varphi} \bar{\varphi}'
\end{align}
and, using \cref{eqn:energy-momentum-scalar-psi} written in terms of the density contrast
$\delta_\chi \equiv \delta \rho_\chi / \bar{\rho}_\chi$,
\begin{subequations}\label{eqn:energy-momentum-chi}
\begin{align}
    \delta_\chi'
    + \psi
    + \partial_i \partial_i \delta u_\chi
    &= \dd{\delta \ln m_\chi}{\tau}
    = \pdd{\ln m_\chi}{\varphi} \bar{\varphi}' \delta \varphi
        + \pd{\ln m_\chi}{\varphi} \delta \varphi'
    \label{eqn:delta-chi-first-order-pressureless}
    \\
    \delta u_\chi'
    + \left( \mathcal{H} + \dd{\ln m_\chi}{\tau} \right)
    \delta u_\chi
    + E / 2
    &= - \delta \ln m_\chi
    = - \pd{\ln m_\chi}{\varphi} \delta \varphi
    \label{eqn:pressureless-momentum-equation}
    .
\end{align}
\end{subequations}
The latter equalities in \cref{eqn:background-continuity-sourced-chi,eqn:energy-momentum-chi} take
$m_\chi$ to be a function of $\varphi$ alone.
\Cref{eqn:energy-momentum-chi} may be combined into a second-order equation of motion for
$\delta_\chi$ of the form
\begin{align}\label{eqn:delta-chi-second-order-pressureless}
\begin{split}
    \delta_\chi''
    + \mathcal{H} \delta_\chi'
    &= \partial_i \partial_i \delta \ln m_\chi
        - \frac{\partial_\tau \left( a \psi \right)}{a}
        + \partial_i \partial_i E / 2
        + \ddd{\delta \ln m_\chi}{\tau}
        + \mathcal{H} \dd{\delta \ln m_\chi}{\tau}
        + \dd{\ln m_\chi}{\tau} \partial_i \partial_i \delta u_\chi
    .
\end{split}
\end{align}

While $\chi$ does exchange energy with $\varphi$, its particle number is still conserved, making the
equations of motion for $\bar{n}_\chi \equiv \bar{\rho}_\chi / \bar{m}_\chi$ and
$\delta_{n_\chi} \equiv \rho_\chi / m_\chi \bar{n}_\chi - \bar{\rho}_\chi / \bar{m}_\chi \bar{n}_\chi
= \delta_\chi - \delta \ln m_\chi$ identical to those for CDM:
\begin{subequations}
\begin{align}
    0
    &= \bar{n}_\chi'
        + 3 \mathcal{H} \bar{n}_\chi
    \\
    0
    &= \delta_{n_\chi}'
        + \partial_i \partial_i \delta u_\chi
        + \psi
    \label{eqn:delta-n-chi-first-order-pressureless}
    .
\end{align}
\end{subequations}
The second-order equation for $\delta_{n_\chi}$ features a much simpler coupling to mediator
perturbations,
\begin{align}
    \delta_{n_\chi}''
    + \mathcal{H} \delta_{n_\chi}'
    &= \partial_i \partial_i \delta \ln m_\chi
        - \frac{\partial_\tau \left( a \psi \right)}{a}
        + \partial_i \partial_i E / 2
        + \dd{\ln m_\chi}{\tau} \partial_i \partial_i \delta u_\chi
    \label{eqn:delta-n-chi-second-order}
    ,
\end{align}
showing that the time derivatives of $\delta \ln m_\chi$ in
\cref{eqn:delta-chi-second-order-pressureless} appear as an artifact of $\delta_\chi$ including mass
fluctuations by definition.
\Cref{eqn:delta-n-chi-second-order} provides a more expedient starting point from which to derive
simplified equations of motion in the subhorizon limit (\cref{app:subhorizon-limit}), as apt to
describe the late-time growth of observable structure.

The standard synchronous gauge used in linear perturbation theory contains a residual gauge symmetry
(which leaves $E$ and $F$ both zero) under coordinate transformations that are time independent (up
to factors of $a$)~\cite{Bardeen:1980kt, Weinberg:2008zzc}.
Setting the fluid velocity of cold dark matter to zero is a convenient means to fix this freedom,
since if initially zero it remains zero at all times.
The velocity of coupled dark matter, however, does not share this property, being sourced by the
mediator.
We therefore follow Refs.~\cite{Archidiacono:2022iuu, Obied:2023clp} in retaining a negligible
amount of CDM that serves to fix the remaining gauge freedom.

\subsection{Mediator dynamics}\label{app:mediator-dynamics}

Variation of the action \cref{eqn:kinetic-theory-action} with respect to $\varphi$ yields
\begin{align}
    \nabla^\mu \nabla_\mu \varphi(x)
    &= \dd{V_\varphi}{\varphi}
        + \pd{\ln m_\chi}{\varphi} \frac{1}{2 \Mpl^2 \sqrt{-g(x)}}
        \sum_p \int \ud \uptau_p \, \delta^4(x - x_p(\uptau_p))
        m_\chi\big( \varphi[x_p(\uptau_p)] \big)
    \label{eqn:varphi-ele}
\end{align}
In deriving \cref{eqn:chi-conservation-app} we identified that the sum over $\chi$ particles in
\cref{eqn:varphi-ele} is proportional to the trace of their stress-energy tensor
\cref{eqn:kinetic-theory-stress-energy}.
Again taking the fluid stress-energy tensor defined in \cref{app:parametrization} yields
\cref{eqn:covariant-klein-gordon}, repeated here:
\begin{align}
    \nabla^\mu \nabla_\mu \varphi
    &= \dd{V_\varphi}{\varphi}
        + \pd{\ln m_\chi}{\varphi} \frac{\rho_\chi - 3 P_\chi}{2 \Mpl^2}
    \equiv \frac{\ud V / \ud \varphi}{2 \Mpl^2}.
    \label{eqn:covariant-klein-gordon-app}
\end{align}
Here $V$ is normalized as the total potential (including interactions) for the canonical field
$\phi = \sqrt{2} \Mpl \varphi$, i.e., $V$ is what appears in the total stress-energy tensor with no
multiplicative factors.

Decomposing the mediator into a background and perturbation as
$\varphi(\tau, \mathbf{x}) = \bar{\varphi}(\tau) + \delta \varphi(\tau, \mathbf{x})$,
\cref{eqn:covariant-klein-gordon-app} reads
\begin{subequations}
\begin{align}
    0
    &= \bar{\varphi}''
        + 2 \mathcal{H} \bar{\varphi}'
        + \frac{a^2}{2 \Mpl^2} \pd{V}{\varphi}
    \\
    0
    &= \delta \varphi''
        + 2 \mathcal{H} \delta \varphi'
        - \partial_i \partial_i \delta \varphi
        + \frac{a^2}{2 \Mpl^2} \pd{\delta V}{\varphi}
        + \bar{\varphi}' \left(
            \psi
            - E' / 2
        \right)
        + \frac{a^2}{2 \Mpl^2} \pd{V}{\varphi} E
    .
    \label{eqn:perturbed-kg-general}
\end{align}
\end{subequations}
Since we take $\chi$ to have negligible pressure and anisotropic stress, the equation of motion
for $\delta \varphi$ expands to
\begin{align}
    \delta \varphi''
    + 2 \mathcal{H} \delta \varphi'
    + \left( - \partial_i \partial_i + a^2 m_\mathrm{eff}^2 \right) \delta \varphi
    &=
        - \frac{a^2 \bar{\rho}_\chi}{2 \Mpl^2} \pd{\ln m_\chi}{\varphi} \delta_\chi
        - \bar{\varphi}' \left( \psi - E' / 2 \right)
        - \frac{a^2}{2 \Mpl^2} \pd{V}{\varphi} E
    \label{eqn:perturbed-kg-app}
    ,
\end{align}
defining the effective mass
\begin{align}
    m_\mathrm{eff}^2
    &\equiv \ddd{V_\varphi}{\varphi}
        + \frac{\bar{\rho}_\chi}{2 \Mpl^2} \pdd{\ln m_\chi}{\varphi}
    .
\end{align}
One could exchange $\delta_\chi$ for $\delta_{n_\chi}$ in \cref{eqn:perturbed-kg-app}; the
incurred term proportional to $\delta \ln m_\chi$ may be absorbed into the effective mass above
via the addition of
$\bar{\rho}_\chi \left( \partial \ln m_\chi / \partial \varphi \right)^2 / 2 \Mpl^2$.
One may perform a similar replacement in the Einstein equation to write the entire system in terms
of $\delta_{n_\chi}$ alone.
For the subhorizon dynamics we seek to study, however, the distinction between the two turns out to
be negligible (as explained in \cref{app:subhorizon-limit}), in which case these contributions to
$m_\mathrm{eff}^2$ are also irrelevant as they are necessarily order $\mathcal{H}^2$ or smaller.

Finally, the perturbed stress-energy components contributed by the scalar are
\begin{subequations}\label{eqn:mediator-stress-energy}
\begin{align}
    \frac{\delta \rho_\varphi}{2 \Mpl^2}
    &= - \frac{E}{2 a^2} \left( \bar{\varphi}' \right)^2
        + \frac{\bar{\varphi}' \delta \varphi'}{a^2}
        + \dd{V_\varphi}{\varphi} \delta \varphi
    \\
    \frac{\delta P_\varphi}{2 \Mpl^2}
    &= - \frac{E}{2 a^2} \left( \bar{\varphi}' \right)^2
        + \frac{\bar{\varphi}' \delta \varphi'}{a^2}
        - \dd{V_\varphi}{\varphi} \delta \varphi
    \\
    \frac{\bar{\rho}_\varphi + \bar{P}_\varphi}{2 \Mpl^2} \delta u_\varphi
    &= - \frac{\bar{\varphi}' \delta \varphi}{a^2}
\end{align}
\end{subequations}
where $\bar{\rho}_\varphi / 2 \Mpl^2 = (\bar{\varphi}')^2 / 2 + V_\varphi(\varphi)$
and $\bar{P}_\varphi / 2 \Mpl^2 = (\bar{\varphi}')^2 / 2 - V_\varphi(\varphi)$.
Only the bare potential appears in the above because we assign the stress-energy in the interaction
term to the dark matter, i.e., $T_{\mu \nu}^\chi$ is written in terms of $m_\chi(\varphi)$.

As a practical point, gauges with $E$ nonzero (like Newtonian gauge) are particularly inconvenient
when solving the Klein-Gordon equation because the Einstein equations do not provide a
straightforward, algebraic expression for $E'$ in terms of other metric perturbations.
In Newtonian gauge, a simple candidate might be \cref{eqn:svt-field-equations-pi-S-psi}, which sets
$E' / 2 = \Psi' = \Phi' - \partial_\tau \left( a^2 \pi^S / \Mpl^2 \right)$; this expression,
however, depends upon the microphysical dynamics of those species that contribute anisotropic
stress.
In addition, the Klein-Gordon equation is only implemented correctly in synchronous gauge in current
versions of \textsf{CLASS}: the inconvenient term $\bar{\varphi}' E' / 2 = \bar{\varphi}' \Psi'$ is
missing entirely, as is the last term of \cref{eqn:perturbed-kg-general} (proportional to $E$).

\subsection{Subhorizon limit}\label{app:subhorizon-limit}

The preceding results fully establish the closed set of equations (less those for SM matter and dark
energy) that specify the dynamics of the metric, dark matter, and the mediator, with the only
approximation being the specialization to linearized scalar perturbations.
We now review further approximations that grant analytic insight into the system (and also explain
the origin of discrepancies and misinterpretations in some prior literature).
Following Ref.~\cite{Weinberg:2002kg, Weinberg:2008zzc}, we aim to decompose solutions into ``fast''
and ``slow'' modes by power counting in $k / a H$; in \LCDM{} and on comoving scales smaller than
the horizon at equality, perturbations in the radiation fluid (plus the tightly coupled baryons) and
in the dark matter density effectively decouple into fast and slow modes, respectively.
Dark matter dominates the slow mode even before equality, despite making a subdominant contribution
to the background density~\cite{Weinberg:2002kg, Weinberg:2008zzc}.
The primary CMB anisotropies on small scales are mostly sourced by the fast mode and are therefore
more sensitive to dark matter's impact on the background than on perturbations (as demonstrated
explicitly for warm dark matter in Ref.~\cite{Voruz:2013vqa}).
We repeat this power counting exercise to verify that the dark force and the mediator itself
(whose Green function does oscillate with frequency $\sim k$) do not impede the gravitational
decoupling of the plasma and dark matter.

\subsubsection{Fast mode}

Taking $\delta u_\chi' \sim k \delta u_\chi$ and that $\ln m_\chi \propto \bar{\varphi}$ varies on
time scales much longer than $1/k$, the momentum equation \cref{eqn:pressureless-momentum-equation}
shows $\delta u_\chi$ to be of order $-\delta \ln m_\chi / k$.
(We take $E = 0$ gauges for simplicity.)
The number conservation equation \cref{eqn:delta-n-chi-first-order-pressureless} then sets
$\delta_{n_\chi}$ of order $\psi / k - k^2 \delta u_\chi \sim \psi / k + k \delta \ln m_\chi$.
Finally, taking $a^2 m_\mathrm{eff}^2 \ll k^2$, the Klein-Gordon equation
\cref{eqn:perturbed-kg-app} sets
$k^2 \delta \varphi \sim - \bar{\varphi}' \psi - a^2 \bar{\rho}_\chi \partial_\varphi \ln m_\chi \delta_{n_\chi} / 2 \Mpl^2$,
dropping the subdominant contribution to $\delta_\chi$ from mass fluctuations $\delta \ln m_\chi$.
In \LCDM{}, radiation perturbations dominate the fast mode of $\psi$, while
$\delta \ln m_\chi = \partial_\varphi \ln m_\chi \delta \varphi$ is only sourced by dark matter.
The mediator is then largely sourced by $\bar{\varphi}' \psi$, since $\delta_{n_\chi}$ is itself
suppressed by a factor of $1/k$---that is, fast-mode perturbations to the dark matter and mediator
decouple in the same sense as for gravity.
The mediator does source dark matter perturbations at the same order in $k$ as gravity, though times
$\bar{\varphi}' \partial_\varphi \ln m_\chi$.
Finally, since the fast mode of the mediator is $\bar{\varphi}'/k^2$ smaller than $\psi$, its
contribution $\propto \bar{\varphi}' \delta \varphi'$ to the equation of motion for $\psi'$
[\cref{eqn:svt-field-equations-trace-psi}] is even more suppressed than dark matter's.
Since the late-time dark matter field is dominated by the slow mode, we expect the effect of the
dark force on the fast mode to have no appreciable impact on the primary CMB nor late-time
structure.

\subsubsection{Slow mode}

In the slow mode, $\delta \varphi'' + 2 \mathcal{H} \delta \varphi'$ is of order
$\mathcal{H}^2 \delta \varphi$ and is therefore negligible in the Klein-Gordon equation (as is
$\bar{\varphi}' E'$), reducing \cref{eqn:perturbed-kg-app} to a nonrelativistic Poisson equation of
the form
\begin{align}\label{eqn:mediator-poisson-app}
    \left( \partial_i \partial_i - a^2 m_\mathrm{eff}^2 \right) \delta \varphi
    &\simeq \frac{a^2 \bar{\rho}_\chi}{2 \Mpl^2} \pd{\ln m_\chi}{\varphi} \delta_\chi
        + \bar{\varphi}' \psi
        + \frac{a^2}{2 \Mpl^2} \pd{V}{\varphi} E.
\end{align}
Note that the term $\bar{\varphi}' \psi$ is \emph{not} negligible; in fact, it modifies the
friction term in the second-order equation of motion for $\delta_{n_\chi}$.\footnote{
    By neglecting this metric term in \cref{eqn:mediator-poisson-app}, the quasistatic approximation
    derived in Refs.~\cite{McDonough:2021pdg, Mirpoorian:2023utj, Wang:2023tjj} for $\delta_\chi''$
    misses the mediator's contribution to the friction term.
    Various references neglect this effect by simply assuming the scalar does not evolve at the
    background level (or does so on time scales negligible compared even to $\mathcal{H}$).
}
Specifically, when inserting \cref{eqn:mediator-poisson-app} into the second-order equation of
motion for $\delta_{n_\chi}$ [\cref{eqn:delta-n-chi-second-order}] with
\cref{eqn:delta-n-chi-first-order-pressureless} substituted for $\psi$,
\begin{align}
\begin{split}
    \delta_{n_\chi}''
    &\simeq - \left(
            \mathcal{H}
            + \frac{
                \ud \ln m_\chi / \ud \tau
            }{
                1 + \left( a m_\mathrm{eff} / k \right)^2
            }
        \right) \delta_{n_\chi}'
        - \frac{\partial_\tau \left( a \psi \right)}{a}
        + \partial_i \partial_i E / 2
        + \frac{a^2 \bar{\rho}_\chi}{2 \Mpl^2}
        \frac{
            \left( \partial \ln m_\chi / \partial \varphi \right)^2
        }{
            1 + \left( a m_\mathrm{eff} / k \right)^2
        }
        \delta_\chi
    \\ &\hphantom{ {}={} }
        - \frac{
            \ud \ln m_\chi / \ud \tau
        }{
            1 + \left( k / a m_\mathrm{eff} \right)^2
        }
        k^2 \delta u_\chi
        + \frac{
            \partial \ln m_\chi / \partial \varphi
        }{
            1 + \left( a m_\mathrm{eff} / k \right)^2
        }
        \left[
            \frac{a^2}{2 \Mpl^2} \pd{V}{\varphi} E
            - \bar{\varphi}' E' / 2
        \right]
    \label{eqn:delta-n-chi-second-order-subhorizon}
    .
\end{split}
\end{align}
The system now depends on both the number and density contrasts, $\delta_{n_\chi}$ and
$\delta_\chi$, which differ precisely by
$- \delta \ln m_\chi = - \partial \ln m_\chi / \partial \varphi \cdot \delta \varphi$.
This relative perturbation in the density that derives from that in the mass, while
formally of order $\left( \partial \ln m_\chi / \partial \varphi \right)^2$, is suppressed by a
factor of $a^2 \bar{\rho}_\chi / 2 \Mpl^2 \left( k^2 + a^2 m_\mathrm{eff}^2 \right)$.
In other words, the time derivatives of $\delta \varphi$ neglected in the quasistatic approximation
of the Klein-Gordon equation (i.e., for the slow mode) are precisely those that differ between the
equations of motion for $\delta_\chi$ and $\delta_{n_\chi}$.
We may therefore replace the latter with the former.

Before proceeding, we further simplify the system by partially fixing the gauge.
For species with no anisotropic stress, the energy-momentum equations decouple from all Einstein
equations other than \cref{eqn:svt-field-equations-trace-psi} in gauges with $E = 0$ (e.g.,
synchronous gauges).
\Cref{eqn:svt-field-equations-trace-psi} itself sets
$\partial_\tau \left( a \psi \right) / a$ in terms of $\delta \rho + 3 \delta P$.
Finally, note that the remaining contribution of $\delta u_\chi$ to the source term
\cref{eqn:delta-n-chi-second-order-subhorizon}, which otherwise
prevents the reduction of the system to a single second-order equation for $\delta_\chi$, is
subleading in the coupling because $\delta u_\chi$ is only sourced by the mediator and is otherwise
zero at all times [per \cref{eqn:pressureless-momentum-equation}].
That is, in the slow mode we may estimate
$k^2 \delta u_\chi \sim k^2 \delta \ln m_\chi / \mathcal{H}$; from \cref{eqn:mediator-poisson-app},
the relevant term in \cref{eqn:delta-n-chi-second-order-subhorizon} is of order
$\left( \partial \ln m_\chi / \partial_\varphi \right)^2 \bar{\varphi}'$ times either
$\partial \ln m_\chi / \partial \varphi$ or $\bar{\varphi}'$.\footnote{
    While $\bar{\varphi}'$ is generally of order $\partial \ln m_\chi / \partial \varphi$, it
    need not be if the mediator's bare potential is nonnegligible, in which case the term
    $\propto \left( \bar{\varphi}' \partial \ln m_\chi / \partial_\varphi \right)^2 \psi$ is
    not necessarily subdominant in the coupling expansion.
    In full generality, of course, three coupled, first-order differential equations cannot be
    reduced to a single second-order system anyway.
}
In addition, its contribution is quadratically suppressed in both the $k \gg a m_\mathrm{eff}$
and $k \ll a m_\mathrm{eff}$ limits.
We therefore drop the entire second line of
\cref{eqn:delta-n-chi-second-order-subhorizon}, yielding
\cref{eqn:delta-chi-second-order-pressureless-subhorizon-synchronous}.

Thus far, we have only made approximations appropriate for solutions that grow on long time scales
without making any other assumptions about the other constituents of the Universe.
The contribution from dark matter dominates in the slow mode over that from not just radiation but
also baryons, since their pressure support before decoupling prevents their
growth~\cite{Weinberg:2002kg, Weinberg:2008zzc}.
The mediator's contribution to the Einstein equations [\cref{eqn:mediator-stress-energy}] is
suppressed both by the coupling and by powers of $k / \mathcal{H}$, since
\cref{eqn:perturbed-kg-app} sets $\bar{\varphi}' \delta \varphi' / \delta \rho_\chi$ of order
$\partial \ln m_\chi / \partial \varphi$ times $\bar{\varphi}' \mathcal{H} / k^2$ for the slow mode
and $\bar{\varphi}' / k$ for the fast mode.\footnote{
    This argument breaks down if the mediator has a sufficiently large homogeneous misalignment and
    (for example) bare mass, but the density perturbations of a free scalar field do not grow
    below its Jeans length anyway~\cite{Hu:2000ke}.
}
\Cref{eqn:delta-n-chi-second-order-subhorizon}, with
$\sum_{I \neq \chi} \left( \delta \rho_I + 3 \delta P_I \right)$ set to zero, thus holds before
decoupling as a generalization of the M\'esz\'aros equation~\cite{Meszaros:1974tb};
with $\sum_{I \neq \chi} \left( \delta \rho_I + 3 \delta P_I \right) = \bar{\rho}_b \delta_b$
after decoupling, it generalizes the equation of motion for density perturbations in \LCDM{},
and changing variables yields \cref{eqn:delta-chi-b-eom} as studied in
\cref{sec:structure-growth-after-decoupling}.
Finally, since the dark force cannot undo the large hierarchy in size between the slow and
fast mode of the dark matter density contrast~\cite{Weinberg:2002kg, Weinberg:2008zzc}, these
results establish the appropriate quasistatic limit that is reached by the system inside the
horizon.
In summary, subhorizon density perturbations of $\chi$ evolve according to
\begin{align}
    \delta_\chi''
    + \left(
        \mathcal{H}
        + \frac{
            \ud \ln m_\chi / \ud \tau
        }{
            1 + \left( a m_\mathrm{eff} / k \right)^2
        }
    \right)
    \delta_\chi'
    &\simeq
        \frac{a^2 \bar{\rho}_\chi}{2 \Mpl^2}
        \left(
            1
            + \frac{
                \left( \partial \ln m_\chi / \partial \varphi \right)^2
            }{
                1 + \left( a m_\mathrm{eff} / k \right)^2
            }
        \right)
        \delta_\chi
        +
        \begin{dcases}
            0, & a \ll a_\star,
            \\
            \frac{a^2 \bar{\rho}_b \delta_b}{2 \Mpl^2}, & a \gg a_\star.
        \end{dcases}
    .
\end{align}

\section{Supplementary results}\label{app:supplementary-results}

\begin{figure}[t!]
\begin{centering}
    \includegraphics[width=\textwidth]{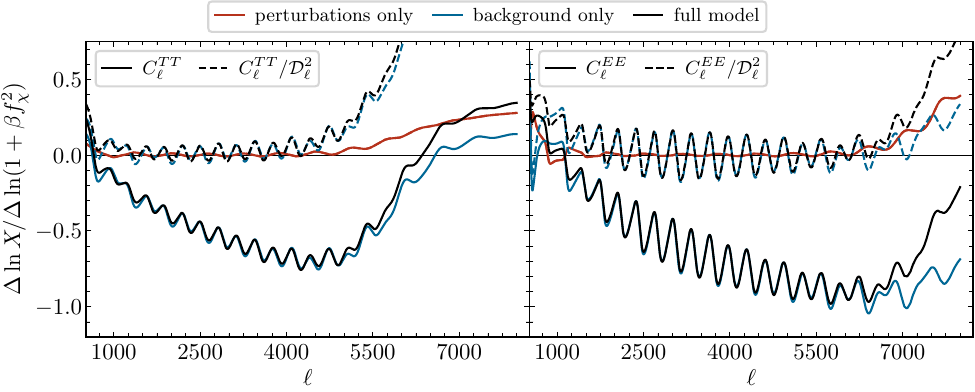}
    \caption{
        Demonstration that the dark-matter mass evolution largely affects the primary CMB
        by modulating diffusion damping.
        Results are presented as in \cref{fig:cmb-sensitivity}, depicting the sensitivity
        of unlensed CMB temperature (left) and polarization (right) anisotropies with (dashed)
        and without (solid) the impact of diffusion removed via the analytic diffusion factor
        $\mathcal{D}_\ell$ [\cref{eqn:diffusion-factor}].
    }
    \label{fig:cmb-sensitivity-undamped}
\end{centering}
\end{figure}

To support the claim in \cref{sec:last-scattering} that the modulation of diffusion is the main
impact of the nonstandard background evolution of dark matter on the generation of small-scale
anisotropies, \cref{fig:cmb-sensitivity-undamped} presents the same sensitivities shown in
\cref{fig:cmb-sensitivity} after dividing out the visibility-averaged damping factor defined
in~\cite{Hu:1994uz},
\begin{align}
    \mathcal{D}(k)
    = \int_0^{a_0} \ud \ln a \, \dd{\kappa}{\ln a} e^{-\kappa(a, a_0)}
        e^{- \left[ k / k_D(a) \right]^2}
    \label{eqn:diffusion-factor}
\end{align}
(with the damping scale $k_D$ defined in, e.g., Refs.~\cite{Kaiser:1983nrc, Zaldarriaga:1995gi}).
Specifically, for each multipole we define
$\mathcal{D}_\ell \equiv \mathcal{D}(\ell / D_{M, \star})$.
The ``undamped'' sensitivities in \cref{fig:cmb-sensitivity-undamped} fully remove the secular drift
between multipoles of $1000$ and $5000$, leaving a small, residual sensitivity that oscillates about
zero.
(The sensitivities diverge at sufficiently small $\ell$ simply because the slow mode of the plasma
eventually dominates over the fast mode, at which point damping is no longer relevant.)

Diffusion is modulated both by the change to the comoving size of the Universe (insofar as it
determines the instantaneous diffusion scale~\cite{Kaiser:1983nrc}) and by slight changes to the
visibility function leading up to recombination, as the diffusion of smaller scale modes is
dominated by earlier times~\cite{Hu:1994uz}.
The visibility function, however, has little sensitivity to the dark force in the interval
where it has substantive support itself.
The amplitude of the polarization spectrum (determined by the interval between last
scatterings~\cite{Zaldarriaga:1995gi}) and the suppression of higher-frequency contributions to the
CMB from averaging over the visibility function are thus largely unchanged.

\begin{figure}[t!]
\begin{centering}
    \includegraphics[width=\textwidth]{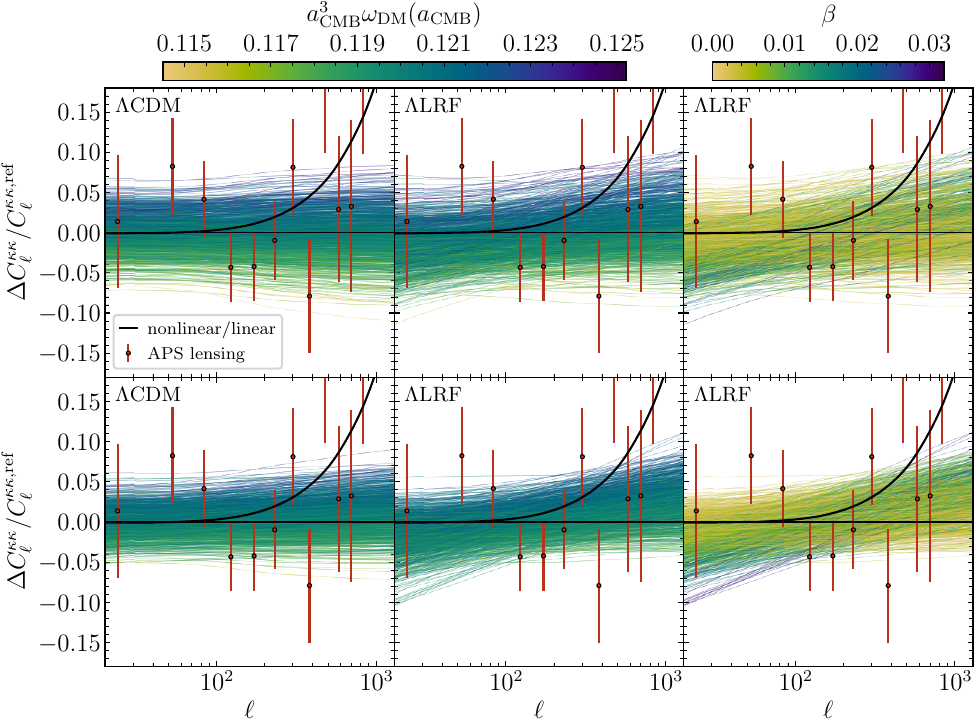}
    \caption{
        Residuals of the CMB lensing convergence relative to the \LCDM{} best fit (to all \Planck{}
        PR3 temperature and polarization data) for samples from posteriors for \LCDM{} (left) and for
        the dark force model (middle and right).
        Results are presented identically to \cref{fig:lensing-spaghetti} but using posteriors
        calibrated to temperature and polarization anisotropies from the full \Planck{} PR3 release
        (top) and from the combination of PR3, ACT DR6, and SPT-3G D1 (bottom).
    }
    \label{fig:lensing-spaghetti-extra}
\end{centering}
\end{figure}

\Cref{fig:lensing-spaghetti-extra} displays posteriors samples of CMB lensing spectra as shown in
\cref{fig:lensing-spaghetti} but including smaller-scale CMB anisotropy data.
Whereas \cref{fig:lensing-spaghetti} includes the subset of \Planck{} PR3 restricted to
$\ell \leq 1000$ in temperature and $\leq 600$ in polarization and temperature-polarization cross
correlation, \cref{fig:lensing-spaghetti-extra} displays results using all PR3 data and also the
previous PR3 subset combined with ACT DR6 and SPT-3G D1.
These two cases therefore include an increasing amount of lensing information via the impact on
two-point statistics, leading to narrower posterior distributions of lensing convergence power (as
is measured via higher-point statistics).
Adding small-scale data reduces the skew of the distributions in \cref{fig:lensing-spaghetti} toward
lower lensing power as calibrated by the large-scale CMB data (i.e., the scales $\ell \lesssim 1000$
that are insensitive to lensing).
\Cref{fig:lensing-spaghetti-extra} displays no clear preference for excess lensing, however, because
neither \LCDM{} nor the dark force model are capable of enhancing the amplitude of structure without
degrading the fit to the CMB on these effectively unlensed scales.

\section{Relic abundance of hyperlight mediators}\label{app:relic-abundance}

In this section, we solve for the evolution of a background mediator field $\bar{\varphi}$ under a
bare potential $V_\varphi(\varphi) = m_\varphi^2 \varphi^2 / 2$ with mass in the hyperlight range
$H_0 \simeq 10^{-33}~\mathrm{eV} < m_\varphi < H_\mathrm{eq} \simeq 10^{-28}~\mathrm{eV}$, i.e., one
that begins oscillating during matter domination.
In the matter era and at leading order in the coupling, the mediator evolves according to
\begin{align}
    \ddot{\bar{\varphi}}
    + \frac{2}{t}{\dot{\bar{\varphi}}}
    + m_\varphi^2 \bar{\varphi}
    &= - \frac{d_{m_\chi}^{(1)} \bar{\rho}_\chi}{2 \Mpl^2}
    \approx - \frac{2 d_{m_\chi}^{(1)} f_\chi}{3 t^2},
    \label{eqn:kg-matter}
\end{align}
where $t$ is the cosmic time coordinate (related to conformal time via $\ud t = a \ud \tau$),
dots denote $t$ derivatives, and $H = 2 / 3 t$ in matter domination.
A \emph{free} massive scalar displaced from its equilibrium point would begin to oscillate with
frequency $m_\varphi$ around $H \simeq m_\varphi$, acting as an ultralight dark matter condensate
whose energy density redshifts like matter ($\propto a^{-3}$)---the standard misalignment
mechanism~\cite{Marsh:2015xka,Hui:2016ltb,Preskill:1982cy,Dine:1982ah}.
At early times when $H \gg m_\varphi$, $\bar{\varphi}(t)$ is well approximated by the massless result
\cref{eqn:massless-mediator-matter-era-bg-soln}, while uncoupled scalars remain frozen at their
initial misalignment.
A scalar's coupling to matter therefore generates an effective misalignment even if its asymptotic
initial condition is zero.

Since the scalar's coupling is linear, \cref{eqn:kg-matter} may be solved exactly by Green function
methods.
The massless result \cref{eqn:massless-mediator-matter-era-bg-soln} (which holds in a
matter-radiation Universe) and the solution to \cref{eqn:kg-matter} (which holds in matter
domination) are simultaneously valid at times between equality and oscillations,
$m_\varphi^{-1} \gtrsim t \gtrsim H_\mathrm{eq}^{-1}$; we therefore match at a time $t_\times$ in
this interval.
The formal solution to \cref{eqn:kg-matter} is then
\begin{align}
    \bar{\varphi}(t)
    &= A(t_{\times}) \frac{\sin(m_\varphi t)}{m_\varphi t}
        + B(t_\times) \frac{\cos(m_\varphi t)}{m_\varphi t}
        - \frac{2 d_{m_\chi}^{(1)} f_\chi}{3}
        \int_{t_\times}^t \ud\tilde{t} \, \frac{G_{R}(t, \tilde{t})}{\tilde{t}^2}
    ,
\end{align}
where the retarded propagator $G_R(t,\tilde{t})$ of the differential operator
$\partial_t^2 + (2 / t) \partial_t + m_\varphi^2$ is
\begin{align}
    G_R(t,t')
    &= \frac{1}{m_\varphi} \frac{\tilde{t}}{t}
        \left[
            \sin(m_\varphi t) \cos(m_\varphi \tilde{t})
            -\cos(m_\varphi t) \sin(m_\varphi \tilde{t})
        \right]
    = \frac{1}{m_\varphi}\frac{\tilde{t}}{t}\sin\left(m_\varphi[t-\tilde{t}]\right)
\end{align}
and the matching coefficients are
$A(t_\times) = \bar{\varphi}(t_\times) + t_\times \dot{\bar{\varphi}}(t_\times)$
and $B(t_\times) = - m_\varphi t_\times^2 \dot{\bar{\varphi}}(t_\times)$ at leading order in small
$m_\varphi t_\times$.

The particular solution integrates to
\begin{align}
    m_\varphi t \int_{t_\times}^t \ud \tilde{t} \, \frac{G_{R}(t, \tilde{t})}{\tilde{t}^2}
    &= \left. \Ci(m_\varphi \tilde{t}) \right\vert_{t_\times}^t \sin(m_\varphi t)
        - \left. \Si(m_\varphi \tilde{t}) \right\vert_{t_\times}^t \cos(m_\varphi t)
    ,
\end{align}
where $\Ci(z)$ and $\Si(z)$ are the trigonometric integral functions.
Because $G_R(t, \tilde{t})$ is oscillatory at source times $\tilde{t} \gtrsim m_\varphi^{-1}$, the
monotonically decaying source becomes inefficient at driving the field, i.e., the integral defining
the particular solution converges for $m_\varphi \tilde{t} \rightarrow \infty$.
In other words, the scalar dynamically decouples from the background source after $H$ drops below
$m_\varphi$.
The late-time solution then reduces to
\begin{align}
    \bar{\varphi}(t)
    &\approx
        \frac{\sqrt{\omega_m} H_{100}}{m_\varphi a(t)^{3/2}}
        \left\{
            \left[
                \frac{3 A(t_\times)}{2}
                - d_{m_\chi}^{(1)} f_\chi
                \left. \Ci \right\vert^\infty_{t_\times}
            \right]
            \sin(m_\varphi t)
            +
            \left[
                \frac{3 B(t_\times)}{2}
                + d_{m_\chi}^{(1)} f_\chi
                \left. \Si \right\vert^\infty_{t_\times}
            \right]
            \cos(m_\varphi t)
        \right\}
    \label{eq:md_relic_solution}
\end{align}
where the asymptotic constants are
\begin{subequations}
\begin{align}
    \left. \Ci \right\vert^\infty_{t_\times}
    &\equiv - \left[ \gamma_\mathrm{E} + \ln(m_\varphi t_\times) \right]
    \\
    \left. \Si \right\vert^\infty_{t_\times}
    &= \frac{\pi}{2} - m_\varphi t_\times
\end{align}
\end{subequations}
with $\gamma_\mathrm{E}$ the Euler-Mascheroni constant.
Because the background varies slowly relative to $m_\varphi$ and we replaced
$m_\varphi t$ with $2 m_\varphi / 3 \sqrt{\bar{\rho}_m / 3 \Mpl^2}$, \cref{eq:md_relic_solution} extends as
written into the dark-energy--dominated era up to a phase shift of the oscillatory arguments.
The late-time ($H \gg m_\varphi$) energy density
$\bar{\rho}_\varphi / \Mpl^2 = \dot{\bar{\varphi}}^2 + m_\varphi^2 \bar{\varphi}^2$, averaged over
oscillations, is then proportional to the sum of the squared coefficients of the sine and cosine
components of \cref{eq:md_relic_solution}:
\begin{align}
    f_\varphi
    = \frac{\bar{\rho}_{\varphi}}{\bar{\rho}_m}
    &\approx \frac{1}{3}
        \left(
            \left[
                \frac{3 A(t_\times)}{2}
                - d_{m_\chi}^{(1)} f_\chi
                \left. \Ci \right\vert^\infty_{t_\times}
            \right]^2
            +
            \left[
                \frac{3 B(t_\times)}{2}
                + d_{m_\chi}^{(1)} f_\chi
                \left. \Si \right\vert^\infty_{t_\times}
            \right]^2
        \right)
    .
\end{align}
The scalar's relic abundance thus redshifts like matter and is manifestly a condensate of ultralight
relic scalar particles, a production mechanism sometimes referred to as a thermal (or in-medium)
misalignment mechanism.

The matching coefficients, derived from the limit $a \gg a_\mathrm{eq}$ of the massless result
\cref{eqn:massless-mediator-matter-era-bg-soln}, reduce to
\begin{subequations}
\begin{align}
    A(t_\times)
    &\approx
        \bar{\varphi}_i - d_{m_\chi}^{(1)} f_\chi
        \left(
            \ln \frac{a_\times}{4 a_\mathrm{eq} / e}
            + t_\times H_\times
        \right)
    \\
    B(t_{\times})
    &\approx d_{m_\chi}^{(1)} f_\chi
        \cdot m_\varphi t_\times
        \cdot t_\times H_\times
    .
\end{align}
\end{subequations}
For times $t_\times$ in matter domination,
$H_\times = \sqrt{\omega_m / a_\times^3} H_{100}$ and $t_\times H_\times = 2/3$.
Writing $H(t_\mathrm{eq}) = \sqrt{2 \omega_m / a_\mathrm{eq}^3} H_{100}$ then sets
$a_\times / a_\mathrm{eq} = \left( H_\mathrm{eq} / \sqrt{2} H_\times \right)^{2/3}$.
The scalar's relative contribution to the matter density is thus
\begin{align}
    f_\varphi
    &\approx \frac{\beta f_\chi^2}{3}
        \left[
            \left(
                \frac{3 \bar{\varphi}_i}{2 d_{m_\chi}^{(1)} f_\chi}
                + \ln \frac{m_\varphi}{H_\mathrm{eq}}
                + \ln \frac{16 \sqrt{2}}{3}
                + \gamma_\mathrm{E}
                - \frac{5}{2}
            \right)^2
            + \frac{\pi^2}{4}
        \right]
    ,
    \label{eqn:hyperlight-relic}
\end{align}
with the term in brackets dominated by the (squared) logarithm for
$m_\varphi \lesssim 10^{-2} H_\mathrm{eq}$.
This result is approximately $4/5$ times that presented in Ref.~\cite{Bottaro:2024pcb}, whose
analytic approximation instantaneously switched off the source at $H \simeq m_\varphi$, whereas the
Green function method retains its contribution through the onset of oscillations.
The logarithmic dependence of the relic energy density on the ratio of the mass
$m_\varphi < H_\mathrm{eq}$ and the Hubble scale $H_\mathrm{eq}$ is characteristic of in-medium
misalignment when the source term is proportional to the total density
$\bar{\rho} \propto H^2$~\cite{Cyncynates:2024bxw, Cyncynates:2024ufu}, as is the case here for a
coupling to (a subcomponent of) matter during the matter era.

\section{Implementation details}\label{app:numerics}

We implement the long-range force in \textsf{CLASS}~\cite{Blas:2011rf, Lesgourgues:2011re} as
described in \cref{app:eoms} in a similar fashion as Ref.~\cite{Archidiacono:2022iuu}, with some
refinements.
\textsf{CLASS}'s input parameter model is based on the present-day density (or fraction) in each
species and the Hubble constant, which is especially inconvenient in models (like those studied
here) for which abundances today are not analytically related to their values at very early times
(long before equality and last scattering).
Rather than adapt \textsf{CLASS}'s root-finding system to solve jointly for the early-time dark
matter density and the density in cosmological constant as a function of $h$, $\omega_\chi(a_0)$ or
$\Omega_\chi(a_0)$, the coupling, and the densities in other species, we instead remove this
functionality entirely and revise \textsf{CLASS} to accept as inputs only parameters that specify
initial conditions (or present-day abundances when trivially related, e.g., the baryon density); we
then compute the Hubble constant and various density fractions after solving the background
dynamics.
This strategy is much simpler to implement and requires only a single numerical integration of the
background equation (rather than the many required for numerical root finding).
Solving for, e.g., the dark energy density $\omega_\Lambda$ that fixes a particular value of
$\theta_s$ is then straightforward to implement in Python using methods from
\textsf{SciPy}~\cite{Virtanen:2019joe}.

One subtle implementation detail relates to an approximation made in
\textsf{HyRec}~\cite{Ali-Haimoud:2010hou, Lee:2020obi} to account for the cosmology dependence of
the Lyman-$\alpha$ net decay rate, which is computed as a Taylor expansion in (effectively) the
hydrogen-to-photon number density ratio, the density in CDM and baryons relative to the photon
number density, and the total radiation density relative to that in photons (i.e.,
$N_\mathrm{eff}$).
Current versions of \textsf{CLASS} set the value of the second expansion parameter in terms of the
present abundance of non--free-streaming matter, which, despite the semantic difference, serves as a
stand-in for species that are matterlike around recombination (e.g., it appropriately excludes
sufficiently light neutrinos).
Prior versions simply set it based on the baryon and CDM densities alone.
Since the $\chi$ fluid is implemented as a new species in \textsf{CLASS}, its density must be
manually accounted for in the correction.
Neglecting it nearly doubles the sensitivity of primary CMB anisotropies on small scales (i.e., as
quantified in \cref{fig:cmb-sensitivity}) to the LRF strength and leads to an unacceptably large
error in \LCDM{} as well.
An exercise similar to \cref{fig:cmb-sensitivity-undamped} shows that much of the error derives from
the propagation of the mismodeled correction to diffusion.

The correction function's dependence on the density in nonbaryonic components that are matterlike at
early times (and $N_\mathrm{eff}$) in reality serves as a stand-in for dependence on the background
expansion rate, since these species do not participate in recombination physics.
As such, computing the correction based on present-day densities is only strictly correct when the
matter components redshift as $a^{-3}$, leaving an order-$\beta$ error in the dark force model.
To be conservative, we therefore dynamically set the value of the correction expansion parameter
based on the instantaneous density in matterlike components, though the impact is smaller than, say,
the difference between results that use \textsf{HyRec} versus \textsf{RECFAST}.
The same concern is likely also relevant for scenarios featuring additional species at early times
(that are presumably not accounted for in the correction at all), such as early dark energy,
especially with recent data from ACT and SPT providing yet more precision deeper in the damping
tail.

In our analyses, we employ the 2018 \Planck{} (PR3) likelihoods via the foreground-marginalized
\texttt{Plik\_lite} variants with \texttt{Commander} and \texttt{SimAll} for
$\ell < 30$~\cite{Planck:2018vyg, Planck:2019nip, clipy}.
We also employ the foreground-marginalized CMB likelihood from ACT DR6~\cite{ACT:2025fju}, as
implemented in \textsf{candl}~\cite{Balkenhol:2024sbv}.
For practical purposes, we truncate the ACT likelihood at $\ell = 4000$, which matches that for
SPT-3G D1~\cite{SPT-3G:2025bzu} (which we also employ); as confirmed by explicit tests, there is
little cosmological information at higher multipoles to warrant the substantial increase in
computational cost of computing theoretical predictions up to $\ell \sim 8000$.
This choice also mitigates some of the sensitivity to (the lack of) nonlinear corrections to the
lensing of primary anisotropies.
When combining ACT or SPT data with \Planck{}, we follow Refs.~\cite{ACT:2025fju, SPT-3G:2025bzu} in
truncating the \Planck{} likelihoods to $\ell \leq 1000$ in temperature and $\ell \leq 600$ in
polarization and temperature-polarization cross correlation.

The numerical precision settings optimized for \Planck{} analyses with \textsf{CLASS} and
\textsf{CAMB}~\cite{Lewis:1999bs} are insufficient for these datasets; Refs.~\cite{AtacamaCosmologyTelescope:2025nti,
SPT-3G:2025bzu} employed recommendations from Ref.~\cite{Hill:2021yec, McCarthy:2021lfp,
Bolliet:2023sst, Jense:2024llt} that were derived without attempting to optimize computational cost
at a fixed target accuracy for likelihood evaluations.
Given that runtime can increase by upwards of an order of magnitude under these precision settings
(especially with massive neutrinos or scalar fields), we derived through iterative testing a set of
reduced precision parameters that remain suitable for these datasets.
(Further refinement is surely possible.)
For reference likelihood evaluations, we use precision settings similar to (or exceeding) those in
Refs.~\cite{AtacamaCosmologyTelescope:2025nti, SPT-3G:2025bzu}.
While we experiment on a single set of cosmological parameters at a time, seeking precision in log
likelihoods for the above experiments of order $0.1$ or better, such a test is not sufficient as the
Boltzmann codes exhibit nonnegligible variability in error with cosmological parameters.
We test the robustness of our choices for a sample of $1000$ parameter sets from posteriors
calibrated to \Planck{} data (merely to provide a sample representative of currently viable models)
that sample over the neutrino mass sum in addition to the standard six \LCDM{} parameters.
We also check cases that include dark forces mediated by a massless scalar.
We seek error distributions comparable to those for the default (i.e., \Planck{}-targeted) precision
settings for the \Planck{} likelihoods, i.e., $68\%$ and $95\%$ of samples having absolute
difference in log likelihoods smaller than $\sim 0.4$ and $\sim 0.7$, respectively.
We also check that errors are not significantly correlated with any cosmological parameter, which is
likely the most important practical requirement.

The \textsf{CLASS} precision parameters we find sufficient for the above requirements are
\begin{lstlisting}
minimal_act_spt_prec_class = dict(
    lmax=5000,
    l_linstep=30,
    transfer_neglect_delta_k_S_t1=0.064,
    transfer_neglect_delta_k_S_e=0.15,
    ur_fluid_trigger_tau_over_tau_k=55,
    radiation_streaming_trigger_tau_c_over_tau=50,
    radiation_streaming_trigger_tau_over_tau_k=60,
    l_max_g=16,
    l_max_ur=25,
    tol_ncdm_synchronous=1e-4,
)
\end{lstlisting}
For each of the ACT, SPT, and (truncated) \Planck{} likelihoods, $95\%$ of samples have absolute
log-likelihood differences smaller than $0.2$-$0.3$, and the distribution of their summed log
likelihood matches the aforementioned target.
In general we find that setting \texttt{lmax} to be $1000$ greater than the cutoff in the likelihood
is sufficient when truncating the ACT DR6 likelihood at higher multipoles.
The parameters beginning with \texttt{transfer\_neglect} determine the interval in wave number above
$k = \ell / D_{M, \star}$ to integrate the transfer functions for particular CMB source terms.
They are specified in units of $\mathrm{Mpc}^{-1}$ in \textsf{CLASS}; in order to make the choices
agnostic to the distance to last scattering (e.g., if the sound horizon is modified) we also
modified the handling of these parameters so that they are specified in units of $1 / D_{M, \star}$.
Finally, for current versions of \textsf{CAMB}~\cite{Lewis:1999bs} we find setting \texttt{kmax} to
$5$ and \texttt{lens\_potential\_accuracy} to $3$ to be sufficient, although we tested
\textsf{CLASS} more extensively given \textsf{CAMB}'s generally better performance.

\bibliography{references,manual}

\end{document}